\documentclass[a4paper,11pt]{article}
\pdfoutput=1

\usepackage{jcappub}
\usepackage[utf8]{inputenc}
\usepackage[T1]{fontenc}
\usepackage{bm}
\usepackage{tabularx}
\usepackage{gensymb}
\usepackage{textcomp}
\usepackage{multirow}
\usepackage{geometry}
\usepackage{pdflscape}
\usepackage{arydshln}

\newcommand{\sv}{\langle \sigma \mathit{v} \rangle}
\newcommand\CellTopTwo{\rule{0pt}{2.8ex}}
\newcolumntype{?}{!{\vrule width 2pt}}

\title{\boldmath Combined dark matter search towards dwarf spheroidal galaxies with \textit{Fermi}-LAT, HAWC, H.E.S.S., MAGIC, and VERITAS}

\collaboration{The \textit{Fermi}-LAT, HAWC, H.E.S.S., MAGIC, and VERITAS Collaborations}

\author[1]{S.~Abdollahi,}
\author[2]{L.~Baldini,}
\author[3]{R.~Bellazzini,}
\author[4]{B.~Berenji,}
\author[5,6]{E.~Bissaldi,}
\author[7,8]{R.~Bonino,}
\author[9]{P.~Bruel,}
\author[10]{S.~Buson,}
\author[11,*]{E.~Charles,}
\author[12]{A.~W.~Chen,}
\author[13,14]{S.~Ciprini,}
\author[15,16]{M.~Crnogorcevic,}
\author[7,8]{A.~Cuoco,}
\author[17]{F.~D'Ammando,}
\author[18]{A.~de~Angelis,}
\author[7,*]{M.~Di~Mauro,}
\author[11]{N.~Di~Lalla,}
\author[6]{L.~Di~Venere,}
\author[19]{A.~Dom\'inguez,}
\author[9]{S.~J.~Fegan,}
\author[2]{A.~Fiori,}
\author[5,6]{P.~Fusco,}
\author[20,21]{V.~Gammaldi,}
\author[6]{F.~Gargano,}
\author[13,14]{D.~Gasparrini,}
\author[13,14]{F.~Giacchino,}
\author[5,6]{N.~Giglietto,}
\author[6,5]{M.~Giliberti,}
\author[5,6]{F.~Giordano,}
\author[17]{M.~Giroletti,}
\author[22]{I.~A.~Grenier,}
\author[23,16]{S.~Guiriec,}
\author[24]{M.~Gustafsson,}
\author[16]{E.~Hays,}
\author[25]{J.W.~Hewitt,}
\author[9]{D.~Horan,}
\author[26]{H.~Katagiri,}
\author[3]{M.~Kuss,}
\author[27,28]{J.~Li,}
\author[29,30]{F.~Longo,}
\author[5,6]{F.~Loparco,}
\author[5,6]{L.~Lorusso,}
\author[31]{G.~Mart\'i-Devesa,}
\author[6]{M.~N.~Mazziotta,}
\author[16,15]{J.~E.~McEnery,}
\author[32,33]{I.~Mereu,}
\author[34]{M.~Meyer,}
\author[11]{P.~F.~Michelson,}
\author[16,35]{N.~Mirabal,}
\author[36]{W.~Mitthumsiri,}
\author[37]{T.~Mizuno,}
\author[13]{A.~Morselli,}
\author[11]{I.~V.~Moskalenko,}
\author[35,16]{M.~Negro,}
\author[11]{N.~Omodei,}
\author[17]{M.~Orienti,}
\author[38,11]{E.~Orlando,}
\author[5,6]{G.~Panzarini,}
\author[30,39]{M.~Persic,}
\author[3]{M.~Pesce-Rollins,}
\author[5,6]{R.~Pillera,}
\author[11]{T.~A.~Porter,}
\author[29,30,17]{G.~Principe,}
\author[5,6]{S.~Rain\`o,}
\author[40,41,42]{R.~Rando,}
\author[2]{M.~Razzano,}
\author[31]{O.~Reimer,}
\author[21,20]{M.~S\'anchez-Conde,}
\author[43]{P.~M.~Saz~Parkinson,}
\author[6]{D.~Serini,}
\author[44]{D.~J.~Suson,}
\author[45,46]{D.~F.~Torres,}
\author[47]{G.~Zaharijas}
\author{(the~\textit{Fermi}-LAT~Collaboration)}

\affiliation[1]{IRAP, Universit\'e de Toulouse, CNRS, UPS, CNES, F-31028 Toulouse, France}
\affiliation[2]{Universit\`a di Pisa and Istituto Nazionale di Fisica Nucleare, Sezione di Pisa I-56127 Pisa, Italy}
\affiliation[3]{Istituto Nazionale di Fisica Nucleare, Sezione di Pisa, I-56127 Pisa, Italy}
\affiliation[4]{California State University, Los Angeles, Department of Physics and Astronomy, Los Angeles, CA 90032, USA}
\affiliation[5]{Dipartimento di Fisica ``M. Merlin" dell'Universit\`a e del Politecnico di Bari, via Amendola 173, I-70126 Bari, Italy}
\affiliation[6]{Istituto Nazionale di Fisica Nucleare, Sezione di Bari, I-70126 Bari, Italy}
\affiliation[7]{Istituto Nazionale di Fisica Nucleare, Sezione di Torino, I-10125 Torino, Italy}
\affiliation[8]{Dipartimento di Fisica, Universit\`a degli Studi di Torino, I-10125 Torino, Italy}
\affiliation[9]{Laboratoire Leprince-Ringuet, CNRS/IN2P3, \'Ecole polytechnique, Institut Polytechnique de Paris, 91120 Palaiseau, France}
\affiliation[10]{Institut f\"ur Theoretische Physik and Astrophysik, Universit\"at W\"urzburg, D-97074 W\"urzburg, Germany}
\affiliation[11]{W. W. Hansen Experimental Physics Laboratory, Kavli Institute for Particle Astrophysics and Cosmology, Department of Physics and SLAC National Accelerator Laboratory, Stanford University, Stanford, CA 94305, USA}
\affiliation[12]{School of Physics, University of the Witwatersrand, Private Bag 3, WITS-2050, Johannesburg, South Africa}
\affiliation[13]{Istituto Nazionale di Fisica Nucleare, Sezione di Roma ``Tor Vergata", I-00133 Roma, Italy}
\affiliation[14]{Space Science Data Center - Agenzia Spaziale Italiana, Via del Politecnico, snc, I-00133, Roma, Italy}
\affiliation[15]{Department of Astronomy, University of Maryland, College Park, MD 20742, USA}
\affiliation[16]{NASA Goddard Space Flight Center, Greenbelt, MD 20771, USA}
\affiliation[17]{INAF Istituto di Radioastronomia, I-40129 Bologna, Italy}
\affiliation[18]{Dipartimento di Fisica, Universit\`a di Udine and Istituto Nazionale di Fisica Nucleare, Sezione di Trieste, Gruppo Collegato di Udine, I-33100 Udine}
\affiliation[19]{Grupo de Altas Energ\'ias, Universidad Complutense de Madrid, E-28040 Madrid, Spain}
\affiliation[20]{Departamento de F\'isica Te\'orica, Universidad Aut\'onoma de Madrid, 28049 Madrid, Spain}
\affiliation[21]{Instituto de F\'isica Te\'orica UAM/CSIC, Universidad Aut\'onoma de Madrid, E-28049 Madrid, Spain}
\affiliation[22]{Universit\'e Paris Cit\'e, Universit\'e Paris-Saclay, CEA, CNRS, AIM, F-91191 Gif-sur-Yvette, France}
\affiliation[23]{The George Washington University, Department of Physics, 725 21st St, NW, Washington, DC 20052, USA}
\affiliation[24]{Georg-August University G\"ottingen, Institute for theoretical Physics - Faculty of Physics, Friedrich-Hund-Platz 1, D-37077 G\"ottingen, Germany}
\affiliation[25]{University of North Florida, Department of Physics, 1 UNF Drive, Jacksonville, FL 32224 , USA}
\affiliation[26]{College of Science, Ibaraki University, 2-1-1, Bunkyo, Mito 310-8512, Japan}
\affiliation[27]{CAS Key Laboratory for Research in Galaxies and Cosmology, Department of Astronomy, University of Science and Technology of China, Hefei 230026, People's Republic of China}
\affiliation[28]{School of Astronomy and Space Science, University of Science and Technology of China, Hefei 230026, People's Republic of China}
\affiliation[29]{Dipartimento di Fisica, Universit\`a di Trieste, I-34127 Trieste, Italy}
\affiliation[30]{Istituto Nazionale di Fisica Nucleare, Sezione di Trieste, I-34127 Trieste, Italy}
\affiliation[31]{Institut f\"ur Astro- und Teilchenphysik, Leopold-Franzens-Universit\"at Innsbruck, A-6020 Innsbruck, Austria}
\affiliation[32]{Istituto Nazionale di Fisica Nucleare, Sezione di Perugia, I-06123 Perugia, Italy}
\affiliation[33]{Dipartimento di Fisica, Universit\`a degli Studi di Perugia, I-06123 Perugia, Italy}
\affiliation[34]{Center for Cosmology and Particle Physics Phenomenology, University of Southern Denmark, Campusvej 55, DK-5230 Odense M, Denmark}
\affiliation[35]{Department of Physics and Center for Space Sciences and Technology, University of Maryland Baltimore County, Baltimore, MD 21250, USA}
\affiliation[36]{Department of Physics, Faculty of Science, Mahidol University, Bangkok 10400, Thailand}
\affiliation[37]{Hiroshima Astrophysical Science Center, Hiroshima University, Higashi-Hiroshima, Hiroshima 739-8526, Japan}
\affiliation[38]{Istituto Nazionale di Fisica Nucleare, Sezione di Trieste, and Universit\`a di Trieste, I-34127 Trieste, Italy}
\affiliation[39]{INAF-Astronomical Observatory of Padova, Vicolo dell'Osservatorio 5, I-35122 Padova, Italy}
\affiliation[40]{Dipartimento di Fisica e Astronomia ``G. Galilei'', Universit\`a di Padova, Via F. Marzolo, 8, I-35131 Padova, Italy}
\affiliation[41]{Istituto Nazionale di Fisica Nucleare, Sezione di Padova, I-35131 Padova, Italy}
\affiliation[42]{Center for Space Studies and Activities ``G. Colombo", University of Padova, Via Venezia 15, I-35131 Padova, Italy}
\affiliation[43]{Santa Cruz Institute for Particle Physics, Department of Physics and Department of Astronomy and Astrophysics, University of California at Santa Cruz, Santa Cruz, CA 95064, USA}
\affiliation[44]{Purdue University Northwest, Hammond, IN 46323, USA}
\affiliation[45]{Institute of Space Sciences (ICE, CSIC), Campus UAB, Carrer de Magrans s/n, E-08193 Barcelona, Spain; and Institut d'Estudis Espacials de Catalunya (IEEC), E-08034 Barcelona, Spain}
\affiliation[46]{Instituci\'o Catalana de Recerca i Estudis Avan\c{c}ats (ICREA), E-08010 Barcelona, Spain}
\affiliation[47]{Center for Astrophysics and Cosmology, University of Nova Gorica, Nova Gorica, Slovenia}

\author[48]{A.~Albert,}
\author[49]{R.~Alfaro,}
\author[50]{C.~Alvarez,}
\author[51]{J.C.~Arteaga-Velázquez,}
\author[49]{D.~Avila Rojas,}
\author[52]{H.A.~Ayala Solares,}
\author[53]{R.~Babu,}
\author[49]{E.~Belmont-Moreno,}
\author[50]{K.S.~Caballero-Mora,}
\author[54]{T.~Capistrán,}
\author[55]{A.~Carramiñana,}
\author[56]{S.~Casanova,}
\author[57]{O.~Chaparro-Amaro,}
\author[51]{U.~Cotti,}
\author[58]{J.~Cotzomi,}
\author[59]{S.~Coutiño de León,}
\author[60]{E.~de la Fuente,}
\author[51]{C.~de León,}
\author[55]{R.~Diaz Hernandez,}
\author[48]{B.L.~Dingus,}
\author[59]{M.A.~DuVernois,}
\author[48]{M.~Durocher,}
\author[60]{J.C.~Díaz-Vélez,}
\author[61]{K.~Engel,}
\author[49]{C.~Espinoza,}
\author[61]{K.L.~Fan,}
\author[54]{N.~Fraija,}
\author[62]{J.A.~García-González,}
\author[54]{F.~Garfias,}
\author[54]{M.M.~González,}
\author[61]{J.A.~Goodman,}
\author[48,*]{J.P.~Harding,}
\author[49]{S.~Hernandez,}
\author[53]{I.~Herzog,}
\author[63]{J.~Hinton,}
\author[61]{D.~Huang,}
\author[50]{F.~Hueyotl-Zahuantitla,}
\author[64]{P.~Hüntemeyer,}
\author[54]{A.~Iriarte,}
\author[65]{V.~Joshi,}
\author[66]{S.~Kaufmann,}
\author[67]{D.~Kieda,}
\author[48]{G.J.~Kunde,}
\author[68]{A.~Lara,}
\author[69]{J.~Lee,}
\author[49]{H.~León Vargas,}
\author[53]{J.T.~Linnemann,}
\author[54]{A.L.~Longinotti,}
\author[66]{G.~Luis-Raya,}
\author[53]{J.~Lundeen,}
\author[48]{K.~Malone,}
\author[58]{O.~Martinez,}
\author[57]{J.~Martínez-Castro,}
\author[70]{H.~Martínez-Huerta,}
\author[71]{J.A.~Matthews,}
\author[72]{P.~Miranda-Romagnoli,}
\author[51]{J.A.~Morales-Soto,}
\author[58]{E.~Moreno,}
\author[52]{M.~Mostafá,}
\author[56]{A.~Nayerhoda,}
\author[73]{L.~Nellen,}
\author[53]{M.U.~Nisa,}
\author[72]{R.~Noriega-Papaqui,}
\author[63]{L.~Olivera-Nieto,}
\author[74]{N.~Omodei,}
\author[53]{A.~Peisker,}
\author[54]{Y.~Pérez Araujo,}
\author[66]{E.G.~Pérez-Pérez,}
\author[69]{C.D.~Rho,}
\author[55]{D.~Rosa-González,}
\author[58]{H.~Salazar,}
\author[53,*]{D.~Salazar-Gallegos,}
\author[49]{A.~Sandoval,}
\author[61]{M.~Schneider,}
\author[49]{J.~Serna-Franco,}
\author[61]{A.J.~Smith,}
\author[69]{Y.~Son,}
\author[67]{R.W.~Springer,}
\author[66]{O.~Tibolla,}
\author[53,*]{K.~Tollefson,}
\author[55]{I.~Torres,}
\author[75]{R.~Torres-Escobedo,}
\author[64]{R.~Turner,}
\author[55]{F.~Ureña-Mena,}
\author[58]{E.~Varela,}
\author[58]{L.~Villaseñor,}
\author[64]{X.~Wang,}
\author[69]{I.J.~Watson,}
\author[52]{K.~Whitaker,}
\author[61]{E.~Willox,}
\author[52]{S.~Yu,}
\author[61]{S.~Yun-Cárcamo,}
\author[75]{H.~Zhou}
\author{(the~HAWC~Collaboration)}

\affiliation[48]{Los Alamos National Laboratory, Los Alamos, NM, USA}
\affiliation[49]{Instituto de F\'{i}sica, Universidad Nacional Autónoma de México, Ciudad de Mexico, Mexico}
\affiliation[50]{Universidad Autónoma de Chiapas, Tuxtla Gutiérrez, Chiapas, México}
\affiliation[51]{Universidad Michoacana de San Nicolás de Hidalgo, Morelia, Mexico}
\affiliation[52]{Department of Physics, Pennsylvania State University, University Park, PA, USA}
\affiliation[53]{Department of Physics and Astronomy, Michigan State University, East Lansing, MI, USA}
\affiliation[54]{Instituto de Astronom\'{i}a, Universidad Nacional Autónoma de México, Ciudad de Mexico, Mexico}
\affiliation[55]{Instituto Nacional de Astrof\'{i}sica, Óptica y Electrónica, Puebla, Mexico}
\affiliation[56]{Institute of Nuclear Physics Polish Academy of Sciences, PL-31342 IFJ-PAN, Krakow, Poland}
\affiliation[57]{Centro de Investigaci\'on en Computaci\'on, Instituto Polit\'ecnico Nacional, M\'exico City, M\'exico}
\affiliation[58]{Facultad de Ciencias F\'{i}sico Matemáticas, Benemérita Universidad Autónoma de Puebla, Puebla, Mexico}
\affiliation[59]{Department of Physics, University of Wisconsin-Madison, Madison, WI, USA}
\affiliation[60]{Departamento de F\'{i}sica, Centro Universitario de Ciencias Exactase Ingenierias, Universidad de Guadalajara, Guadalajara, Mexico}
\affiliation[61]{Department of Physics, University of Maryland, College Park, MD, USA}
\affiliation[62]{Tecnologico de Monterrey, Escuela de Ingenier\'{i}a y Ciencias, Ave. Eugenio Garza Sada 2501, Monterrey, N.L., Mexico, 64849}
\affiliation[63]{Max-Planck Institute for Nuclear Physics, 69117 Heidelberg, Germany}
\affiliation[64]{Department of Physics, Michigan Technological University, Houghton, MI, USA}
\affiliation[65]{Erlangen Centre for Astroparticle Physics, Friedrich-Alexander-Universit\"at Erlangen-N\"urnberg, Erlangen, Germany}
\affiliation[66]{Universidad Politecnica de Pachuca, Pachuca, Hgo, Mexico}
\affiliation[67]{Department of Physics and Astronomy, University of Utah, Salt Lake City, UT, USA}
\affiliation[68]{Instituto de Geof\'{i}sica, Universidad Nacional Autónoma de México, Ciudad de Mexico, Mexico}
\affiliation[69]{University of Seoul, Seoul, Rep. of Korea}
\affiliation[70]{Departamento de Física y Matemáticas, Universidad de Monterrey, Av. Morones Prieto 4500, San Pedro Garza García 66238, Nuevo León, Mexico}
\affiliation[71]{Dept of Physics and Astronomy, University of New Mexico, Albuquerque, NM, USA}
\affiliation[72]{Universidad Autónoma del Estado de Hidalgo, Pachuca, Mexico}
\affiliation[73]{Instituto de Ciencias Nucleares, Universidad Nacional Autónoma de Mexico, Ciudad de Mexico, Mexico}
\affiliation[74]{Department of Physics, Stanford University: Stanford, CA 94305–4060, USA}
\affiliation[75]{Tsung-Dao Lee Institute and School of Physics and Astronomy, Shanghai Jiao Tong University, Shanghai, China}

\author[76,77]{F.~Aharonian,}
\author[78]{F.~Ait~Benkhali,}
\author[79,*]{C. Armand,}
\author[80]{J.~Aschersleben,}
\author[81,82]{M.~Backes,}
\author[83]{V.~Barbosa~Martins,}
\author[84]{R.~Batzofin,}
\author[85,86]{Y.~Becherini,}
\author[86,87]{D.~Berge,}
\author[88]{B.~Bi,}
% ...
\author{(the~H.E.S.S.~Collaboration)}

\affiliation[76]{Dublin Institute for Advanced Studies, 31 Fitzwilliam Place, Dublin 2, Ireland}
\affiliation[77]{Max-Planck-Institut f\"ur Kernphysik, P.O. Box 103980, D 69029 Heidelberg, Germany}
\affiliation[78]{Landessternwarte, Universit\"at Heidelberg, K\"onigstuhl, D 69117 Heidelberg, Germany}
\affiliation[79]{Université Savoie Mont Blanc, CNRS, Laboratoire d'Annecy de Physique des Particules - IN2P3, 74000 Annecy, France}
\affiliation[80]{Kapteyn Astronomical Institute, University of Groningen, Landleven 12, 9747 AD Groningen, The Netherlands}
\affiliation[81]{University of Namibia, Department of Physics, Private Bag 13301, Windhoek 10005, Namibia}
\affiliation[82]{Centre for Space Research, North-West University, Potchefstroom 2520, South Africa}
\affiliation[83]{DESY, D-15738 Zeuthen, Germany}
\affiliation[84]{Institut f\"ur Physik und Astronomie, Universit\"at Potsdam, Karl-Liebknecht-Strasse 24/25, D 14476 Potsdam, Germany}
\affiliation[85]{Université de Paris, CNRS, Astroparticule et Cosmologie, F-75013 Paris, France}
\affiliation[86]{Department of Physics and Electrical Engineering, Linnaeus University, 351 95 V\"axj\"o, Sweden}
\affiliation[87]{Institut f\"ur Physik, Humboldt-Universit\"at zu Berlin, Newtonstr. 15, D 12489 Berlin, Germany}
\affiliation[88]{Institut f\"ur Astronomie und Astrophysik, Universit\"at T\"ubingen, Sand 1, D 72076 T\"ubingen, Germany}
% ...

\author[112]{H.~Abe,}
\author[112]{S.~Abe,}
\author[113]{V.~A.~Acciari,}
\author[114]{I.~Agudo,}
\author[115]{T.~Aniello,}
\author[116,154]{S.~Ansoldi,}
\author[115]{L.~A.~Antonelli,}
\author[117]{A.~Arbet Engels,}
\author[118]{C.~Arcaro,}
\author[119]{M.~Artero,}
% ...
\author{(the~MAGIC~Collaboration)}

\affiliation[112]{Japanese MAGIC Group: Institute for Cosmic Ray Research (ICRR), The University of Tokyo, Kashiwa, 277-8582 Chiba, Japan}
\affiliation[113]{Instituto de Astrof\'isica de Canarias and Dpto. de Astrof\'isica, Universidad de La Laguna, E-38200, La Laguna, Tenerife, Spain}
\affiliation[114]{Instituto de Astrof\'isica de Andaluc\'ia-CSIC, Glorieta de la Astronom\'ia s/n, 18008, Granada, Spain}
\affiliation[115]{National Institute for Astrophysics (INAF), I-00136 Rome, Italy}
\affiliation[116]{Universit\`a di Udine and INFN Trieste, I-33100 Udine, Italy}
\affiliation[117]{Max-Planck-Institut f\"ur Physik, D-80805 M\"unchen, Germany}
\affiliation[118]{Universit\`a di Padova and INFN, I-35131 Padova, Italy}
\affiliation[119]{Institut de F\'isica d'Altes Energies (IFAE), The Barcelona Institute of Science and Technology (BIST), E-08193 Bellaterra (Barcelona), Spain}
% ...
\affiliation[154]{also at International Center for Relativistic Astrophysics (ICRA), Rome, Italy}
% ...

\author[163]{A.~Acharyya,}
\author[164]{C.~B.~Adams,}
\author[165]{A.~Archer,}
\author[166]{P.~Bangale,}
\author[167]{J.~T.~Bartkoske,}
\author[168]{P.~Batista,}
\author[169]{W.~Benbow,}
\author[170]{J.~H.~Buckley,}
\author[171]{Y.~Chen,}
\author[172]{J.~L.~Christiansen,}
\author[169]{A.~J.~Chromey,}
\author[170]{M.~Errando,}
\author[173]{M.~Escobar~Godoy,}
\author[174]{A.~Falcone,}
\author[171]{S.~Feldman,}
\author[167]{Q.~Feng,}
\author[175]{J.~P.~Finley,}
\author[166]{G.~M.~Foote,}
\author[176]{L.~Fortson,}
\author[173]{A.~Furniss,}
\author[177]{G.~Gallagher,}
\author[168,*]{C.~Giuri,}
\author[169]{W.~Hanlon,}
\author[173]{O.~Hervet,}
\author[169,178]{C.~E.~Hinrichs,}
\author[173]{J.~Hoang,}
\author[166]{J.~Holder,}
\author[170]{Z.~Hughes,}
\author[179,180]{T.~B.~Humensky,}
\author[171]{W.~Jin,}
\author[173]{M.~N.~Johnson,}
\author[181]{P.~Kaaret,}
\author[165]{M.~Kertzman,}
\author[182]{M.~Kherlakian,}
\author[167]{D.~Kieda,}
\author[168]{T.~K.~Kleiner,}
\author[166]{N.~Korzoun,}
\author[183]{F.~Krennrich,}
\author[179]{S.~Kumar,}
\author[184]{M.~Lundy,}
\author[168]{G.~Maier,}
\author[185]{C.~E~McGrath,}
\author[181]{M.~J.~Millard,}
\author[177]{J.~Millis,}
\author[166]{C.~L.~Mooney,}
\author[186]{P.~Moriarty,}
\author[187]{R.~Mukherjee,}
\author[188]{D.~Nieto,}
\author[184,189]{S.~O'Brien,}
\author[171]{R.~A.~Ong,}
\author[190,168]{M.~Pohl,}
\author[182,*]{E.~Pueschel,}
\author[185]{J.~Quinn,}
\author[170]{P.~L.~Rabinowitz,}
\author[184]{K.~Ragan,}
\author[191]{P.~T.~Reynolds,}
\author[176]{D.~Ribeiro,}
\author[169]{E.~Roache,}
\author[171]{J.~L.~Ryan,}
\author[168]{I.~Sadeh,}
\author[169]{L.~Saha,}
\author[192]{M.~Santander,}
\author[175]{G.~H.~Sembroski,}
\author[187]{R.~Shang,}
\author[173]{M.~Splettstoesser,}
\author[193]{D.~Tak,}
\author[176]{A.~K.~Talluri,}
\author[194]{J.~V.~Tucci,}
\author[171]{V.~V.~Vassiliev,}
\author[183]{A.~Weinstein,}
\author[173]{D.~A.~Williams,}
\author[184]{S.~L.~Wong}
\author{(the~VERITAS~Collaboration)}

\affiliation[163]{CP3-Origins, University of Southern Denmark, Campusvej 55, 5230 Odense M, Denmark}
\affiliation[164]{Physics Department, Columbia University, New York, NY 10027, USA}
\affiliation[165]{Department of Physics and Astronomy, DePauw University, Greencastle, IN 46135-0037, USA}
\affiliation[166]{Department of Physics and Astronomy and the Bartol Research Institute, University of Delaware, Newark, DE 19716, USA}
\affiliation[167]{Department of Physics and Astronomy, University of Utah, Salt Lake City, UT 84112, USA}
\affiliation[168]{DESY, Platanenallee 6, 15738 Zeuthen, Germany}
\affiliation[169]{Center for Astrophysics $|$ Harvard \& Smithsonian, Cambridge, MA 02138, USA}
\affiliation[170]{Department of Physics, Washington University, St. Louis, MO 63130, USA}
\affiliation[171]{Department of Physics and Astronomy, University of California, Los Angeles, CA 90095, USA}
\affiliation[172]{Physics Department, California Polytechnic State University, San Luis Obispo, CA 94307, USA}
\affiliation[173]{Santa Cruz Institute for Particle Physics and Department of Physics, University of California, Santa Cruz, CA 95064, USA}
\affiliation[174]{Department of Astronomy and Astrophysics, 525 Davey Lab, Pennsylvania State University, University Park, PA 16802, USA}
\affiliation[175]{Department of Physics and Astronomy, Purdue University, West Lafayette, IN 47907, USA}
\affiliation[176]{School of Physics and Astronomy, University of Minnesota, Minneapolis, MN 55455, USA}
\affiliation[177]{Department of Physics and Astronomy, Ball State University, Muncie, IN 47306, USA}
\affiliation[178]{Department of Physics and Astronomy, Dartmouth College, 6127 Wilder Laboratory, Hanover, NH 03755 USA}
\affiliation[179]{Department of Physics, University of Maryland, College Park, MD, USA}
\affiliation[180]{NASA GSFC, Greenbelt, MD 20771, USA}
\affiliation[181]{Department of Physics and Astronomy, University of Iowa, Van Allen Hall, Iowa City, IA 52242, USA}
\affiliation[182]{Fakult\"at f\"ur Physik \& Astronomie, Ruhr-Universit\"at Bochum, D-44780 Bochum, Germany}
\affiliation[183]{Department of Physics and Astronomy, Iowa State University, Ames, IA 50011, USA}
\affiliation[184]{Physics Department, McGill University, Montreal, QC H3A 2T8, Canada}
\affiliation[185]{School of Physics, University College Dublin, Belfield, Dublin 4, Ireland}
\affiliation[186]{School of Natural Sciences, University of Galway, University Road, Galway, H91 TK33, Ireland}
\affiliation[187]{Department of Physics and Astronomy, Barnard College, Columbia University, NY 10027, USA}
\affiliation[188]{Institute of Particle and Cosmos Physics, Universidad Complutense de Madrid, 28040 Madrid, Spain}
\affiliation[189]{Arthur B. McDonald Canadian Astroparticle Physics Research Institute, 64 Bader Lane, Queen$'$s University, Kingston, ON Canada, K7L 3N6}
\affiliation[190]{Institute of Physics and Astronomy, University of Potsdam, 14476 Potsdam-Golm, Germany}
\affiliation[191]{Department of Physical Sciences, Munster Technological University, Bishopstown, Cork, T12 P928, Ireland}
\affiliation[192]{Department of Physics and Astronomy, University of Alabama, Tuscaloosa, AL 35487, USA}
\affiliation[193]{SNU Astronomy Research Center, Seoul National University, Seoul 08826, Republic of Korea}
\affiliation[194]{Department of Physics, Indiana University-Purdue University Indianapolis, Indianapolis, IN 46202, USA}

\affiliation[*]{Corresponding authors: C. Armand, E. Charles, M. di Mauro, C. Giuri, J. P. Harding, D. Kerszberg, T. Miener, E. Moulin, L. Oakes, V. Poireau, E. Pueschel, J. Rico, L. Rinchiuso, D. Salazar-Gallegos, K. Tollefson.}

\emailAdd{echarles@slac.stanford.edu}
\emailAdd{dimauro.mattia@gmail.com}
\emailAdd{jpharding@lanl.gov}
\emailAdd{salaza82@msu.edu}
\emailAdd{tollefs2@msu.edu}
\emailAdd{contact.hess@hess-experiment.eu}
\emailAdd{contact.magic@mpp.mpg.de}
\emailAdd{chiara.giuri@desy.de}
\emailAdd{elisa.pueschel@astro.rub.de}

\abstract{Dwarf spheroidal galaxies (dSphs) are excellent targets for indirect dark matter (DM) searches using gamma-ray telescopes because they are thought to have high DM content and a low astrophysical background. The sensitivity of these searches is improved by combining the observations of dSphs made by different gamma-ray telescopes. We present the results of a combined search by the most sensitive currently operating gamma-ray telescopes, namely: the satellite-borne \textit{Fermi}-LAT telescope; the ground-based imaging atmospheric Cherenkov telescope arrays H.E.S.S., MAGIC, and VERITAS; and the HAWC water Cherenkov detector. Individual datasets were analyzed using a common statistical approach. Results were subsequently combined via a global joint likelihood analysis. We obtain constraints on the velocity-weighted cross section $\sv$ for DM self-annihilation as a function of the DM particle mass. This five-instrument combination allows the derivation of up to 2-3 times more constraining upper limits on $\sv$ than the individual results over a wide mass range spanning from 5~GeV to 100~TeV. Depending on the DM content modeling, the 95\% confidence level observed limits reach $1.5\times$10$^{-24}$~cm$^3$s$^{-1}$ and $3.2\times$10$^{-25}$~cm$^3$s$^{-1}$, respectively, in the $\tau^+\tau^-$ annihilation channel for a DM mass of 2 TeV.}

\begin{document}
\maketitle
\flushbottom

\section{Introduction\label{sec:introduction}}

The observation of gamma rays from dark matter (DM) interactions would answer key questions about the nature of the Universe. Such a discovery would also shed light on our understanding of physics beyond the Standard Model (SM). According to the latest measurements of the energy content of the Universe, baryonic matter is observed to make up about 5\% of the total energy content, with the rest being Cold Dark Matter (CDM) (27\%) and dark energy (68\%)~\cite{2020A&A...641A...6P}. Experimental evidence for DM can be found in observations of gravitational lensing~\cite{2019A&A...631A..40S}, the large-scale structure of galaxies~\cite{2006Natur.440.1137S}, the power spectrum of the cosmic microwave background~\cite{2016A&A...594A..13P}, and the rotational velocity of galaxies~\cite{1985ApJ...295..305V}, among other probes~\cite{2018RPPh...81f6201R}. Assuming DM to be made of elementary particles, in most models they are considered to be electrically neutral, stable on cosmological time scales, and cold (i.e., non-relativistic). A well-motivated class of CDM candidate particles are the Weakly Interacting Massive Particles (WIMPs): elementary particles not contained in the SM, which may have been produced in the early Universe and that would annihilate or decay into SM particles. If we assume that a GeV--TeV mass scale particle, with weak-scale couplings, was decoupled from thermal equilibrium in the early Universe, such particles could reproduce the observed DM relic density~\cite{2020A&A...641A...6P,2018ipap.book.....D}. This coincidence between the predicted thermal abundance of WIMPs and the observed DM density is known as the ``WIMP miracle''~\cite{2014IJMPS..3060256S}. A precise computation of the thermally-averaged velocity-weighted annihilation cross section provides $\langle \sigma v \rangle_{\rm th} \simeq$ 2$\times$10$^{-26}$ cm$^3$s$^{-1}$~\cite{2012PhRvD..86b3506S}.

The primary WIMP DM search modes are: \textit{direct detection} searches for signatures of the interaction of local DM particles in underground detectors; searches for DM particle \textit{production} in accelerators such as the Large Hadron Collider at CERN; and the \textit{indirect search} for DM decay or annihilation to SM products by ground-based and spaceborne observatories. In indirect searches, one looks for spatial and/or spectral signatures of DM annihilation or decay in the astrophysical fluxes of SM particles. Among those, gamma rays are detectable over a broad energy range with both spaceborne and ground-based astronomical observatories. They point back to their sources and are not subject to significant energy losses in the local universe. The aforementioned properties make gamma rays excellent probes for DM searches. Among the privileged targets for gamma-ray observations are the center of the Milky Way and its nearby satellite galaxies.

Dwarf spheroidal galaxies (dSphs) are groups of gravitationally bound stars with a typical radius of 10$^2$--10$^3$\,pc. They host small amounts of visible mass (10$^3$--10$^7$\,M$_\odot$) and show consequently low optical luminosities (10$^3$--10$^7$\,L$_\odot$)~\cite{2018RPPh...81e6901S}. The dSph satellites of the Milky Way are located at $\mathcal{O}$(100 kpc) Galactocentric distance, and mainly at high Galactic latitudes, although some of them lie closer to the Galactic plane, such as the Sagittarius dSph. Their kinematics tend to be dominated by random stellar motion, and the amplitude of the star motion is driven by the gravitational potential of the galaxy~\cite{1998gaas.book.....B,2008gady.book.....B}. The gas or dust content of dSphs is extremely low, and therefore cannot be used to trace the galactic dynamics~\cite{2014ApJ...795L...5S}. The low luminosity of dSphs and their high mass-to-light ratio, compared to Milky-Way like galaxies, indicate that they are DM-dominated galaxies with negligible astrophysical background from gamma-ray emitters~\cite{2004PhRvD..69l3501E,2016ApJ...832L...6W}. The dSphs are therefore among the most promising targets for indirect DM searches using gamma-ray signals.

In this paper, we combine observations of dSphs performed with five different gamma-ray instruments, including satellite, water Cherenkov detector, and imaging atmospheric Cherenkov telescopes (IACTs) --- \textit{Fermi}-LAT, HAWC, H.E.S.S., MAGIC, and VERITAS --- to search for DM signals. Our goal is to increase the sensitivity of indirect DM searches with gamma rays by using all available data on the selected objects over the widest possible DM mass range. Each of the instruments involved in the combination has accumulated large datasets of dSph observations. Each corresponding collaboration carried out analyses of the individual datasets. Then, a combined likelihood analysis with all their respective data was performed. This new combined study aims to extract more information from already independently analyzed data and to increase our chance of a possible DM detection. For this work, each collaboration analysis of their own datasets was done using common DM spectral and morphological models and common conventions on the statistical analysis, including the treatment of the relevant statistical uncertainties. This standardization avoids the need to share low-level experimental data (event lists) and instrument response functions (IRFs). Each collaboration computed a likelihood function versus the velocity-weighted cross section $\sv$, scanning over the DM particle masses for several annihilation channels and each observed target. The obtained likelihood values were subsequently shared and combined into a global joint likelihood function, from which constraints on $\sv$ were derived. This methodology was applied before in the previous combined DM search towards dSph galaxies by MAGIC and \textit{Fermi}-LAT~\cite{2016JCAP...02..039M}. Those data constitute a subset of the dataset included in the present work.

This article is organized as follows: in Section~\ref{sec:experiments}, we present the five collaborations that have participated in the combination and their observational datasets. In Section~\ref{sec:dm} we recall the signal expected from DM annihilation, its dependence on the DM distribution in dSphs, and the associated uncertainties. In Section~\ref{sec:likelihood} we describe the likelihood analysis technique and the procedure to combine the results. We present the results in Section~\ref{sec:results} and discuss and conclude in Section~\ref{sec:conclusions}.

\section{Observations and data\label{sec:experiments}}

In this section, we provide details about the sources observed, the detectors, and their associated datasets, which are also summarized in Table~\ref{tab:summary}. In general, we restricted the dSph observations included from \textit{Fermi}-LAT and HAWC based on the intersection of dSphs studied in the two independent sets of $J$-factor calculations used in this work (see Section~\ref{sec:dm} for more details). Even more stringent exclusion limits, especially at the lowest DM masses, could be obtained by including all known dSphs. However, since through our selection the known dSphs with the largest expected DM-induced gamma-ray flux are already included, we expect this improvement to be minor with respect to the results presented in Section~\ref{sec:results}.

\subsection{\textit{Fermi}-LAT}

The Large Area Telescope onboard the NASA \textit{Fermi} satellite (\textit{Fermi}-LAT) is a pair conversion telescope orbiting the Earth at an altitude of $\sim550$~km~\cite{2009ApJ...697.1071A}. It is sensitive to gamma rays in the energy range from 20~MeV to $>1$~TeV. \textit{Fermi}-LAT covers the lowest energy region of this study. \textit{Fermi}-LAT has a wide field of view covering about 20\% of the sky, and scans the whole sky approximately every 3~hours. The instrument detects gamma rays via pair conversion of the incoming photon to an electron and positron within the detector. Similar to particle accelerator experiments, the direction of these particles is recorded in a silicon strip tracker and the energy deposited in a cesium iodide calorimeter. Detailed descriptions of the detector and its performance can be found in~\cite{2012ApJS..203....4A}. The energy resolution of \textit{Fermi}-LAT at 1~GeV is 10\% and it reaches its minimum at 10~GeV, where it is 5\%. The 68\% containment radius angular resolution for one photon is $1\degree$ ($0.2\degree$) at 1~GeV ($>10$~GeV). An overview of the dSphs observed by \textit{Fermi}-LAT and a DM search using these observations is given in~\cite{2015PhRvL.115w1301A}.

We analyze almost ten years of Pass~8~\cite{2013arXiv1303.3514A,2018arXiv181011394B} data (data processing {\tt P8R2}) from August 2008 to March 2018. We select {\tt SOURCE} class events, passing the basic quality filter cuts, and their corresponding {\tt P8R2\_SOURCEVETO\_V2} IRFs~\cite{2015PhRvD..91l2002A}. We choose energies between 300~MeV and 1~TeV and apply a cut to zenith angles $<100\degree$ between 300~MeV and 1~GeV and $<105\degree$ above 1~GeV in order to exclude contamination from the Earth's limb. We model the background with sources reported in the \textit{Fermi}-LAT 8-year source catalog (4FGL)~\cite{2020ApJS..247...33A}. We also include in the model the latest released interstellar emission model (IEM), namely {\tt gll\_iem\_v07.fits}\footnote{A complete discussion about this new IEM can be found at \url{https://fermi.gsfc.nasa.gov/ssc/data/analysis/software/aux/4fgl/Galactic_Diffuse_Emission_Model_for_the_4FGL_Catalog_Analysis.pdf}}, and its corresponding isotropic template {\tt iso\_P8R2\_SOURCEVETO\_V2\_v1.txt}. We analyze the $12\degree\times12\degree$ regions of interest centered on the locations of the dSphs and choose a pixel size of $0.08\degree$. We include in the background model sources located in a region $14\degree\times14\degree$ in order to include also sources at most $1\degree$ outside our region of interest. We apply the same analysis performed on dSphs in~\cite{2015PhRvL.115w1301A,2017ApJ...834..110A,2021PhRvD.103l3005D} by employing versions {\tt 0.18.0} of {\tt Fermipy} and {\tt 1.2.3} of the {\tt Fermitools}\footnote{\url{https://fermi.gsfc.nasa.gov/ssc/data/analysis/}}.

\subsection{HAWC}

The High-Altitude Water Cherenkov (HAWC) observatory is a high-energy gamma-ray detector located at Sierra Negra, Mexico. The site is $4100$~m above sea level, at latitude and longitude ($19.0\degree$ N, $97.3\degree$ W). HAWC is a survey instrument with a wide field of view, and observes two thirds of the sky every day within declinations of $-20.0\degree$ to $60.0\degree$~\cite{2020ApJ...905...76A}. It is sensitive to gamma-ray energies between $300$~GeV and \textgreater 100 TeV~\cite{2017ApJ...843...39A}, with 30--100\% energy resolution. HAWC is an array of 300 water Cherenkov detectors (WCD) covering a $22{\scriptstyle,}000$~m$^2$ area. When a cosmic particle enters the atmosphere and interacts, it initiates an air shower, the particles of which travel through the HAWC WCDs, emitting Cherenkov light in the water. The Cherenkov light is detected by the photo-multiplier tubes (PMTs) in the WCDs~\cite{2017ApJ...843...39A}. The timing and intensity of the Cherenkov light detected by the different PMTs across the HAWC array is used to determine the incident angle, energy, and species of the initiating cosmic particle~\cite{2017ApJ...843...39A}. The HAWC sensitivity in this analysis depends on the spectrum of the source, and its angular resolution (68\% containment) varies from 1\degree\ at 1~TeV to 0.2\degree\ at \textgreater 30~TeV~\cite{2017ApJ...843...39A}. 

For this analysis, HAWC reconstructs gamma-ray energies from the fraction of PMTs hit on the array~\cite{2017ApJ...843...39A}. HAWC has been operated as a partial detector since August 2013 and with the full detector since March 2015. HAWC provides data corresponding to 1038 days of livetime starting on November 26, 2014. The exposure to each dSph depends on the declination of the source with respect to the latitude of HAWC (see HAWC's exposure to all dSphs included in this analysis in Table~\ref{tab:summary}). The background is estimated through HAWC's direct integration method~\cite{2017ApJ...843...39A} with the region of interest around the dSphs set at 4\degree. HAWC used {\tt HAL} and {\tt threeML} public software for its analysis framework~\cite{2015arXiv150708343V}. In this work, the dSphs are treated as extended sources in HAWC data and analyzed following the methods presented in~\cite{2018ApJ...853..154A}.

\subsection{H.E.S.S.}

The High Energy Stereoscopic System (H.E.S.S.) is an array of five IACTs situated at an altitude of 1800~m ($23.3\degree$ S, $16.5\degree$ E) in the Khomas Highland of Namibia, sensitive to gamma rays from $\sim$ 50~GeV to $\sim$ 100~TeV. Four 12-m diameter telescopes (CT1-4) are positioned at the corners of a 120-m side square, and one 28-m diameter telescope (CT5) is located at the center of the array. The observations used in this work are taken in wobble mode~\cite{1994APh.....2..137F} where the telescope pointing alternates directions with an offset of $0.7\degree$ from the target position. Standard quality selection procedures are applied to the data~\cite{2006A&A...457..899A}. After the calibration of the raw data, the events are reconstructed using a template-fitting technique in which the images recorded in the cameras are compared to shower templates computed from a semi-analytical model~\cite{2009APh....32..231D}. This technique achieves an energy resolution of 10\% and an angular resolution of 0.06\degree at 68\% containment radius for gamma-ray energies above 200 GeV for the CT1-4 array, with a field of view of 5\degree. The point-source 5$\sigma$ flux sensitivity is about 1\% of the Crab Nebula flux for 25~h of observations near zenith~\cite{2006A&A...457..899A}.

H.E.S.S. data include observations of the dSphs Carina, Coma Berenices, Fornax, and Sculptor. Data were collected with the CT1-4 telescope array, requiring at least two telescopes to detect the same air shower. For each dSph, the signal region, referred to as the ON region, is defined as a disk centered at the nominal position of the dSph with a radius of 0.1\degree. We estimate the background contribution in the ON regions using several background-control (or OFF) regions of the same size, shape and angular distance from the pointing position as the corresponding ON region~\cite{2007A&A...466.1219B}. This method guarantees almost identical acceptances in the ON and OFF regions, such that no further offline background normalization is necessary. H.E.S.S. follows the recommended statistical approach of~\cite{1983ApJ...272..317L} to derive the excess significances provided in Table~\ref{tab:summary}.

Sculptor and Carina~\cite{2011APh....34..608H,2014PhRvD..90k2012A,2018JCAP...11..037A} were observed between January 2008 and December 2009. Coma Berenices and Fornax~\cite{2014PhRvD..90k2012A,2018JCAP...11..037A} were observed from 2010 to 2013 and in 2010, respectively. The pointing offset in the case of Fornax is larger, since it was not the primary target of the corresponding dataset. The observation time of the four dSphs analyzed in this paper amounts to 54~h. The data are independently analyzed with the H.E.S.S. official software packages \texttt{ParisAnalysis}~\cite{2009APh....32..231D}, and \texttt{HAP}~\cite{2014APh....56...26P} for analysis result crosschecks.

\subsection{MAGIC}

The \textit{Florian Goebel} Major Atmospheric Gamma-ray Imaging Cherenkov (MAGIC) telescopes are located at the Roque de los Muchachos Observatory ($ 28.8\degree$ N, $ 17.9\degree$ W, 2200~m above sea level) on the Canary Island of La Palma, Spain. MAGIC consists of two IACTs, each with a 17-m reflector diameter and equipped with fast imaging cameras with a $3.5\degree$ field of view. MAGIC is sensitive to gamma rays above $\gtrsim 30$~GeV. The 5$\sigma$ sensitivity above 220~GeV for 50~h of observations of a point-like source is $\sim0.7\%$ of the flux of the Crab Nebula, with an associated energy resolution of $\sim$ 16\% and a $0.07\degree$ angular resolution measured as the 68\% containment radius of the gamma-ray excess~\cite{2016APh....72...76A}.

MAGIC provides data from four dSphs observed with the full two-telescope MAGIC system: Segue~1~\cite{2014JCAP...02..008A}, Ursa Major II~\cite{2018JCAP...03..009A}, Draco, and Coma Berenices~\cite{2022PDU....3500912A}. In total, a 354~h dataset was collected in wobble mode~\cite{1994APh.....2..137F} with two pointing positions, offset by $0.4\degree$ from the center of the target, used for each source. In the MAGIC analysis, the size of the ON and the respective OFF regions are optimized for each source to increase the sensitivity to a potential DM signal (see~\cite{2018JCAP...03..009A} for more details). Observations of Segue~1 were taken between 2011 and 2013 and represent the deepest IACT observation of any dSph to date. Ursa Major II, Draco, and Coma Berenices were observed between 2014 and 2016, March to September 2018, and January to June 2019, respectively. MAGIC hardware upgrades such as~\cite{2016APh....72...61A} and changes in the telescope reflectivity define distinct periods of data taking, each of which is analyzed using an individual set of IRFs obtained from dedicated Monte Carlo simulations. The data are reduced with the standard \texttt{MARS} analysis software~\cite{2013ICRC...33.2937Z}.

\subsection{VERITAS}

The Very Energetic Radiation Imaging Telescope Array System (VERITAS) instrument is an array of four ground-based IACTs with 12-m focal length placed $\sim 100$~m from each other. VERITAS is situated at the Fred Lawrence Whipple Observatory in southern Arizona, (31.7\degree N, 111.0\degree W), 1300~m above sea level. VERITAS has been operational since 2007. The total field of view is $\sim 3.5\degree$. VERITAS is most sensitive in the very-high-energy band, from $\sim 85$~GeV up to $\sim 30$~TeV. Its energy resolution is $\sim 15$\%, and its angular resolution in terms of the 68\% containment radius is $<0.1\degree$ per event at 1~TeV. Its sensitivity is such that a point-like source with 1\% of the flux of the Crab Nebula can be detected within 25~h at zenith angles smaller than 30\degree~\cite{2015ICRC...34..771P}.

For this analysis, VERITAS provides data for four dSphs in the Northern Hemisphere: Bo\"{o}tes I, Draco, Segue~1, and Ursa Minor. The targets were observed from 2009 to 2013 for $\sim 216$~h~\cite{2017PhRvD..95h2001A}. Data were collected in wobble pointing mode~\cite{1994APh.....2..137F}, with $0.5\degree$ between the camera center and the source position.

Loose event selection criteria are used in order to keep the energy threshold as low as possible, increasing the sensitivity to a potential DM signal at low energy. The analysis incorporated into this work adopted a $\theta < 0.17\degree$ cut on the angular distance of the reconstructed shower arrival position with respect to the target position to define the ON region. A novel background estimation method is used~\cite{2013ICRC...33.2604Z}, in which OFF events are extracted from an annular region centered on the tracking position rather than the target position. Gradients across the camera are taken into account through a zenith-dependent acceptance function~\cite{2003A&A...410..389R}. The \texttt{VEGAS} software pipeline for analysis and reconstruction~\cite{2008ICRC....3.1385C} is used for this analysis.

\newgeometry{margin=2.5cm}
\begin{landscape}
\begin{table}
\caption{Summary of dSph observations by each experiment used in this work. A `$-$' indicates the experiment did not observe this dSph or did not include its data on it for this study. For \textit{Fermi}-LAT, the exposure at 1~GeV is given. For HAWC, $|\Delta\phi|$ is the absolute difference between the source declination and HAWC latitude, i.e. the zenith angle of the source at culmination. HAWC is more sensitive to sources with smaller $|\Delta\phi|$. For IACTs, we show the zenith angle range, the total exposure, the energy of the lowest and highest energy ON event considered for unbinned analysis and the lowest and highest energy bin edges considered for binned analysis, the angular radius $\theta_{\rm{sig}}$ of the signal or ON region, the ratio $\tau$ of exposures between the background-control (OFF) and signal (ON) regions, and the significance of gamma-ray excess, S($\sigma$), in standard deviations.} 
\centering 
{\begin{tabular}{c ? c ? c ? c c c c c c c}
\hline
\hline
& \multicolumn{1}{c?}{\textit{Fermi}-LAT}  & \multicolumn{1}{c?}{HAWC} & \multicolumn{7}{c}{H.E.S.S., MAGIC, VERITAS} \\
\hline
Source name & Exposure ($10^{11}$ s cm$^2$) & $|\Delta\phi|$ (\degree) & IACTs & Zenith (\degree) & Time exposure (h) & Energy range (TeV) & $\theta_{\rm{sig}}$ (\degree) & $\tau$ & S ($\sigma$) \\ 
\hline\hline 
Bo\"{o}tes I            & $2.6$ & $4.5$ & VERITAS  & $15-30$ & $14.0$ & $0.10 - 41$ & $0.17$ & $8.6$ & $-1.0$ \\
Canes Venatici I        & $2.9$ & $14.6$ & $-$ & $-$ & $-$ & $-$ & $-$ & $-$ & $-$ \\
Canes Venatici II       & $2.9$ & $15.3$ & $-$ & $-$ & $-$ & $-$ & $-$ & $-$ & $-$ \\
Carina                  & $3.1$ & $-$ & H.E.S.S. & $27-46$ & $23.7$ & $0.31 - 70$ & $0.10$ & $18.0$ & $-0.3$ \\
\hdashline
\multirow{2}{*}{Coma Berenices}          & \multirow{2}{*}{$2.7$} & \multirow{2}{*}{$4.9$} & H.E.S.S. & $47-49$ & $11.4$ & $0.55 - 70$ & $0.10$ & $14.4$ & $-0.4$ \\
                                         & & & MAGIC & $5 - 37 $ & $49.5$ & $0.06 - 10$ & $0.17$ & $1.0$ & $0.8$\\
\hdashline
\multirow{2}{*}{Draco}                   & \multirow{2}{*}{$3.8$} & \multirow{2}{*}{$38.1$} & MAGIC & $29 - 45$ & $52.1$ & $0.07 - 10$ & $0.22$ & $1.0$ & $-0.7$ \\
                                         & & & VERITAS & $25-40$ & $49.8$ & $0.12 - 70$ & $0.17$ & $9.0$ & $-1.0$ \\
\hdashline
Fornax                  & $2.7$ & $-$ & H.E.S.S. & $11-25$ & $6.8$ & $0.23 - 70$ & $0.10$ & $45.5$ & $-1.5$ \\
Hercules                & $2.8$ & $6.3$ & $-$ & $-$ & $-$ & $-$ & $-$ & $-$ & $-$ \\
Leo I                   & $2.5$ & $6.7$ & $-$ & $-$ & $-$ & $-$ & $-$ & $-$ & $-$ \\
Leo II                  & $2.6$ & $3.1$ & $-$ & $-$ & $-$ & $-$ & $-$ & $-$ & $-$ \\
Leo IV                  & $2.4$ & $19.5$ & $-$ & $-$ & $-$ & $-$ & $-$ & $-$ & $-$ \\
Leo V                   & $2.4$ & $-$ & $-$ & $-$ & $-$ & $-$ & $-$ & $-$ & $-$ \\
Leo T                   & $2.6$ & $-$ & $-$ & $-$ & $-$ & $-$ & $-$ & $-$ & $-$ \\
Sculptor                & $2.7$ & $-$ & H.E.S.S. & $10-46$ & $11.8$ & $0.20 - 70$ & $0.10$ & $19.8$ & $-2.2$ \\
\hdashline
\multirow{2}{*}{Segue 1}& \multirow{2}{*}{$2.5$} & \multirow{2}{*}{$2.9$} & MAGIC & $13-37$ & $158.0$ & $0.06 - 10$ & $0.12$ & $1.0$ & $-0.5$ \\
                        & & & VERITAS & $15-35$ & $92.0$ & $0.08 - 50$ & $0.17$ & $7.6$ & $0.7$ \\
\hdashline
Segue 2                 & $2.7$ & $-$ & $-$ & $-$ & $-$ & $-$ & $-$ & $-$ & $-$ \\
Sextans                 & $2.4$ & $20.6$ & $-$ & $-$ & $-$ & $-$ & $-$ & $-$ & $-$ \\
Ursa Major I            & $3.4$ & $32.9$ & $-$ & $-$ & $-$ & $-$ & $-$ & $-$ & $-$ \\
Ursa Major II           & $4.0$ & $44.1$ & MAGIC & $35-45$ & $94.8$ & $0.12 - 10$ & $0.30$ & $1.0$ & $-2.1$ \\
Ursa Minor              & $4.1$ & $-$ &  VERITAS & $35-45$ & 60.4 & $0.16-93$ & 0.17 & 8.4 & $-0.1$ \\
\hline
\end{tabular}}
\label{tab:summary} 
\end{table}
\end{landscape}
\restoregeometry

\section{Dark Matter signal\label{sec:dm}}

The differential flux of gamma rays of self-annihilating Majorana DM particles from an astrophysical source is~\cite{2005PhR...405..279B}:
\begin{equation}
    \frac{\text{d}^2\Phi \left(\sv, J \right)}{\text{d}E \text{d}\Omega} = \left(\frac{\sv}{2m^{2}_{\textrm{DM}}} \sum_{f} \text{BR}_{f}\frac{\text{d}N_{f}}{\text{d}E}\right) \times \left(\frac{1}{4\pi}\frac{\text{d}J}{\text{d}\Omega} \right).
    \label{eq:dm-flux}
\end{equation}
The first term of Eq.~\ref{eq:dm-flux} is the {\it particle physics factor} for DM annihilation, and depends on the thermally-averaged velocity-weighted cross section, $\sv$, the mass of the DM particle, $m_{\textrm{DM}}$, and average differential spectrum of gamma rays per annihilation, $\text{d}N_{f}/\text{d}E$, for each final state $f$ weighted by its branching ratio $\text{BR}_{f}$. We study seven annihilation channels with final states $W^+W^-$, $ZZ$, $b\bar{b}$, $t\bar{t}$, $e^+e^-$, $\mu^+\mu^-$, and $\tau^+\tau^-$ separately by assuming a 100\% branching ratio in each case. The values considered for $m_{\textrm{DM}}$ range from 5 GeV to 100 TeV, except for the $W^+W^-$, $ZZ$, and $t\bar{t}$ channels, where the range is from the corresponding kinematic threshold to 100 TeV. The $\text{d}N_{f}/\text{d}E$ spectrum is calculated according to~\cite{2011JCAP...03..051C} and includes electroweak corrections of the final state products (see Appendix~\ref{app:spectra}).

The $\text{d}J/\text{d}\Omega$ term in Eq.~\ref{eq:dm-flux} is the {\it astrophysical factor}, or differential $J$-factor (see Appendix~\ref{app:j-factors}). The differential $J$-factor is proportional to the intensity of the expected DM annihilation gamma-ray signal from an astrophysical source as seen from Earth. The differential $J$-factor is defined as the integral of the square of the DM density distribution $\rho_{\rm{DM}}$ along the line-of-sight (l.o.s.):
\begin{equation}
    \frac{\text{d}J}{\text{d}\Omega} = \displaystyle{\int_\mathrm{l.o.s.} \: \rho_{\textrm{DM}}^2(r(s,\theta)) \:ds},
    \label{eq:differential-j-factor}
\end{equation}
where $\rho_{\textrm{DM}}$ is assumed to be spherically symmetric and depends on the radial distance $r$ from the center of the dSph. This distance can also be expressed in terms of the distance $s$ from Earth along the l.o.s., and the angular distance, $\theta$, with respect to the center of the dSph as $r^2(s, \theta) = s^2 +d^2 -2sd\cos\theta$, where $d$ is the distance between the Earth and the dSph nominal position. The total $J$-factor can then be obtained by integrating over the solid angle $\Delta \Omega$ corresponding to the dSph extension. We note that we use $J$-factors in the integrated form $J(\Delta \Omega)$ at a given angular radius $\theta$ for the remainder of this paper (see Table~\ref{tab:j-factor}), which reads:
\begin{equation}
    J(\Delta \Omega) = \displaystyle{\int_{\Delta \Omega} \int_\mathrm{l.o.s.} \: \rho_{\textrm{DM}}^2(r(s,\theta)) \:ds \: d\Omega}.
    \label{eq:total-j-factor}
\end{equation}
The derivation of $\rho_{\textrm{DM}}$ is performed through the Jeans analysis with the formalism of the spherical Jeans equation~\cite{2015MNRAS.446.3002B,2015ApJ...801...74G,2008gady.book.....B,2015MNRAS.453..849B}. This method allows the reconstruction of galactic dynamics based on spectroscopic data, assuming in the case of dSphs that the systems are in statistical equilibrium, non-rotating, and have a spherical symmetry. 

\begin{table}[t]
\centering
\caption{Summary of the relevant properties of the dSphs considered in the present work. Column 1 lists the dSphs. Columns 2 and 3 present their heliocentric distance and Galactic coordinates, respectively. Columns 4 and 5 report the $J$-factors of each object given by the independent $\mathcal{GS}$ and $\mathcal{B}$ studies and their estimated $\pm 1\sigma$ uncertainties. The values $\log_{10}J$~($\mathcal{GS}$ set) correspond to the mean $J$-factor values for a dSph extension truncated at the outermost observed star. The values $\log_{10}J$~($\mathcal{B}$ set) are provided for an extension ending at the tidal radius of each dSph.}
\small{\begin{tabular}{ccccc}
\hline
\hline
\CellTopTwo{}
Name & Distance & $l, b$ & $\log_{10}J$~($\mathcal{GS}$ set) & $\log_{10}J$~($\mathcal{B}$ set)\\
& \scriptsize{(kpc)} &  \scriptsize{($\degree$)} & \scriptsize{$\log_{10}(\rm{GeV}^2 \rm{cm}^{-5})$} & \scriptsize{$\log_{10}(\rm{GeV}^2 \rm{cm}^{-5})$} \\
\hline
\CellTopTwo{}
Bo\"otes I & $66$ & $358.08,\: 69.62$ & $18.24^{+0.40}_{-0.37}$ & $18.85^{+1.10}_{-0.61}$ \\
\CellTopTwo{}
Canes Venatici I & $218$ & $74.31,\: 79.82$ & $17.44^{+0.37}_{-0.28}$ & $17.63^{+0.50}_{-0.20}$ \\
\CellTopTwo{}
Canes Venatici II & $160$ & $113.58,\: 82.70$ & $17.65^{+0.45}_{-0.43}$ & $18.67^{+1.54}_{-0.97}$ \\
\CellTopTwo{}
Carina & $105$ & $260.11,\: -22.22$ & $17.92^{+0.19}_{-0.11}$ & $18.02^{+0.36}_{-0.15}$ \\
\CellTopTwo{}
Coma Berenices & $44$ & $241.89,\: 83.61$ & $19.02^{+0.37}_{-0.41}$ & $20.13^{+1.56}_{-1.08}$ \\
\CellTopTwo{}
Draco & $76$ & $86.37,\: 34.72$ & $19.05^{+0.22}_{-0.21}$ &  $19.42^{+0.92}_{-0.47}$ \\
\CellTopTwo{}
Fornax & $147$ & $237.10,\: -65.65$ & $17.84^{+0.11}_{-0.06}$ &  $17.85^{+0.11}_{-0.08}$ \\
\CellTopTwo{}
Hercules & $132$ & $28.73,\: 36.87$ & $16.86^{+0.74}_{-0.68}$ &  $17.70^{+1.08}_{-0.73}$ \\
\CellTopTwo{}
Leo I & $254$ & $225.99,\: 49.11$ & $17.84^{+0.20}_{-0.16}$ &  $17.93^{+0.65}_{-0.25}$ \\
\CellTopTwo{}
Leo II & $233$ & $220.17,\: 67.23$ & $17.97^{+0.20}_{-0.18}$ &  $18.11^{+0.71}_{-0.25}$ \\
\CellTopTwo{}
Leo IV & $154$ & $265.44,\: 56.51$ & $16.32^{+1.06}_{-1.70}$ & $16.36^{+1.44}_{-1.65}$ \\
\CellTopTwo{}
Leo V & $178$ & $261.86,\: 58.54$ & $16.37^{+0.94}_{-0.87}$ & $16.30^{+1.33}_{-1.16}$ \\
\CellTopTwo{}
Leo T & $417$ & $214.85,\: 43.66$ & $17.11^{+0.44}_{-0.39}$ & $17.67^{+1.01}_{-0.56}$ \\
\CellTopTwo{}
Sculptor & $86$ & $287.53,\: -83.16$ & $18.57^{+0.07}_{-0.05}$ &  $18.63^{+0.14}_{-0.08}$ \\
\CellTopTwo{}
Segue 1 & $23$ & $220.48,\: 50.43$ & $19.36^{+0.32}_{-0.35}$ &  $17.52^{+2.54}_{-2.65}$ \\
\CellTopTwo{}
Segue 2 & $35$ & $149.43,\: -38.14$ & $16.21^{+1.06}_{-0.98}$ & $19.50^{+1.82}_{-1.48}$ \\
\CellTopTwo{}
Sextans & $86$ & $243.50,\: 42.27$ & $17.92^{+0.35}_{-0.29}$ &  $18.04^{+0.50}_{-0.28}$ \\
\CellTopTwo{}
Ursa Major I & $97$ & $159.43,\: 54.41$ & $17.87^{+0.56}_{-0.33}$ & $18.84^{+0.97}_{-0.43}$ \\
\CellTopTwo{}
Ursa Major II & $32$ & $152.46,\: 37.44$ & $19.42^{+0.44}_{-0.42}$ & $20.60^{+1.46}_{-0.95}$ \\
\CellTopTwo{}
Ursa Minor & $76$ & $104.97,\: 44.80$ & $18.95^{+0.26}_{-0.18}$ & $19.08^{+0.21}_{-0.13}$ \\
\hline
\hline
\CellTopTwo{}
\end{tabular}}
\label{tab:j-factor}
\end{table}

In this study, we compute the limits on $\sv$ from the combination of the dSph observations using two independent sets of $J$-factors produced by Geringer-Sameth et al.~\cite{2015ApJ...801...74G} (or $\mathcal{GS}$ set), and by Bonnivard et al.~\cite{2015MNRAS.446.3002B,2015MNRAS.453..849B} (or $\mathcal{B}$ set), respectively. We produce results from both $\mathcal{GS}$ and $\mathcal{B}$ sets to estimate the impact of the systematic uncertainty in the $J$-factor derivation. More details about the assumptions of each $J$-factor set can be found in Appendix~\ref{app:j-factors}, which also presents a comparison between the two sets of differential $J$-factors versus the angular radius of the integration region $\Delta \Omega$, measured from the dSph nominal position, for each of the twenty dSphs. A detailed review of the computation of $J$-factors can be found in~\cite{2020PhRvD.102b3029B}. The majority of $J$-factor estimations, especially for the DM-dominated dSphs, are compatible within uncertainties for the different sets of $J$-factors in the literature. Comparing the $\mathcal{GS}$ and $\mathcal{B}$ $J$-factors, the largest discrepancy occurs for the $J$-factor of Segue~1 (see Table~\ref{tab:j-factor}), because the analysis of the $\mathcal{B}$ set~\cite{2015MNRAS.453..849B} is extremely sensitive to the inclusion or exclusion of a few marginal member stars~\cite{2016MNRAS.462..223B}. Given that MAGIC and VERITAS have accumulated a wealth of observation time on Segue~1 and that the DM content of Segue~1 has been an active subject of debate~\cite{2016MNRAS.462..223B}, the usage of two different sets of $J$-factors in this work is well-motivated\footnote{We further note that the use of different $J$-factor calculations would result in different extracted limits on the velocity-weighted annihilation cross section. For instance, the $J$-factors derived in~\cite{2020PhRvD.102f1302A} weaken limits derived with \textit{Fermi}-LAT data as well as those derived with VERITAS data~\cite{2024PhRvD.110f3034A}.}.

Observations carried out by H.E.S.S. towards Sagittarius dSph~\cite{2008APh....29...55A,2014PhRvD..90k2012A} amount to 90~hours. However, Sagittarius's relative proximity ($\sim$ 17~kpc) to the Galactic Center subjects it to tidal effects from the Milky Way. Determination of the DM distribution in this object is challenging~\cite{2012ApJ...746...77V}, therefore H.E.S.S. observations of Sagittarius are not included in this study. 

Ultra-faint Milky Way satellites have been recently unveiled by optical surveys in the southern hemisphere sky like PanSTARRS and DES. Of these, Reticulum II, Tucana II, Tucana III, Tucana IV and Grus II have been observed by H.E.S.S.~\cite{2020PhRvD.102f2001A}. The uncertainty on the DM distribution for ultra-faint systems is challenging to measure or predict. In addition to the uncertainty on the $J$-factor coming from the limited number of stellar tracers, the assumptions made to derive the $J$-factor value, such as a single stellar population, spherical symmetry, constant velocity anisotropy, or absence of tidal stripping, are expected to significantly impact the $J$-factor estimate for these systems. The selection of member stars for these ultra-faint systems is also complex due to the difficulty in distinguishing member stars from foreground stars. Therefore, the DES dSphs observed by H.E.S.S. are not considered in this study. Similarly, the ultra-faint satellite Triangulum II has been observed by HAWC~\cite{2018ApJ...853..154A} and MAGIC~\cite{2020PDU....2800529A} but its seemingly high $J$-factor~\cite{2015ApJ...814L...7K} may be overestimated due to the contamination of velocity dispersion measurements by binary systems. Further observations showed that the system may not be in equilibrium and may suffer from tidal disruption~\cite{2017AJ....154..267C}, which makes the determination of the DM distribution complex. Observations of Triangulum II are thus not considered in this study.

\section{Joint likelihood analysis\label{sec:likelihood}}

We conduct our DM search by means of a maximum likelihood analysis, in which we take advantage of the different spectral and morphological features expected for signal (see Section~\ref{sec:dm}) and background. Note that morphological features are treated differently in the analysis of each instrument and will be described later in this section. 

We perform the derivation of our upper limits on $\sv$ using the test statistic, $\mathrm{TS}$, for each considered annihilation channel and DM mass:
\begin{equation}
    \mathrm{TS} = -2 \ln{\lambda(\sv)},
    \label{eq:ts}
\end{equation}
where $\lambda(\sv)$ is the profile likelihood ratio defined as a function of the annihilation cross section $\sv$, which reads~\cite{2011EPJC...71.1554C}:
\begin{equation}
    \lambda \left( \sv \mid \bm{\mathcal{D}_{\text{dSphs}}} \right) = \frac{\mathcal{L} \left( \sv ; \bm{\hat{\hat{\nu}}} \mid \bm{\mathcal{D}_{\text{dSphs}}} \right)}{\mathcal{L} \left( \widehat{\sv} ; \bm{\hat{\nu}} \mid \bm{\mathcal{D}_{\text{dSphs}}} \right)},
    \label{eq:likelihood-profile}
\end{equation}
where $\bm{\mathcal{D}_{\text{dSphs}}}$ is our set of observations, $\bm{\nu}$ represents the nuisance parameters, $\widehat{\sv}$ and $\bm{\hat{\nu}}$ are the values that maximize $\mathcal{L}$ globally, and $\bm{\hat{\hat{\nu}}}$ is the set of values that maximize $\mathcal{L}$ for a given value of $\sv$.

The \textit{total} joint likelihood function $\mathcal{L}$ that describes all observations from all instruments and dSphs is factored into \textit{partial} joint likelihood functions corresponding to each dSph $l$ ($\mathcal{L}_{\text{dSph},l}$) and its corresponding $J$-factor ($\mathcal{J}_l$):
\begin{equation}
    \mathcal{L} \left( \sv ; \bm{\nu} \mid \bm{\mathcal{D}_{\text{dSphs}}} \right) = \prod_{l=1}^{N_{\text{dSphs}}} \mathcal{L}_{\text{dSph},l} \left( \sv ; J_{l}, \bm{\nu_{l}} \mid \bm{\mathcal{D}_{l}} \right) \times \mathcal{J}_l \left( J_{l} \mid J_{l,\text{obs}}, \sigma_{\log{J_l}} \right).
    \label{eq:dSph-combination}
\end{equation}
The quantity $ N_{\text{dSphs}} = 20 $ is the number of dSphs considered in this work. $ \bm{\mathcal{D}_{l}} $ is the set of gamma-ray observations for the $l$-th dSph; $ \bm{\nu_{l}} $ is the set of nuisance parameters affecting the gamma-ray observations of the $l$-th dSph, excluding $ J_{l} $. $ J_{l} $ is the total $J$-factor of the $l$-th dSph as defined in Eq.~\ref{eq:total-j-factor}; we treat it as an additional nuisance parameter. The quantities $\log_{10} J_{l,\text{obs}}$ and $\sigma_{\log{J_l}}$ are obtained from fitting a log-normal function of $J_{l,\text{obs}}$ to the posterior distribution of $J_{l}$~\cite{2015PhRvL.115w1301A}; their values are listed in Table~\ref{tab:j-factor}. Note that when $\sigma_{\log{J_l}}$ is asymmetric, the negative value was chosen as this results in a more conservative limit. The likelihood term $\mathcal{J}_l$ constraining the value of $ J_{l} $ can thus be written as:
\begin{equation}
    \mathcal{J}_l \left( J_l \mid J_{l,\text{obs}}, \sigma_{\log{J_l}} \right) = \frac{1}{\ln{(10)} J_{l,\text{obs}} \sqrt{2\pi} \sigma_{\log{J_l}}} \exp{\left(-\frac{\left( \log_{10} J_l - \log_{10} J_{l,\text{obs}} \right)^{2}}{2\sigma_{\log{J_l}}^{2}}\right)}.
    \label{eq:lkl-j-factor}
\end{equation}
Both sets of $J$-factors in Table~\ref{tab:j-factor} will be used independently in our analysis. Note that the right-hand side of Eq.~\ref{eq:lkl-j-factor} is normalized such that it can be interpreted both as the likelihood function for $J_l$ and as the probability density function (PDF) for the associated random variable $J_{l,\text{obs}}$. Furthermore, the quantities $\sv$ and $J_{l}$ are degenerate in the computation of $\mathcal{L}_{\text{dSph},l}$, which depends on $\frac{d\Phi}{dE}$ (see Eq.~\ref{eq:dm-flux}). Therefore, as noted in~\cite{2016JCAP...02..039M}, it is sufficient to compute $\mathcal{L}_{\text{dSph},l}$ versus $\sv$ for a fixed value of $J_{l}$. We use the $J_{l,\text{obs}}(\mathcal{GS})$ reported in Table~\ref{tab:j-factor} to perform the profiling of $\mathcal{L}$ with respect to $J_l$. The degeneracy implies that for any $J'_l \neq J_{l,\text{obs}}$ (in practice in our case we used $J'_l = J_{l,\text{obs}}(\mathcal{B})$ to compute results from a different set of $J$-factors as explained in Section~\ref{sec:dm}): 
\begin{equation}
    \mathcal{L}_{\text{dSph},l}\left( \sv ; J'_l, \bm{\nu_l} \mid \bm{\mathcal{D}_{l}} \right) = \mathcal{L}_{\text{dSph},l}\left( \frac{J'_l}{J_{l,\text{obs}}}\sv ; J_{l,\text{obs}}, \bm{\nu_l} \mid \bm{\mathcal{D}_{l}} \right),
    \label{eq:j-factor-scaling}
\end{equation}
which is a straightforward rescaling operation that reduces the computational needs of the profiling operation since:
\begin{equation}
    \mathcal{L} \left( \sv ; \bm{\hat{\nu}} \mid \bm{\mathcal{D}_{\text{dSphs}}} \right) = \prod_{l=1}^{N_{\text{dSphs}}} \text{max}_{J_l} \bigg[ \mathcal{L}_{\text{dSph},l} \left( \sv ; J_{l}, \bm{\hat{\nu}_{l}} \mid \bm{\mathcal{D}_{l}} \right) \times \mathcal{J}_l \left( J_{l} \mid J_{l,\text{obs}}, \sigma_{\log_{J_l}} \right) \bigg].
    \label{eq:lkl-profile-over-j}
\end{equation}
In addition, Eq.~\ref{eq:j-factor-scaling} enables the combination of data from different gamma-ray instruments and observed dSphs via tabulated values of $\mathcal{L}_{\text{dSph},l}$ versus $\sv$. In this work, we equivalently use tabulated values of $\lambda$ from Eq.~\ref{eq:likelihood-profile} versus $\sv$. The likelihood $\mathcal{L}_{\text{dSph},l}$ is computed for a fixed value of $J_l$ and profiled with respect to all instrumental nuisance parameters $\bm{\nu_l}$. These nuisance parameters are discussed in more detail below. These values are produced by each detector independently and therefore there is no need to share low-level information used to produce them, such as event lists or IRFs.

\begin{figure}
\centering{
\includegraphics[scale=0.85]{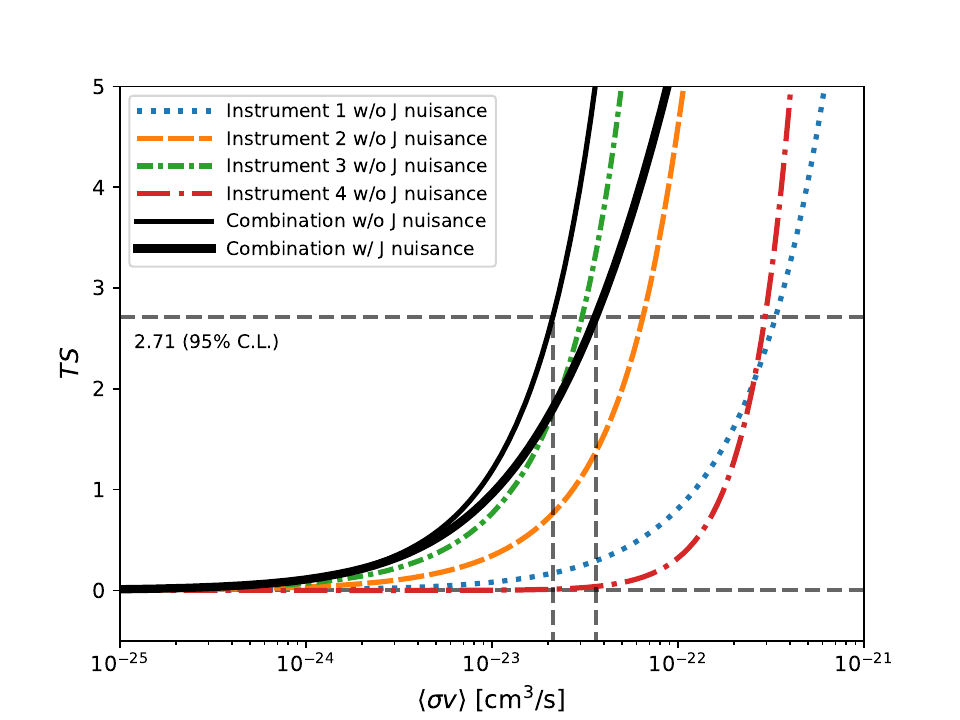}}
\caption{Illustration of a real data combination showing a comparison between $\mathrm{TS}$ provided by four instruments (non-solid colored lines) from the observation of the same dSph without any $J$-factor nuisance parameter included and their sum, i.e. the resulting combined likelihood (thin black line), for a DM particle mass of 20~TeV. The intersection of the likelihood profiles with the line $\mathrm{TS}$ = 2.71 indicates the 95\% C.L. upper limit on $\langle \sigma v \rangle$ (see text for more details). The combined likelihood (thin black line) shows a smaller value of upper limit on $\langle \sigma v \rangle$ than those derived by individual instruments. We also show how the uncertainty on the $J$-factor affects the combined likelihood and degrades the upper limit on $\langle \sigma v \rangle$ (thick black line). All likelihood profiles are normalized so that the global minimum $\widehat{\langle \sigma v \rangle}$ is 0 to facilitate data handling. We note that each profile depends on the observational conditions under which a target object was observed. The sensitivity of a given instrument can be degraded and the upper limits less constraining if the observations suffer from non-optimal conditions such as a large zenith angle of observation or a short exposure time.}
\label{fig:exemple-combination}
\end{figure}

Figure~\ref{fig:exemple-combination} illustrates the multi-instrument combination technique used in this study with a comparison of the upper limit on $\sv$ obtained from the combination of the observations of four experiments towards one dSph versus the upper limit from individual instruments. It also shows graphically the effect of the $J$-factor uncertainty on the combined observations.

The \textit{partial} joint likelihood function for gamma-ray observations of each dSph ($\mathcal{L}_{\text{dSph},l}$) is written as the product of the likelihood terms describing the $N_{\text{exp},l}$ observations performed with any of our observatories:
\begin{equation}
    \mathcal{L}_{\text{dSph},l}\left( \sv ; J_l, \bm{\nu_l} \mid \bm{\mathcal{D}_{l}} \right) = \prod_{k=1}^{N_{\text{exp},l}} \mathcal{L}_{lk} \left( \sv ; J_l, \bm{\nu_{lk}} \mid \bm{\mathcal{D}_{lk}} \right),
    \label{eq:dSph-lkl}
\end{equation}
where each $\mathcal{L}_{lk}$ term refers to an observation of the $l$-th dSph with associated $k$-th IRFs. The quantity $ N_{\text{exp},l} $ varies from dSph to dSph and can be retrieved from Table~\ref{tab:summary}.

Each collaboration separately analyzes their data for $\bm{\mathcal{D}_{lk}}$ corresponding to dSph $l$ and gamma-ray detector $k$, using as many common assumptions as possible in the analysis (see the next paragraph and further down in this section for additional details). We compute the values for the likelihood functions $\mathcal{L}_{lk}$ (see Eq.~\ref{eq:dSph-lkl}) for a fixed value of $J_l$ and profile over the rest of the nuisance parameters $\bm{\nu_{lk}}$. Then, values of $\lambda$ from Eq.~\ref{eq:likelihood-profile} are computed as a function of $\sv$ and shared using a common format. Results are computed for seven annihilation channels, $W^+W^-$, $ZZ$, $b\bar{b}$, $t\bar{t}$, $e^+e^-$, $\mu^+\mu^-$, and $\tau^+\tau^-$ over the 62 $m_{\textrm{DM}}$ values between 5 GeV and 100 TeV provided in~\cite{2011JCAP...03..051C}. The $\sv$ range is defined between $10^{-28}$ and $10^{-18}\rm{cm}^{3}\cdot \rm{s^{-1}}$, with 1001 logarithmically spaced values. The likelihood combination, i.e. Eq.~\ref{eq:dSph-combination}, and profile over the $J$-factor to compute the profile likelihood ratio $\lambda$, i.e. Eq.~\ref{eq:likelihood-profile}, are carried out with two different public analysis software packages, namely \texttt{gLike}~\cite{rico_2022_7342721} and \texttt{LklCom}~\cite{tjark_miener_2021_4597500}, which provide the same results~\cite{2021arXiv211201818M}.

As mentioned previously, each experiment computes the $\mathcal{L}_{lk}$ from Eq.~\ref{eq:dSph-lkl} differently. The remainder of this section will describe the main differences between the approaches of each experiment. Four experiments, namely \textit{Fermi}-LAT, HAWC, H.E.S.S. and MAGIC, use a binned likelihood to compute the $\mathcal{L}_{lk}$. VERITAS uses an unbinned likelihood approach described in detail at the end of this section. For the four experiments using a binned likelihood, for each observation $\bm{\mathcal{D}_{lk}}$ of a given dSph $l$ carried out using a given gamma-ray detector $k$, the binned likelihood function is: 
\begin{equation}
    \begin{split}
    \mathcal{L}_{lk} \left( \sv ; J_l, \bm{\nu_{lk}} \mid \bm{\mathcal{D}_{lk}} \right) = \prod_{i=1}^{N_{\text{E'}}} \prod_{j=1}^{N_{\text{P'}}} \Big[ \mathcal{P} (s_{lk,ij}(\sv, J_l, \bm{\nu_{lk}}) + b_{lk,ij}(\bm{\nu_{lk}}) \mid N_{lk,ij}) \Big]\\ \times \mathcal{L}_{lk,\bm{\nu}} \left( \bm{\nu_{lk}} \mid \bm{\mathcal{D}_{\nu_{lk}}} \right),
    \end{split}
    \label{eq:binned-lkl}
\end{equation}
where $N_{\text{E'}}$ and $N_{\text{P'}}$ are the number of considered bins in reconstructed energy and arrival direction, respectively. $\mathcal{P}$ represents a Poisson PDF for the number of gamma-ray candidate events $ N_{lk,ij} $ observed in the $i$-th bin in energy and $j$-th bin in arrival direction, where the expected number is the sum of the expected mean number of signal events $ s_{lk,ij} $ (produced by DM annihilation) and of background events $ b_{lk,ij} $. $\mathcal{L}_{lk,\bm{\nu}}$ is the likelihood term for the extra $ \bm{\nu_{lk}} $ nuisance parameters that vary from one instrument $k$ to another. The expected counts for signal events $s_{ij}$ for a given dSph $l$ and detector $k$ is given by:
\begin{equation}
    \begin{split}
    s_{ij}(\sv, J) = \int_{E'_{\text{min},i}}^{E'_{\text{max},i}} \text{d}E' \int_{P'_{\text{min},j}}^{P'_{\text{max},j}} \text{d}\Omega' \int_{0}^{\infty} \text{d}E \int_{\Delta\Omega_{tot}} \text{d}\Omega \int_{t_{\text{min}}}^{t_{\text{max}}} \text{d}t \frac{\text{d}^{2}\Phi (\sv, J)}{\text{d}E \text{d}\Omega}\\ \times \text{IRF} \left( E', P' \mid E, P, t \right),
    \end{split}
    \label{eq:signal-events}
\end{equation}
where $E'$ and $E$ are the reconstructed and true energies, $P'$ and $P$ the reconstructed and true arrival directions; $E'_{\text{min},i}$, $P'_{\text{min},j}$, $E'_{\text{max},i}$, and $P'_{\text{max},j}$ are their lower and upper limits of the $i$-th energy bin and the $j$-th arrival direction bin; $t_{\text{min}}$ and $t_{\text{max}}$ delimit the time interval during which the dSph $l$ was observed by the detector $k$; $t$ is the time along the observations; $\text{d}^{2}\Phi/\text{d}E\text{d}\Omega$ is the gamma-ray flux in the source region (see Eq.~\ref{eq:dm-flux}); and $\text{IRF} \left( E', P' \mid E, P, t \right)$ is the time-dependent IRF (IRF=0 in periods when there are no observations or during the detectors' deadtime). The IRF can be factored as the product of the effective collection area of the detector, $A_{\mathrm{eff}} (E, P, t)$, the PDF for the energy estimator, $f_{E} (E' \mid E,t)$, and the PDF for the arrival direction estimator, $f_{P} (P' \mid E,P,t)$: $\text{IRF} ( E', P' \mid E, P, t ) = A_{\mathrm{eff}} (E, P, t) \times f_{E} (E' \mid E,t) \times f_{P} (P' \mid E,P,t)$. Note that for \textit{Fermi}-LAT, HAWC, MAGIC, and VERITAS the effect of the finite angular resolution is taken into account through the convolution of $\text{d}^2\Phi/\text{d}E \text{d}\Omega$ with $f_{P}$ in Eq.~\ref{eq:signal-events}, whereas in the case of H.E.S.S., $f_{P}$ is approximated by a delta function. This approximation has been made in order to maintain compatibility of the result with what has been previously published. The difference introduced by this approximation in the resulting s$_{ij}$ is $<5\%$ for all considered dSphs.

As mentioned above, the extra nuisance parameters designated by the $\mathcal{L}_{lk,\bm{\nu}}$ term in Eq.~\ref{eq:binned-lkl} vary from one detector to another. One such difference is the treatment of the background. Analyses with \textit{Fermi}-LAT and HAWC data model the background using a template fitting approach; therefore there is no $\mathcal{L}_{lk,\bm{\nu}}$ term in Eq.~\ref{eq:binned-lkl} for either detector. On the other hand, IACTs (H.E.S.S., MAGIC, VERITAS) rely on ON/OFF measurements to estimate the background, which implies the use of an additional likelihood term $\mathcal{L}_{lk,\bm{\nu}}$ to describe this extra nuisance parameter. In practice, this additional likelihood corresponds to a Poisson term for the OFF observations $\mathcal{P} (\tau b \mid N_{\text{OFF}})$, where $N_{\text{OFF}}$ is the number of observed events in the OFF region, $b$ is the number of expected background events, and the quantity $\tau$ is the acceptance-corrected exposure ratio between the ON and OFF regions. OFF regions are taken sufficiently far away from the signal region such that negligible DM signal is expected in the OFF regions\footnote{The DM halos of several targets (Carina, Draco, Fornax) are expected to be sufficiently extended that defining a signal-free background region is challenging. However, this leads to conservative estimation of the signal significance and the limits on the velocity-weighted annihilation cross section.}. This latter quantity, $\tau$, is another nuisance parameter for this approach of background estimation and is therefore described by another likelihood term $\mathcal{T}$ including the statistical and systematic uncertainties on $\tau$. 

In this work, VERITAS uses a different likelihood than Eq.~\ref{eq:binned-lkl} (see below for more details). Therefore, in practice, only H.E.S.S. and MAGIC use in their analysis the likelihood ($\mathcal{L}_{lk,\bm{\nu}}$) with the extra nuisance parameters. In general, it is written as:
\begin{equation}
    \mathcal{L}_{lk,\bm{\nu}} \left( \bm{\nu_{lk}} \mid \bm{\mathcal{D}_{\nu_{lk}}} \right) = \mathcal{P} (\tau b_{lk,ij} \mid N_{\text{OFF},lk,ij}) \times \mathcal{T} \left( \tau \mid \tau_{\text{obs}}, \sigma_{\tau} \right),
    \label{eq:lkl-nuisance-iacts}
\end{equation}
where $N_{\text{OFF},lk,ij} $ and $b_{lk,ij}$ are respectively the number of observed events in the OFF region and the number of expected background events in the $i$-th bin of reconstructed energy and the $j$-th bin of reconstructed arrival direction. The PDF $\mathcal{T}$ for $\tau$ consists of a Gaussian function with mean $\tau_{\text{obs}}$ and standard deviation $\sigma_{\tau} = \sqrt{\sigma_{\tau, \text{stat}}^{2} + \sigma_{\tau, \text{syst}}^{2}}$, which includes statistical and systematic uncertainties, $\sigma_{\tau, \text{stat}}$ and $\sigma_{\tau, \text{syst}}$ respectively~\cite{2018JCAP...03..009A}. In practice, only the MAGIC analysis includes the treatment of $\tau$ as a nuisance parameter due to its deep exposure, and therefore high statistics, which implies the possibility of reaching the systematic ceiling of its sensitivity. In the analysis, MAGIC considers a systematic uncertainty of $ \sigma_{\tau, \text{syst}} = 0.015 \cdot \tau_{\text{obs}}$ in the background estimation based on a dedicated performance study~\cite{2016APh....72...76A}. Note that the value of $\tau$ is the result of the profiling over the likelihood $\mathcal{T}$ and is always close to the mean value $\tau_{\text{obs}}$ computed as the ratio of the number of events in regions adjacent to the ON and OFF regions. H.E.S.S. finds this systematic effect to be negligible compared to the statistical uncertainties for their given datasets. We also note that Eq.~\ref{eq:lkl-nuisance-iacts} and the above discussion concerns nuisance parameters related to the background estimate but there are other instrumental nuisance parameters. In particular, here we are overlooking uncertainties in the IRFs that may arise from discrepancies between the Monte Carlo simulations and the data.

Instead of the binned likelihood from Eq.~\ref{eq:binned-lkl}, VERITAS adopted an unbinned likelihood approach as described in~\cite{2017ICRC...35..904Z}, testing a null hypothesis (only background) against an alternative one (signal plus background hypothesis). The likelihood function is composed of the product of two Poisson probabilities in the ON and OFF regions, including the expected spectral shape information (known for a given DM model) in order to maximize the sensitivity to DM-originated gamma-ray signals. This likelihood function is defined in the following way:
\begin{equation}
    \mathcal{L}_{lk} \left( \sv ; J_l, \bm{\nu_{lk}} \mid \bm{\mathcal{D}_{lk}} \right) = \frac{(s(\sv,J_l)+b)^{N_{\mathrm{ON}}}e^{-(s(\sv,J_l)+ b)}}{N_{\mathrm{ON}}!}\frac{(\tau b)^{N_{\mathrm{OFF}}}e^{-\tau b}}{N_{\mathrm{OFF}}!}\prod_{i=1}^{N_{\mathrm{ON}}}{P_{i}(E_{i})}
    \label{eq:unbined-lkl}
\end{equation}
where $N_{\mathrm{ON}}$ and $N_{\mathrm{OFF}}$ are, respectively, the number of counts in the ON and OFF region, $\tau$ is the background normalization, $b$ is the number of expected background events, $P_{i}$ is the likelihood of the $i$-th event in the ON region to be reconstructed with measured energy $E_{i}$, and $s(\sv,J_l)$ is the number of expected signal events. No extra nuisance parameters were considered in this unbinned likelihood approach; therefore there is no $\mathcal{L}_{lk,\bm{\nu}}$ term in Eq.~\ref{eq:unbined-lkl}. Additionally, the VERITAS likelihood considers an energy-dependent point spread function with which the \textit{J}-profile is convolved. To approximate the $J$-factor treatment of the other instruments, a single $J$-factor is used, derived from the energy-averaged convolved \textit{J}-profile. Compared to the binned likelihood approach, the unbinned likelihood method used by VERITAS results in comparable limits. A more comprehensive review of the differences between the analyses of different instruments can be found in~\cite{2020Galax...8...25R}.

\section{Results\label{sec:results}}

No significant DM emission was detected with the combined analysis of data from the five instruments. We present the upper limits on $\sv$ assuming seven independent DM self-annihilation channels, namely $W^+W^-$, $ZZ$, $b\bar{b}$, $t\bar{t}$, $e^+e^-$, $\mu^+\mu^-$, and $\tau^+\tau^-$. Our upper limits are determined by solving $\mathrm{TS}= 2.71 $, where $2.71$ corresponds to a one-sided 95\% confidence level assuming the test statistic behaves like a $\chi^2$ distribution with one degree of freedom. This assumption is not entirely accurate because of the degeneracy between $\sv$ and $J$ in the gamma-ray flux computation (see Eq.~\ref{eq:dm-flux}) and the fact that $J$ is considered as a nuisance parameter with a log-normal PDF. Yet, using simulations we verified that solving $\mathrm{TS}= 2.71 $ produces an over-coverage and not an under-coverage. We therefore choose to compute upper limits using this definition as it is widely used, which allows easier comparison to previous results, and because over-coverage corresponds to more conservative limits.

The 68\% and 95\% containment bands are produced from 300 Poisson realizations of the null hypothesis corresponding to each of the combined datasets. These 300 realizations are combined identically to the data. The containment bands and the median are extracted from the distribution of resulting limits on the null hypothesis. These 300 realizations are obtained either by Poisson simulations of the OFF observations, for HAWC, H.E.S.S., MAGIC, and VERITAS, or taken from real observations of empty fields of view in the case of \textit{Fermi}-LAT~\cite{2015PhRvL.115w1301A,2017ApJ...834..110A,2021PhRvD.103l3005D}.

\begin{figure}
\centering{
\vspace{-3.45cm}
\begin{tabular}{cc}
\includegraphics[width=0.49\textwidth]{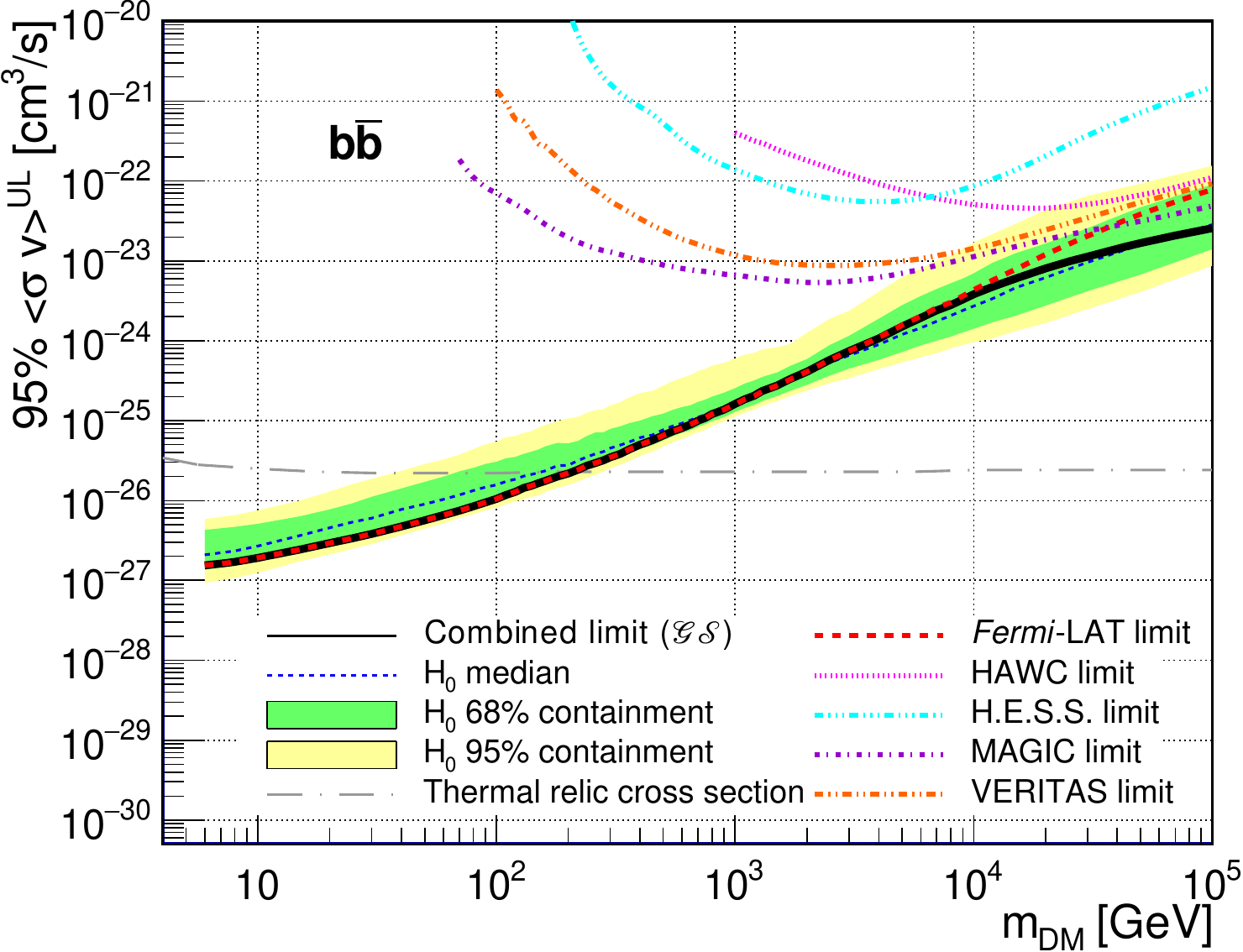} &
\includegraphics[width=0.49\textwidth]{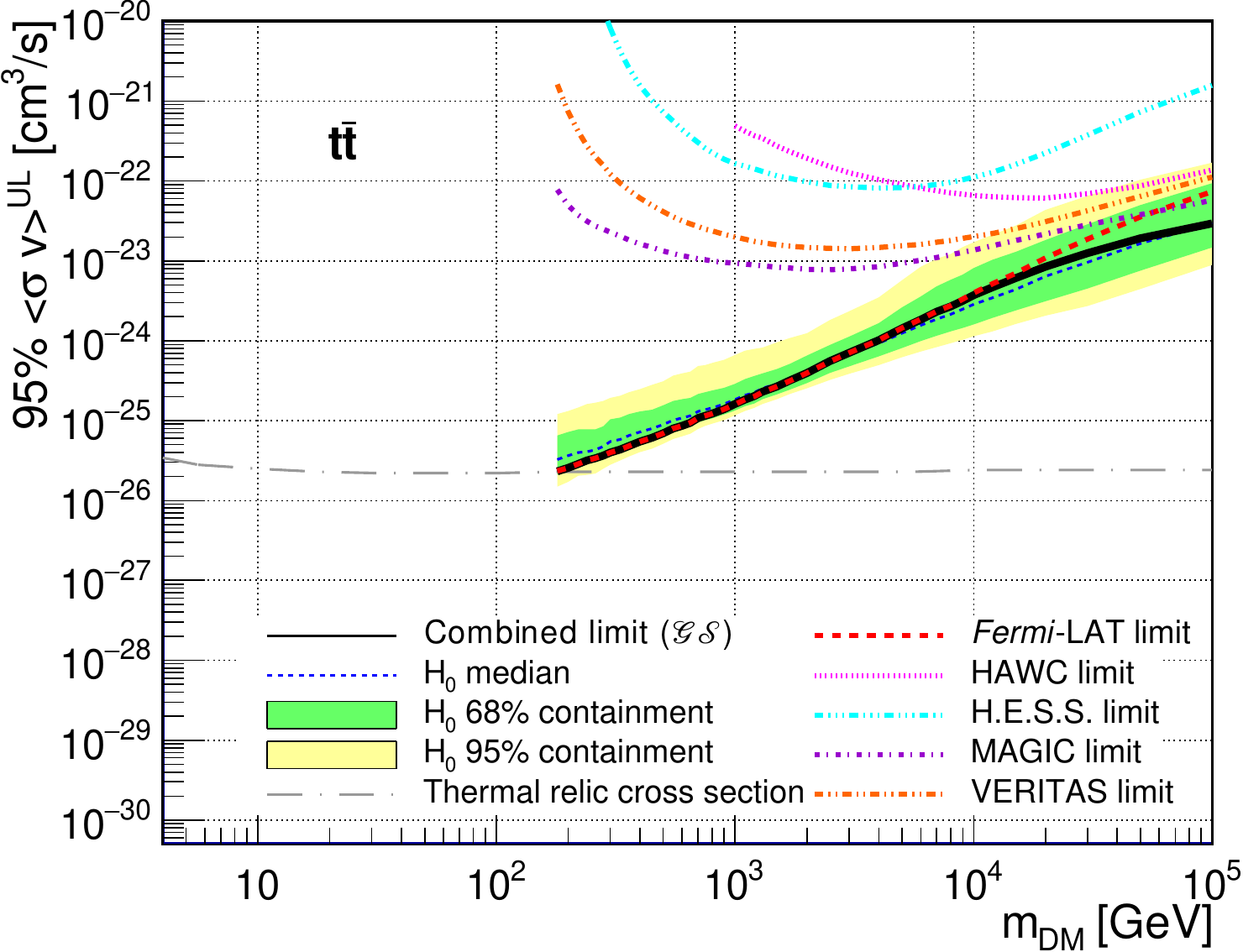} \\
\includegraphics[width=0.49\textwidth]{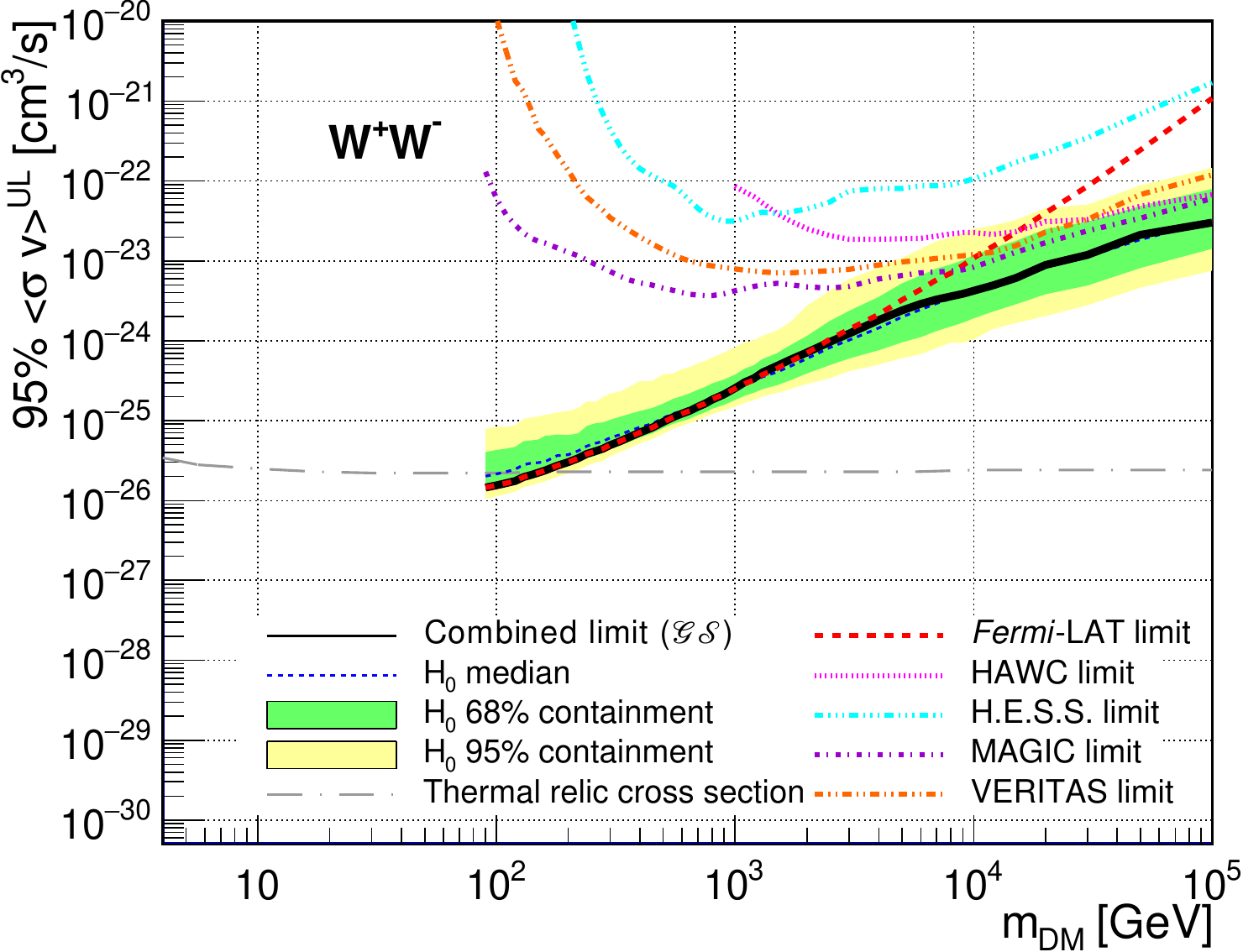} &
\includegraphics[width=0.49\textwidth]{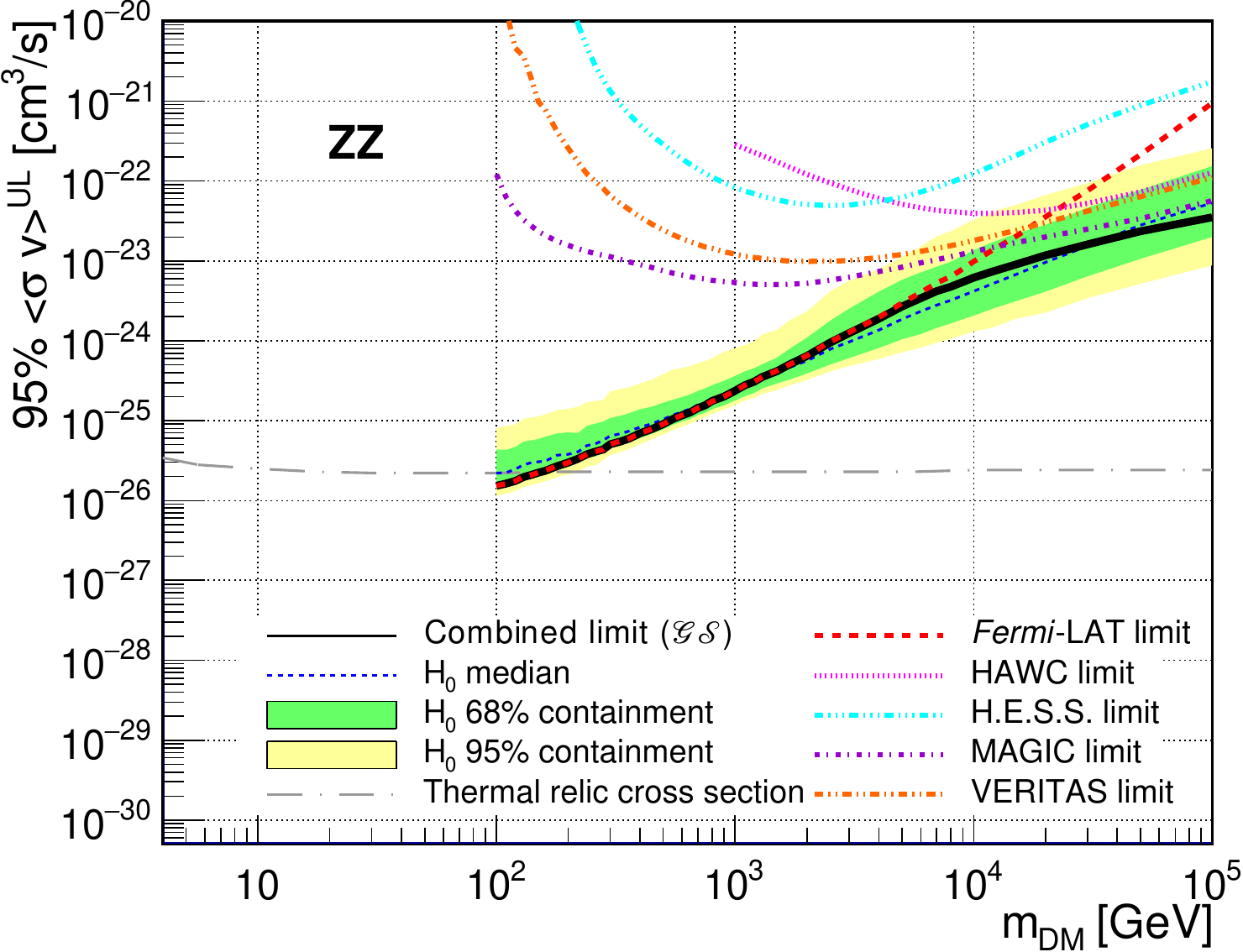} \\
\includegraphics[width=0.49\textwidth]{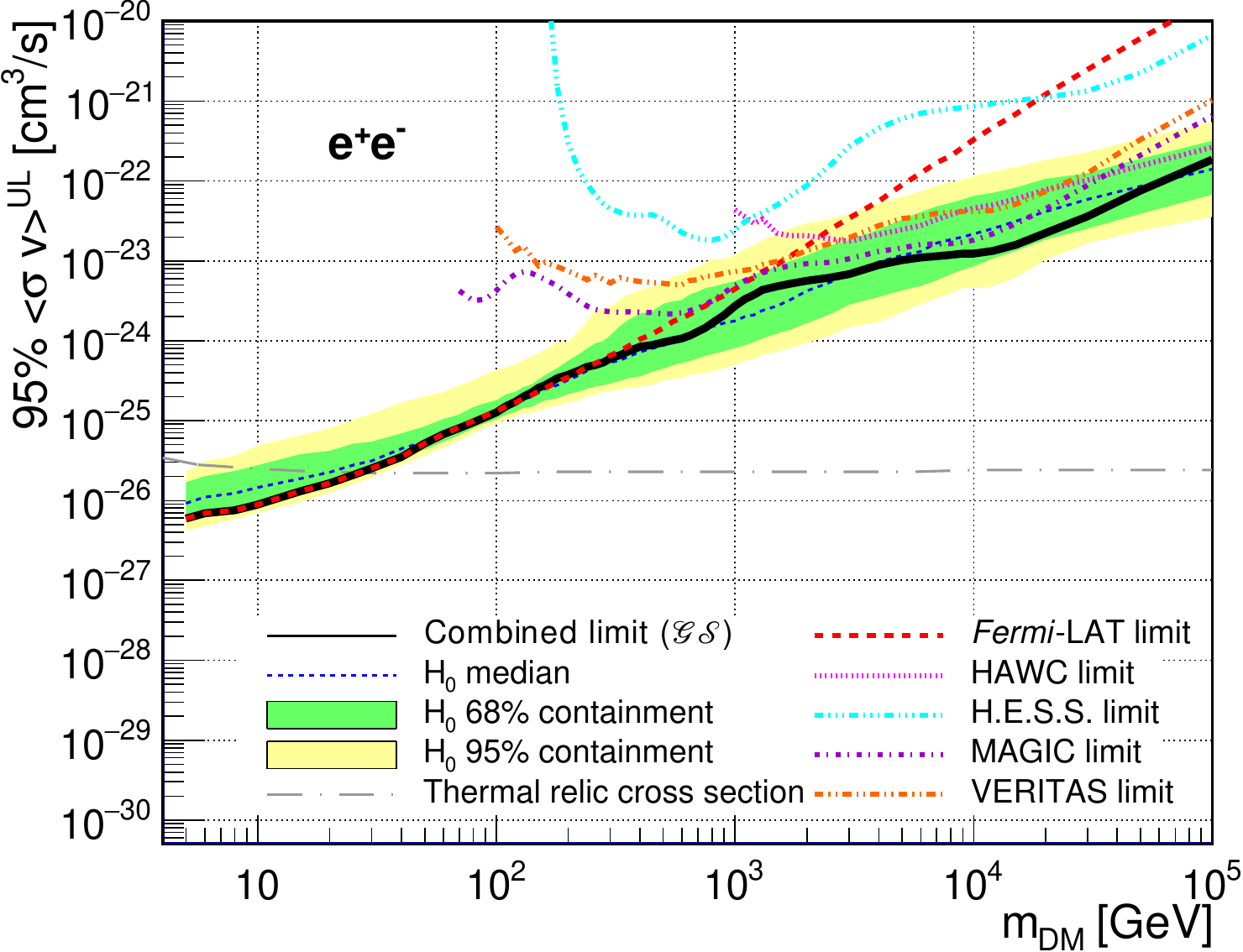} &
\includegraphics[width=0.49\textwidth]{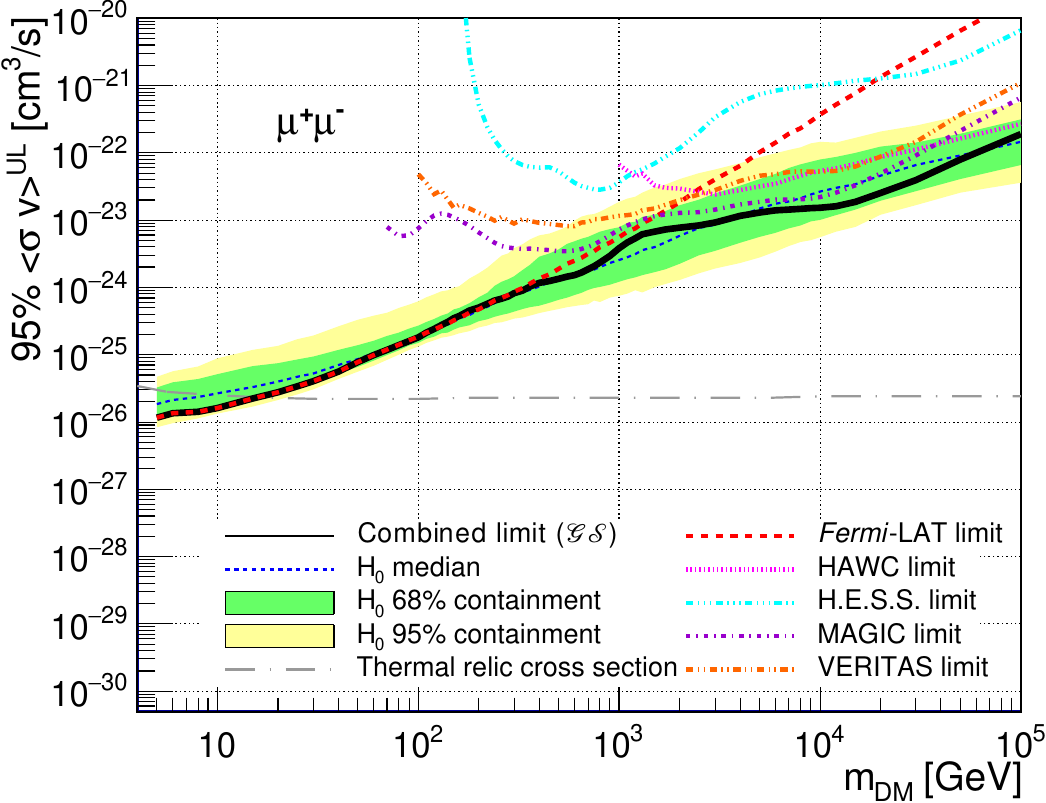} \\
\multicolumn{2}{c}{\includegraphics[width=0.49\textwidth]{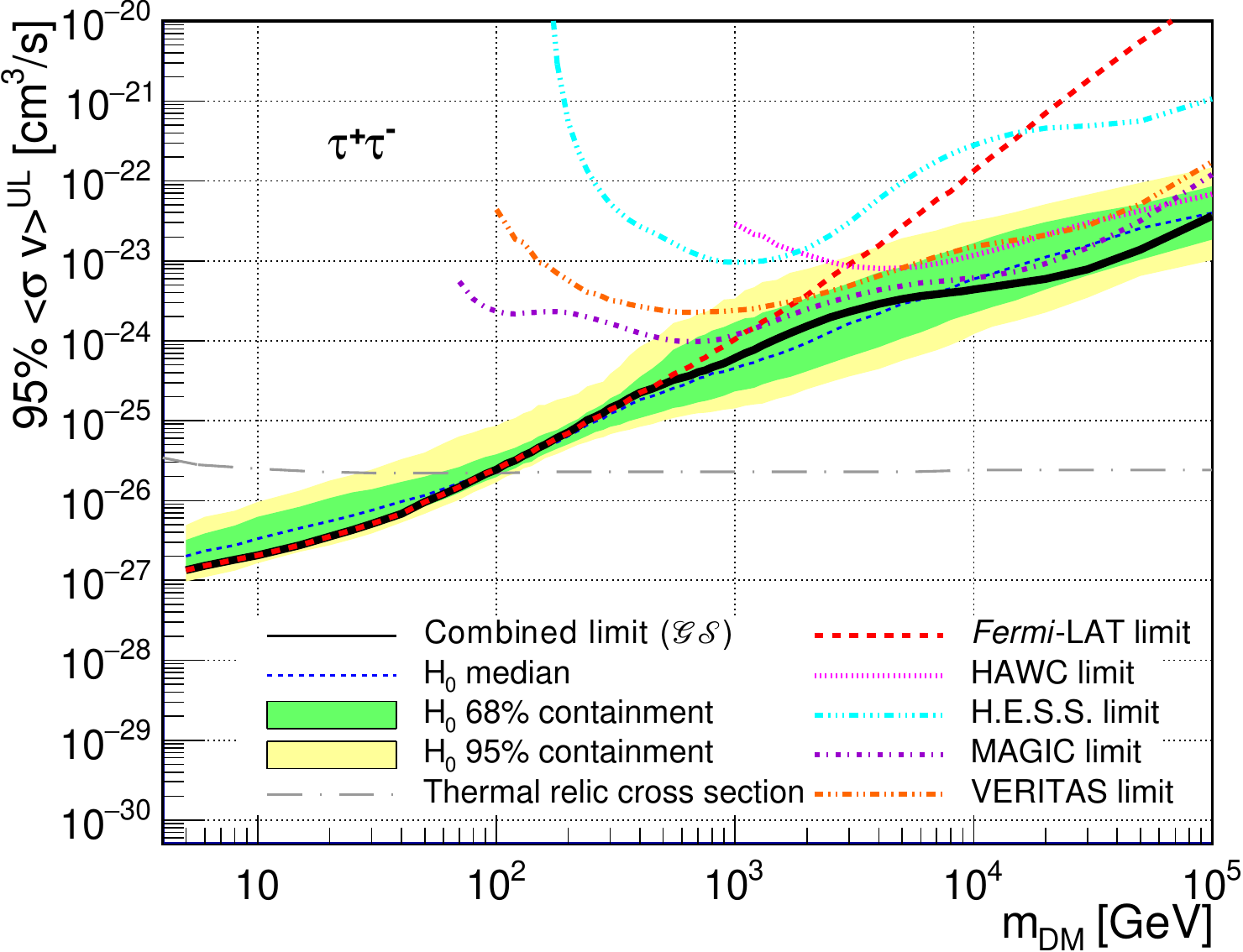}}
\end{tabular}
}
\vspace{-0.45cm}
\caption{Upper limits at 95\% confidence level on $\sv$ as a function of the DM mass for seven annihilation channels, using the set of $J$-factors from~\cite{2015ApJ...801...74G} ($\mathcal{GS}$ set in Table~\ref{tab:j-factor}). The black solid line represents the observed combined limits obtained for the 20 dSphs included in this work, the blue dashed line is the median of the null hypothesis (H$_0$) corresponding to the expected limits with no DM signal ($\langle \sigma v \rangle = 0 $), while the green and yellow bands show the 68\% and 95\% containment bands. Upper limits for each individual instrument are also indicated. The value of the thermal relic cross section as a function of the DM mass is given as the gray dotted-dashed line~\cite{2012PhRvD..86b3506S}.}
\label{fig:limits-geringer-sameth}
\end{figure}

\begin{figure}
\centering{
\vspace{-2cm}
\begin{tabular}{cc}
\includegraphics[width=0.49\textwidth]{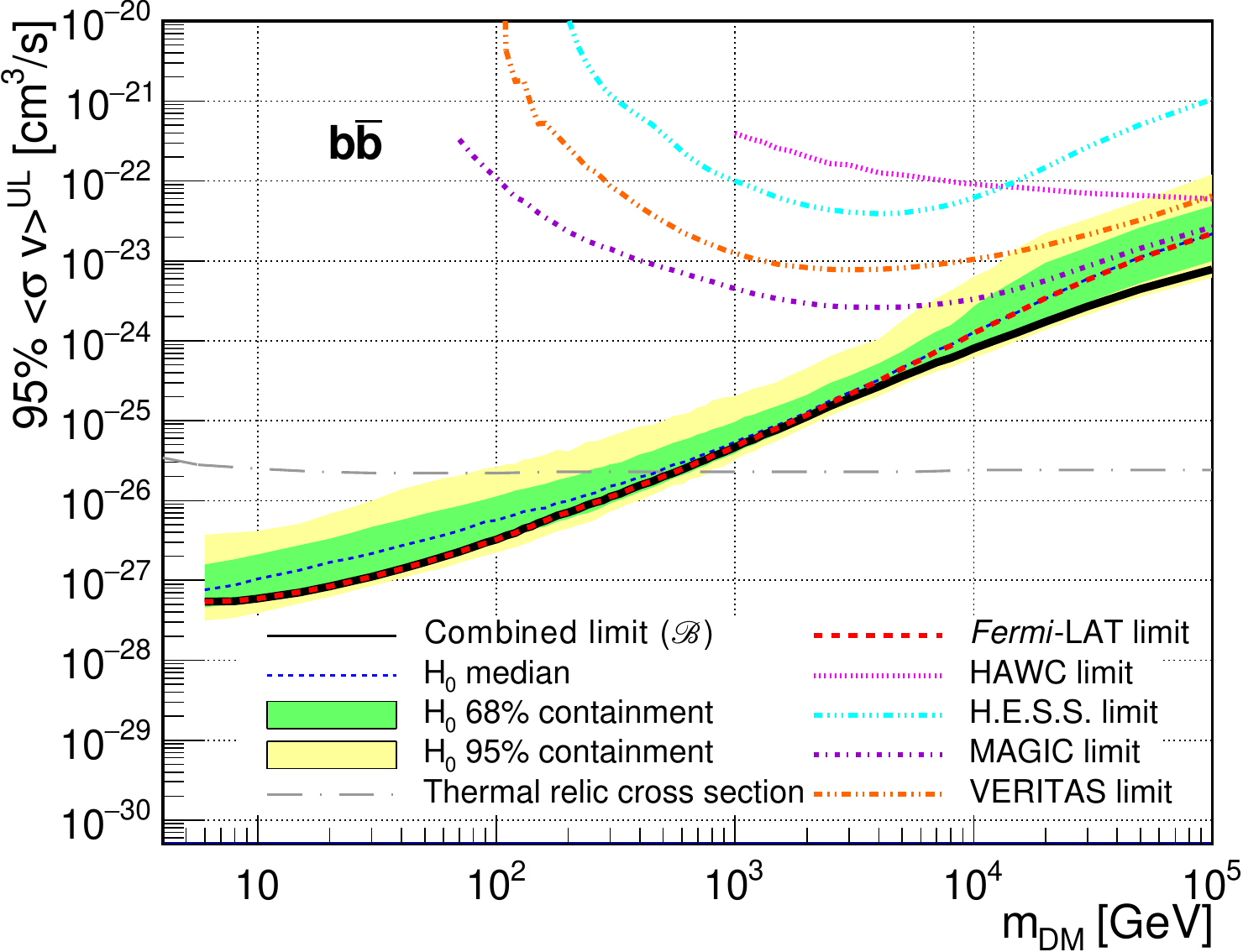} &
\includegraphics[width=0.49\textwidth]{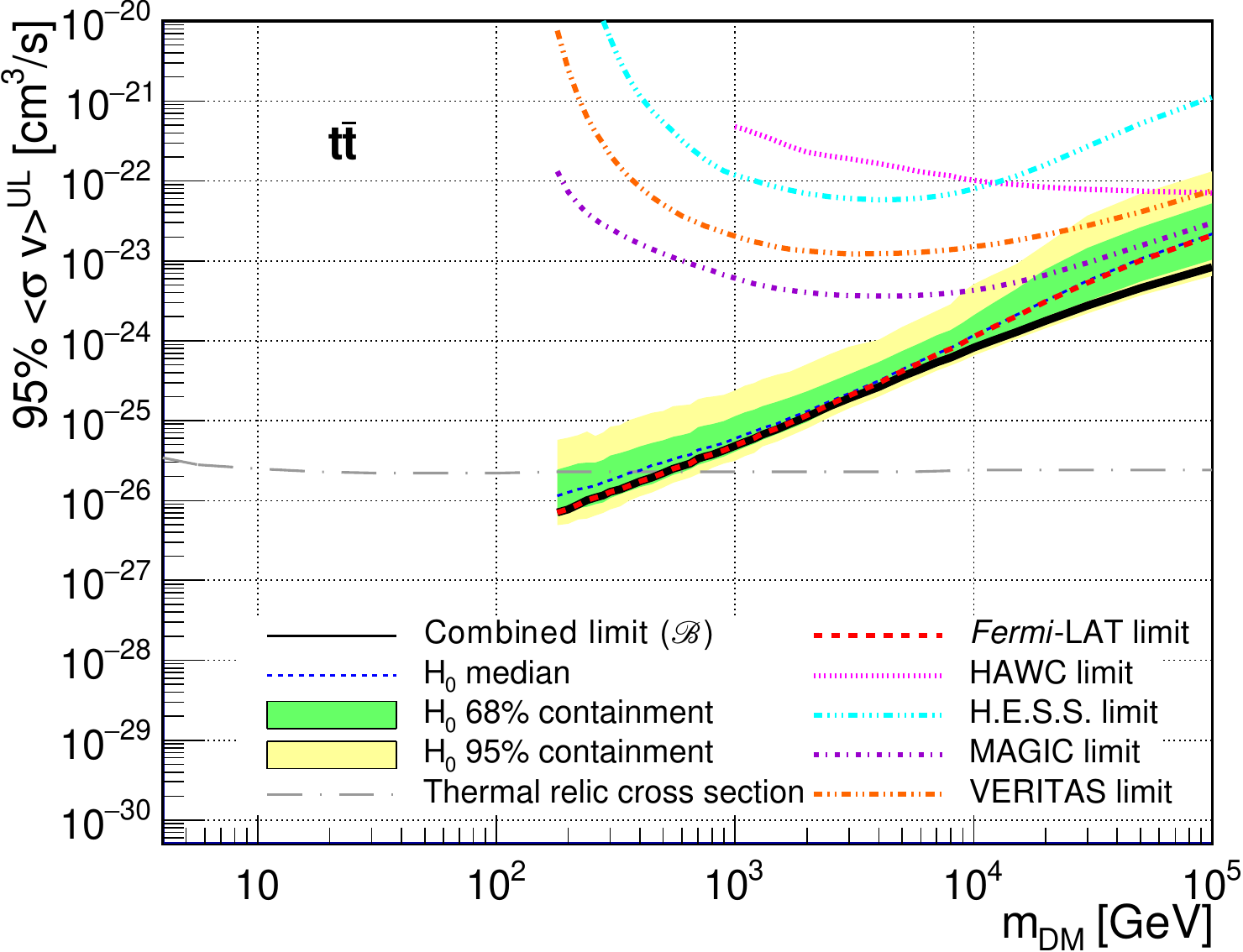} \\
\includegraphics[width=0.49\textwidth]{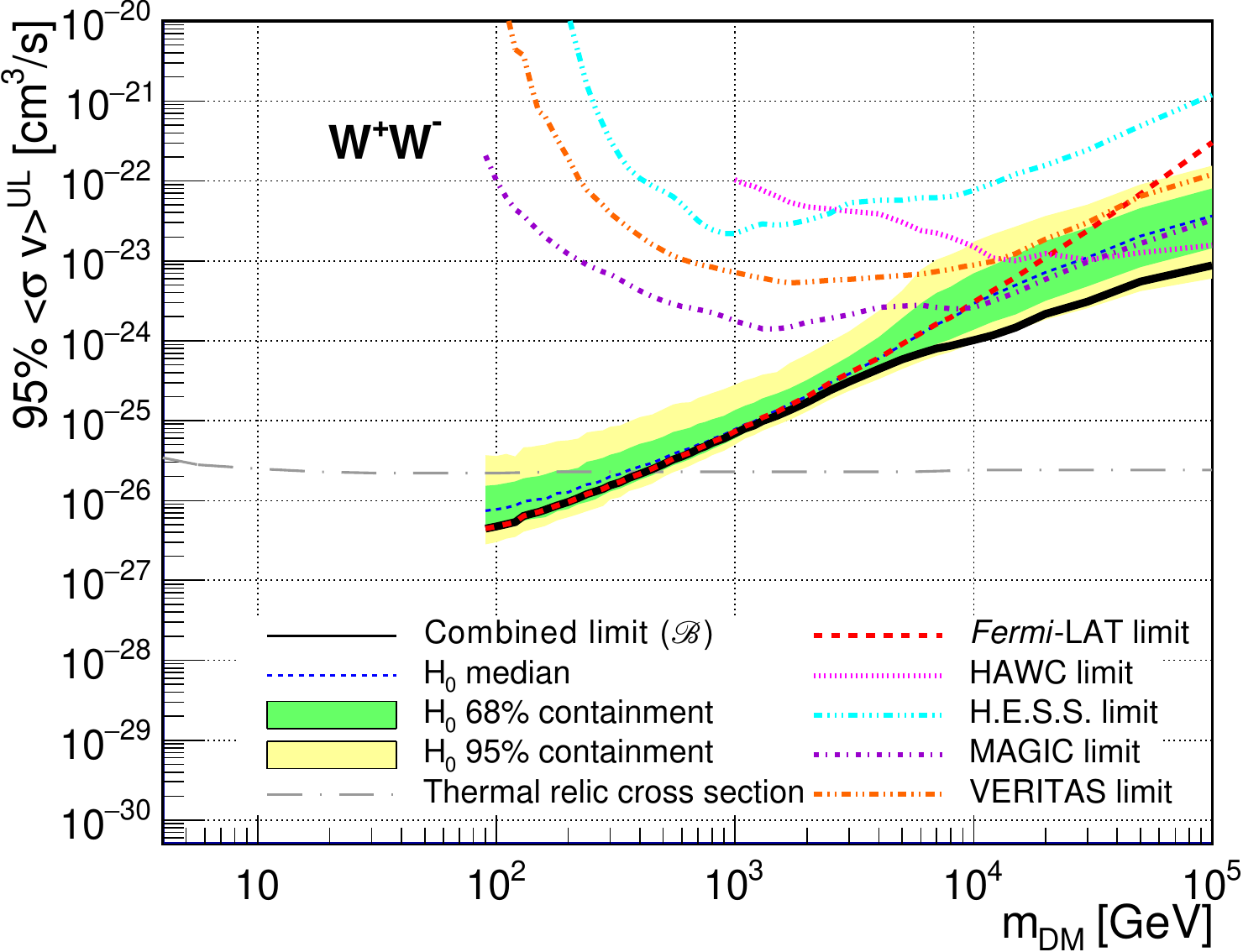} &
\includegraphics[width=0.49\textwidth]{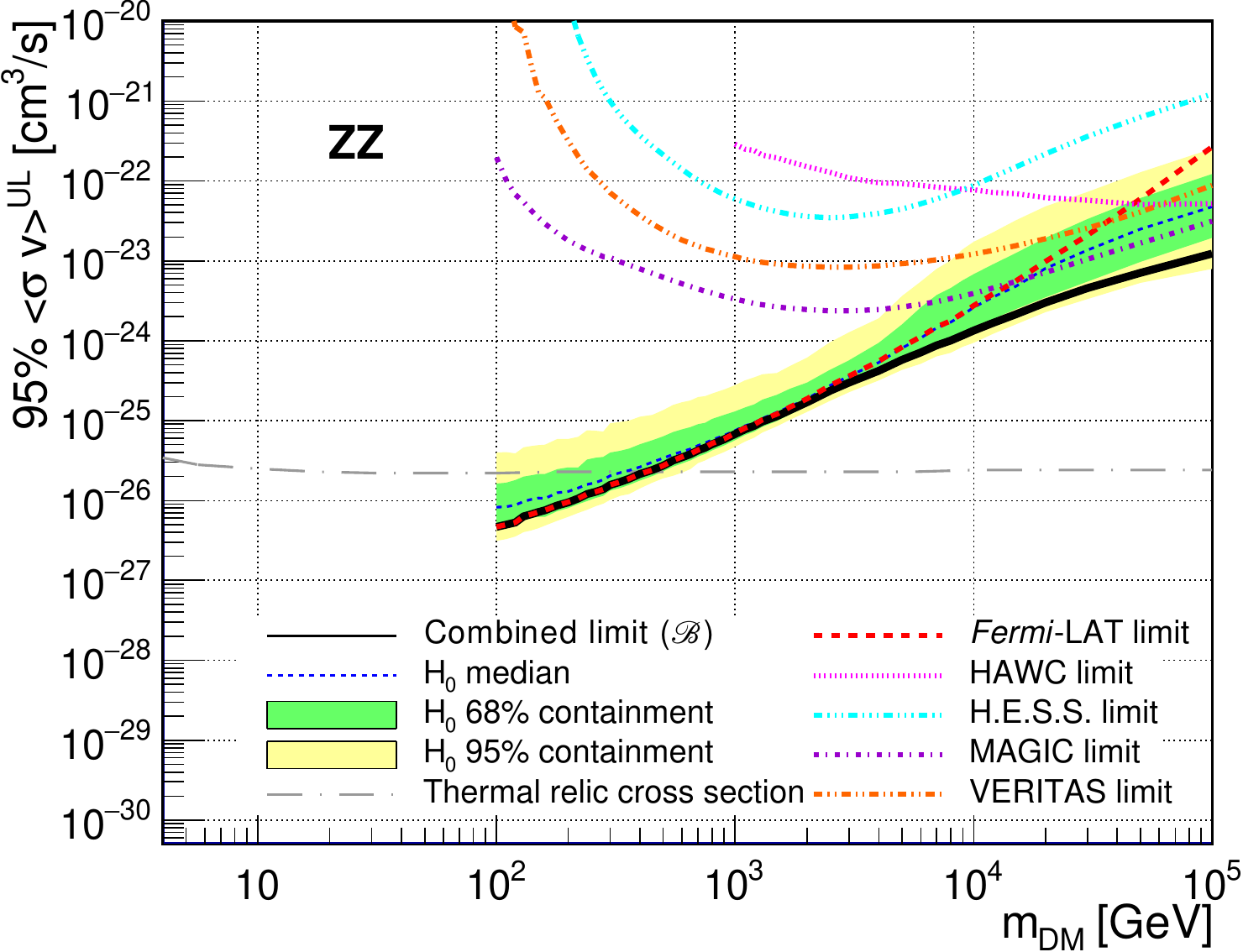} \\
\includegraphics[width=0.49\textwidth]{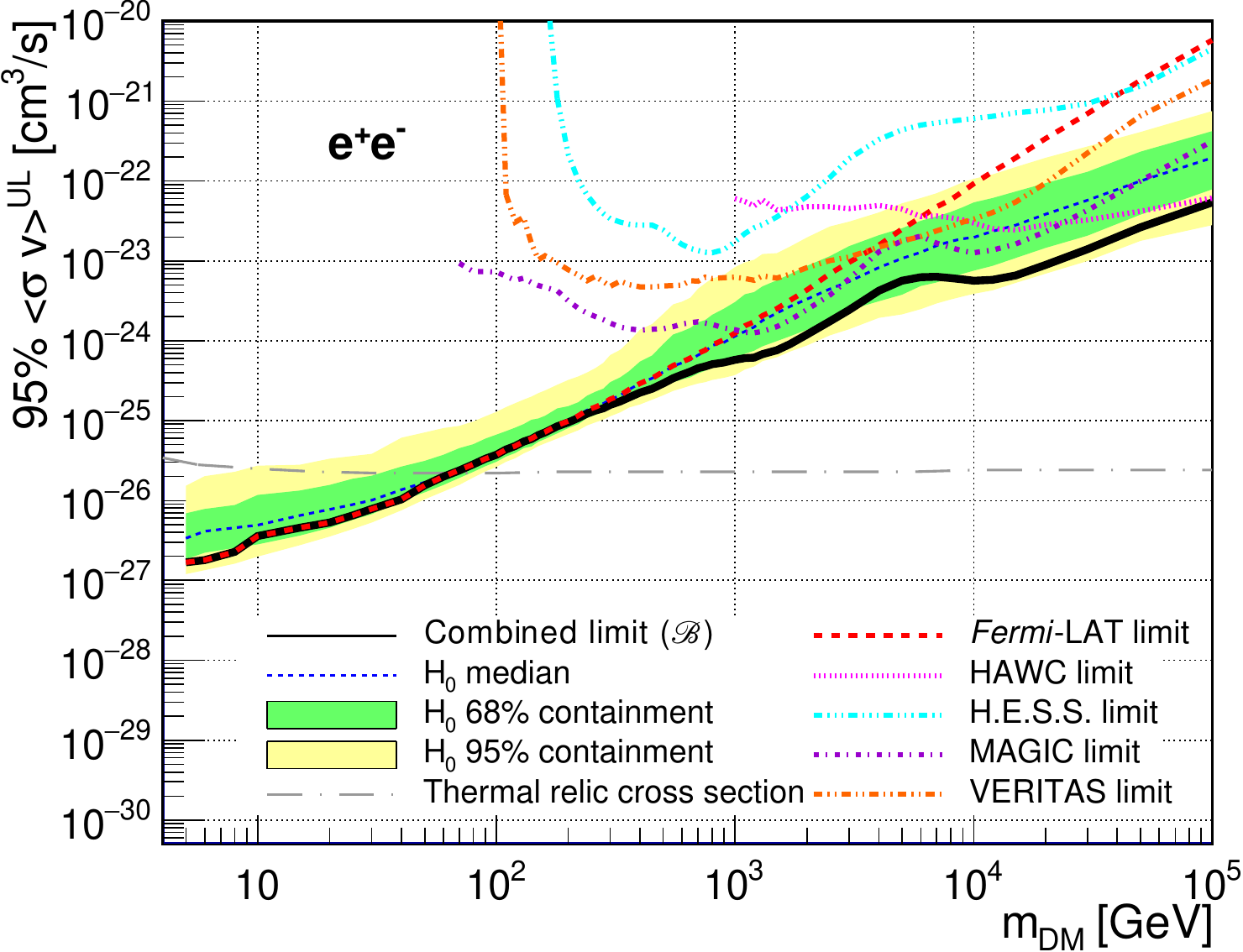} &
\includegraphics[width=0.49\textwidth]{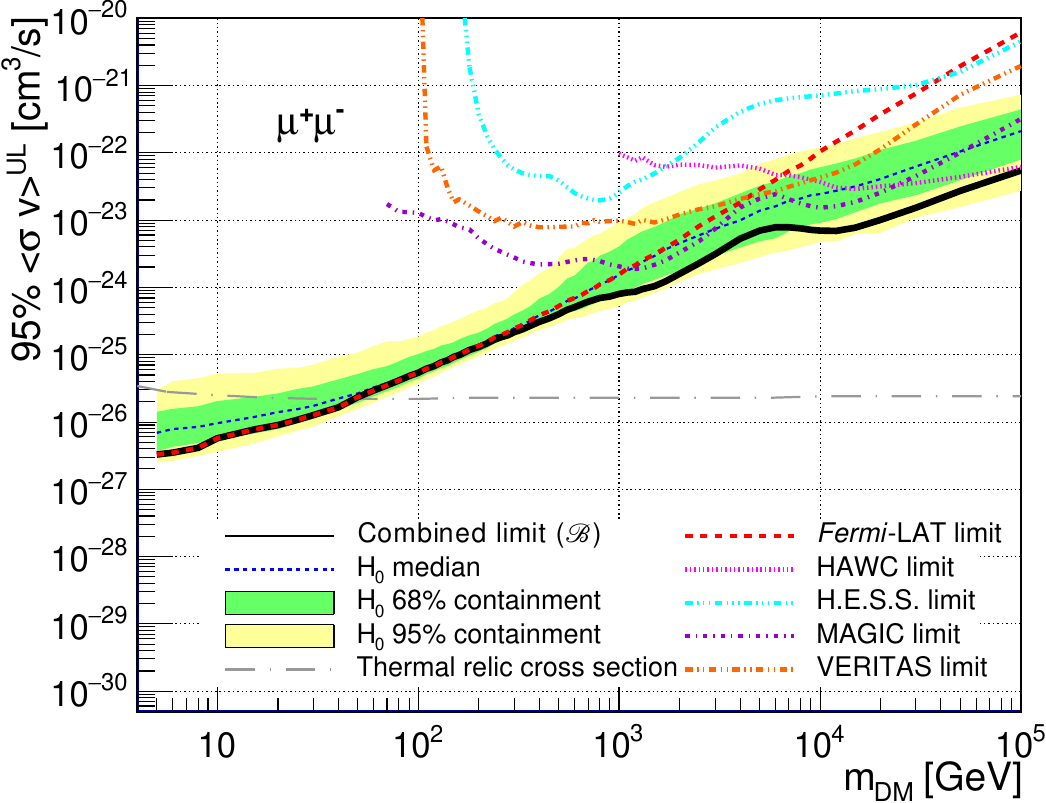} \\
\multicolumn{2}{c}{\includegraphics[width=0.49\textwidth]{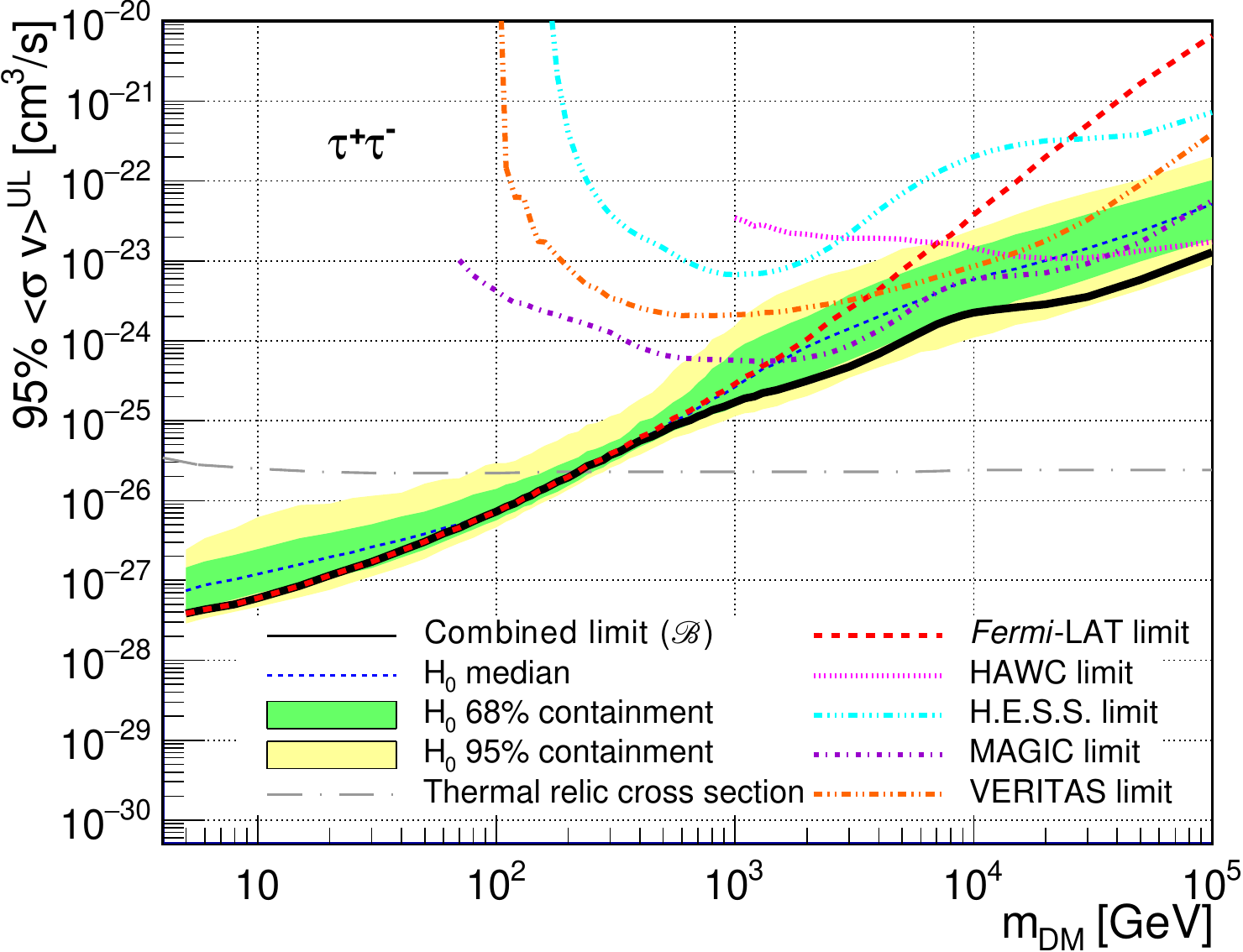}}
\end{tabular}
}
\caption{Same as Fig.~\ref{fig:limits-geringer-sameth}, using the set of $J$-factors from~\cite{2015MNRAS.446.3002B,2015MNRAS.453..849B} ($\mathcal{B}$ set in Table~\ref{tab:j-factor}).}
\label{fig:limits-bonnivard}
\end{figure}

The obtained limits are shown in Fig.~\ref{fig:limits-geringer-sameth} for the $\mathcal{GS}$ set of $J$-factors~\cite{2015ApJ...801...74G} and in Fig.~\ref{fig:limits-bonnivard} for the $\mathcal{B}$ set of $J$-factors~\cite{2015MNRAS.446.3002B,2015MNRAS.453..849B}. The combined limits are presented with their 68\% and 95\% containment bands, and are expected to be close to the median limit when no signal is present. We observe agreement with the null hypothesis for all channels, within two standard deviations, between the observed limits and the expectations given by the median limits. Limits obtained from each detector are also indicated in the figures, where limits for all dSphs observed by the specific instrument have been combined. 

Below $\sim 300$~GeV, the limits obtained by \textit{Fermi}-LAT dominate for all annihilation channels. From $\sim 300$~GeV to $\sim 2$~TeV, \textit{Fermi}-LAT results continue to dominate for the hadronic ($b\bar{b}$ and $t\bar{t}$) and bosonic ($W^+W^-$ and $ZZ$) DM channels, yet the IACTs (H.E.S.S., MAGIC, and VERITAS) and \textit{Fermi}-LAT all contribute to the limit for leptonic ($e^+e^-$, $\mu^+\mu^-$, and $\tau^+\tau^-$) DM channels. For DM masses between $\sim 2$~TeV and $\sim 10$~TeV, the IACTs dominate leptonic DM annihilation channels, whereas both \textit{Fermi}-LAT and the IACTs contribute strongly to the gauge boson and the quark DM annihilation channels. From $\sim 10$~TeV to $\sim 100$~TeV, both the IACTs and HAWC contribute significantly to the lepton annihilation channel limit. For the quark and gauge boson annihilation channels, the IACTs and \textit{Fermi}-LAT both contribute strongly.

We note that there is a consistent and sizable under fluctuation at high masses in the limits we obtain in Fig.~\ref{fig:limits-bonnivard} with the $\mathcal{B}$ set of $J$-factors. It is due to the Ursa Major II dSph, whose $J$-factor increases by more than one order of magnitude compared to the $\mathcal{GS}$ set (see Table~\ref{tab:j-factor}), which makes it a dominant dSph for the combined limit. In particular in this range the MAGIC limit contributes significantly with 94.8 hours of observations while observing Ursa major II with a -2.1 $\sigma$ excess. We further note that the limits computed using the $\mathcal{B}$ set of $J$-factors are always more constraining compared to the ones calculated with the $\mathcal{GS}$ set. For the $W^+W^-$, $ZZ$, $b\bar{b}$, and $t\bar{t}$ channels, the ratio between the limits computed with the two sets of $J$-factors varies by factors of 3--5 depending on the DM particle mass, with the largest ratio around 10~TeV. For the channels $e^+e^-$, $\mu^+\mu^-$, and $\tau^+\tau^-$, the ratio lies between 2 and 6, peaking around 1~TeV. Examining Figs.~\ref{fig:comparison-j-factors-1} and~\ref{fig:comparison-j-factors-2} in Appendix~\ref{app:j-factors}, these differences are explained by the fact that the $\mathcal{B}$ set provides higher $J$-factors for the majority of the studied dSphs, with the notable exception of Segue~1~\cite{2016MNRAS.462..223B}. The variation on the ratio of the limits for the two sets is due to different dSphs dominating the limits at different DM particle masses. This comparison demonstrates the magnitude of systematic uncertainties associated with the choice of the $J$-factor calculation.

\section{Discussion and conclusions\label{sec:conclusions}}

In this multi-instrument analysis, we used observations of 20 dSphs by the gamma-ray instruments \textit{Fermi}-LAT, HAWC, H.E.S.S., MAGIC, and VERITAS to perform a collective search for DM annihilation signals. The data were combined across sources and detectors to increase the sensitivity of the search. We do not observe any significant deviation from the null hypothesis of no gamma-ray signal from DM. Therefore, we present our results in terms of upper limits on the thermally-averaged velocity-weighted annihilation cross section for seven potential DM annihilation channels.

\textit{Fermi}-LAT data results in the most stringent constraints for all channels considered in this work for DM masses below approximately 1~TeV. The other telescopes contribute significantly at higher DM masses. Overall, for multi-TeV DM masses, the combined DM constraints from all five telescopes are 2--3 times stronger than for any individual telescope for multi-TeV DM, except for the hadronic channels, where the improvement of the combined analysis appears for DM masses above $\sim 10$~TeV. Similarly, a detected DM signal in the multi-TeV domain would have been 2--3 times more significant in this study than in any single-instrument search.

Assuming the DM content of the dSphs is relatively well constrained, our results produce robust limits derived from observations of various dSphs. Our combined analysis improves the sensitivity over previously published results from each individual detector and provides the most stringent limits on DM annihilation from dSphs. Our results are based on deep exposures of promising known dSphs with the most sensitive gamma-ray instruments currently in operation. The limits obtained span the largest mass range of any WIMP DM search. Therefore, we believe our results could constitute a legacy of a generation of gamma-ray instruments on WIMP DM searches towards dSphs. Nevertheless, these results can still be gradually improved as more data is collected, in particular from the large field of view instruments that are \textit{Fermi}-LAT and HAWC for which the limits will decrease by at least one over the square root of total observation time, depending on the strength of the background and analysis approach~\cite{2024PhRvD.109f3024M}. Significant and more abrupt enhancements will require more sensitive instruments to be operational or new dSphs with significantly higher $J$-factors to be discovered.

While the limits shown extend to 100~TeV, the datasets used in this work are also suitable for setting limits on DM annihilation above 100~TeV, up to several PeV. However, a different input photon spectrum must be used in this mass range, reflecting the dominant particle physical processes at these energies~\cite{2021JHEP...06..121B}, and the allowed theoretical benchmarks in this mass range differ from the simple thermal WIMP scenario. We limited ourselves to the mass range that has been previously studied by all instruments; so far only VERITAS has published limits in this mass range using the same data~\cite{2024PhRvD.110f3034A}. We leave a dedicated effort on the $>100$~TeV mass range for future work.

This analysis additionally serves as a proof of concept for future multi-instrument and multi-messenger combination analyses. With this collaborative effort, we have managed to sample over four orders in magnitude in gamma-ray energies with distinct observational techniques. Determining the nature of DM continues to be an elusive and difficult problem. Larger datasets with diverse measurement techniques can be essential in the DM search. Our collaborative work can be extended in the future to include other gamma-ray detectors such as LHAASO~\cite{2021ChPhC..45b5002A}, CTA~\cite{2019scta.book.....C}, and SWGO~\cite{2019BAAS...51g.109H}. A future collaboration using similar techniques to the ones described in this paper could be applicable even beyond gamma rays. For example, the models we used for this study include annihilation channels with neutrinos in the final state. Advanced studies could aim to merge our results with those from neutrino observatories with large datasets such as IceCube~\cite{2017JInst..12P3012A}, ANTARES~\cite{2011NIMPA.656...11A}, and KM3NeT~\cite{2016JPhG...43h4001A}. Finally, an expansion of this analysis to include DM decay could also be considered.

Our analysis combines data from multiple instruments to produce strong constraints on astrophysical objects. From this perspective, these methods can be applied beyond DM searches. Almost every astrophysical study can benefit from multi-instrument, multi-wavelength gamma-ray studies, which make it possible to study the cosmos with greater precision and detail.

\acknowledgments

The \textit{Fermi}-LAT Collaboration acknowledges generous ongoing support from a number of agencies and institutes that have supported both the development and the operation of the LAT as well as scientific data analysis. These include the National Aeronautics and Space Administration and the Department of Energy in the United States, the Commissariat \`a l'Energie Atomique and the Centre National de la Recherche Scientifique / Institut National de Physique Nucl\'eaire et de Physique des Particules in France, the Agenzia Spaziale Italiana and the Istituto Nazionale di Fisica Nucleare in Italy, the Ministry of Education, Culture, Sports, Science and Technology (MEXT), High Energy Accelerator Research Organization (KEK) and Japan Aerospace Exploration Agency (JAXA) in Japan, and the K.~A.~Wallenberg Foundation, the Swedish Research Council and the Swedish National Space Board in Sweden. Additional support for science analysis during the operations phase is gratefully acknowledged from the Istituto Nazionale di Astrofisica in Italy and the Centre National d'\'Etudes Spatiales in France. This work performed in part under DOE Contract DE-AC02-76SF00515.

The HAWC collaboration acknowledges the support from: the US National Science Foundation (NSF); the US Department of Energy Office of High-Energy Physics; the Laboratory Directed Research and Development (LDRD) program of Los Alamos National Laboratory; Consejo Nacional de Ciencia y Tecnolog\'{i}a (CONACyT), M\'{e}xico, grants LNC-2023-117, 271051, 232656, 260378, 179588, 254964, 258865, 243290, 132197, A1-S-46288, A1-S-22784, CF-2023-I-645, c\'{a}tedras 873, 1563, 341, 323, Red HAWC, M\'{e}xico; DGAPA-UNAM grants IG101323, IN111716-3, IN111419, IA102019, IN106521, IN114924, IN110521 , IN102223; VIEP-BUAP; PIFI 2012, 2013, PROFOCIE 2014, 2015; the University of Wisconsin Alumni Research Foundation; the Institute of Geophysics, Planetary Physics, and Signatures at Los Alamos National Laboratory; Polish Science Centre grant, 2024/53/B/ST9/02671; Coordinaci\'{o}n de la Investigaci\'{o}n Cient\'{i}fica de la Universidad Michoacana; Royal Society - Newton Advanced Fellowship 180385; Gobierno de España and European Union -
NextGenerationEU, grant CNS2023- 144099; The Program Management Unit for Human Resources \& Institutional Development, Research and Innovation, NXPO (grant number B16F630069); Coordinaci\'{o}n General Acad\'{e}mica e Innovaci\'{o}n (CGAI-UdeG), PRODEP-SEP UDG-CA-499; Institute of Cosmic Ray Research (ICRR), University of Tokyo. H.F. acknowledges support by NASA under award number 80GSFC21M0002. C.R. acknowledges support from National Research Foundation of Korea (RS-2023-00280210). We also acknowledge the significant contributions over many years of Stefan Westerhoff, Gaurang Yodh and Arnulfo Zepeda Dom\'inguez, all deceased members of the HAWC collaboration. Thanks to Scott Delay, Luciano D\'{i}az and Eduardo Murrieta for technical support.

The support of the Namibian authorities and of the University of Namibia in facilitating the construction and operation of H.E.S.S. is gratefully acknowledged, as is the support by the German Ministry for Education and Research (BMBF), the Max Planck Society, the German Research Foundation (DFG), the Helmholtz Association, the Alexander von Humboldt Foundation, the French Ministry of Higher Education, Research and Innovation, the Centre National de la Recherche Scientifique (CNRS/IN2P3 and CNRS/INSU), the Commissariat à l’énergie atomique et aux énergies alternatives (CEA), the U.K. Science and Technology Facilities Council (STFC), the Irish Research Council (IRC) and the Science Foundation Ireland (SFI), the Knut and Alice Wallenberg Foundation, the Polish Ministry of Education and Science, agreement no. 2021/WK/06, the South African Department of Science and Technology and National Research Foundation, the University of Namibia, the National Commission on Research, Science \& Technology of Namibia (NCRST), the Austrian Federal Ministry of Education, Science and Research and the Austrian Science Fund (FWF), the Australian Research Council (ARC), the Japan Society for the Promotion of Science, the University of Amsterdam and the Science Committee of Armenia grant 21AG-1C085. We appreciate the excellent work of the technical support staff in Berlin, Zeuthen, Heidelberg, Palaiseau, Paris, Saclay, Tübingen and in Namibia in the construction and operation of the equipment. This work benefited from services provided by the H.E.S.S. Virtual Organisation, supported by the national resource providers of the EGI Federation.

The MAGIC collaboration would like to thank the Instituto de Astrof\'{\i}sica de Canarias for the excellent working conditions at the Observatorio del Roque de los Muchachos in La Palma. The financial support of the German BMBF, MPG and HGF; the Italian INFN and INAF; the Swiss National Fund SNF; the grants PID2019-104114RB-C31, PID2019-104114RB-C32, PID2019-104114RB-C33, PID2019-105510GB-C31, PID2019-107847RB-C41, PID2019-107847RB-C42, PID2019-107847RB-C44, PID2019-107988GB-C22, PID2022-136828NB-C41, PID2022-137810NB-C22, PID2022-138172NB-C41, PID2022-138172NB-C42, PID2022-138172NB-C43, PID2022-139117NB-C41, PID2022-139117NB-C42, PID2022-139117NB-C43, PID2022-139117NB-C44 funded by the Spanish MCIN/AEI/ 10.13039/501100011033 and “ERDF A way of making Europe”; the Indian Department of Atomic Energy; the Japanese ICRR, the University of Tokyo, JSPS, and MEXT; the Bulgarian Ministry of Education and Science, National RI Roadmap Project DO1-400/18.12.2020 and the Academy of Finland grant nr. 320045 is gratefully acknowledged. This work was also been supported by Centros de Excelencia ``Severo Ochoa'' y Unidades ``Mar\'{\i}a de Maeztu'' program of the Spanish MCIN/AEI/ 10.13039/501100011033 (CEX2019-000920-S, CEX2019-000918-M, CEX2021-001131-S) and by the CERCA institution and grants 2021SGR00426 and 2021SGR00773 of the Generalitat de Catalunya; by the Croatian Science Foundation (HrZZ) Project IP-2022-10-4595 and the University of Rijeka Project uniri-prirod-18-48; by the Deutsche Forschungsgemeinschaft (SFB1491) and by the Lamarr-Institute for Machine Learning and Artificial Intelligence; by the Polish Ministry Of Education and Science grant No. 2021/WK/08; and by the Brazilian MCTIC, CNPq and FAPERJ. This work was supported by the Grant RYC2021-032552-I funded by MCIN/AEI/10.13039/501100011033 and by the European Union NextGenerationEU/PRTR.

The research of VERITAS is supported by grants from the U.S. Department of Energy Office of Science, the U.S. National Science Foundation and the Smithsonian Institution, by NSERC in Canada, and by the Helmholtz Association (including the Young Investigators Program of the Helmholtz Association) in Germany. This research used resources provided by the Open Science Grid, which is supported by the National Science Foundation and the U.S. Department of Energy's Office of Science, and resources of the National Energy Research Scientific Computing Center (NERSC), a U.S. Department of Energy Office of Science User Facility operated under Contract No. DE-AC02-05CH11231. We acknowledge the excellent work of the technical support staff at the Fred Lawrence Whipple Observatory and at the collaborating institutions in the construction and operation of the instrument.

\bibliographystyle{JHEP}
\bibliography{main}

\providecommand{\href}[2]{#2}\begingroup\raggedright\begin{thebibliography}{10}

\bibitem{2020A&A...641A...6P}
{Planck Collaboration}, \emph{{Planck 2018 results. VI. Cosmological
  parameters}}, \href{https://doi.org/10.1051/0004-6361/201833910}{\emph{\aap}
  {\bfseries 641} (2020) A6}
  [\href{https://arxiv.org/abs/1807.06209}{{\ttfamily 1807.06209}}].

\bibitem{2019A&A...631A..40S}
S.~{Schuldt}, G.~{Chiriv{\`\i}}, S.~H. {Suyu}, A.~{Y{\i}ld{\i}r{\i}m},
  A.~{Sonnenfeld}, A.~{Halkola} et~al., \emph{{Inner dark matter distribution
  of the Cosmic Horseshoe (J1148+1930) with gravitational lensing and
  dynamics}}, \href{https://doi.org/10.1051/0004-6361/201935042}{\emph{\aap}
  {\bfseries 631} (2019) A40}
  [\href{https://arxiv.org/abs/1901.02896}{{\ttfamily 1901.02896}}].

\bibitem{2006Natur.440.1137S}
V.~{Springel}, C.~S. {Frenk} and S.~D.~M. {White}, \emph{{The large-scale
  structure of the Universe}},
  \href{https://doi.org/10.1038/nature04805}{\emph{\nat} {\bfseries 440} (2006)
  1137} [\href{https://arxiv.org/abs/astro-ph/0604561}{{\ttfamily
  astro-ph/0604561}}].

\bibitem{2016A&A...594A..13P}
{Planck Collaboration}, \emph{{Planck 2015 results. XIII. Cosmological
  parameters}}, \href{https://doi.org/10.1051/0004-6361/201525830}{\emph{\aap}
  {\bfseries 594} (2016) A13}
  [\href{https://arxiv.org/abs/1502.01589}{{\ttfamily 1502.01589}}].

\bibitem{1985ApJ...295..305V}
T.~S. {van Albada}, J.~N. {Bahcall}, K.~{Begeman} and R.~{Sancisi},
  \emph{{Distribution of dark matter in the spiral galaxy NGC 3198.}},
  \href{https://doi.org/10.1086/163375}{\emph{\apj} {\bfseries 295} (1985)
  305}.

\bibitem{2018RPPh...81f6201R}
L.~{Roszkowski}, E.~M. {Sessolo} and S.~{Trojanowski}, \emph{{WIMP dark matter
  candidates and searches{\textemdash}current status and future prospects}},
  \href{https://doi.org/10.1088/1361-6633/aab913}{\emph{Reports on Progress in
  Physics} {\bfseries 81} (2018) 066201}
  [\href{https://arxiv.org/abs/1707.06277}{{\ttfamily 1707.06277}}].

\bibitem{2018ipap.book.....D}
A.~{De Angelis} and M.~{Pimenta}, \emph{{Introduction to Particle and
  Astroparticle Physics}}, {Undergraduate Lecture Notes in Physics}. {Springer
  Nature}, 2018,
  \href{https://doi.org/10.1007/978-3-319-78181-5}{10.1007/978-3-319-78181-5}.

\bibitem{2014IJMPS..3060256S}
P.~{Salati}, \emph{{Dark matter annihilation in the universe}},  in
  \emph{International Journal of Modern Physics Conference Series}, vol.~30 of
  \emph{International Journal of Modern Physics Conference Series}, p.~1460256,
  May, 2014, \href{https://doi.org/10.1142/S2010194514602567}{DOI}
  [\href{https://arxiv.org/abs/1403.4495}{{\ttfamily 1403.4495}}].

\bibitem{2012PhRvD..86b3506S}
G.~{Steigman}, B.~{Dasgupta} and J.~F. {Beacom}, \emph{{Precise relic WIMP
  abundance and its impact on searches for dark matter annihilation}},
  \href{https://doi.org/10.1103/PhysRevD.86.023506}{\emph{\prd} {\bfseries 86}
  (2012) 023506} [\href{https://arxiv.org/abs/1204.3622}{{\ttfamily
  1204.3622}}].

\bibitem{2018RPPh...81e6901S}
L.~E. {Strigari}, \emph{{Dark matter in dwarf spheroidal galaxies and indirect
  detection: a review}},
  \href{https://doi.org/10.1088/1361-6633/aaae16}{\emph{Reports on Progress in
  Physics} {\bfseries 81} (2018) 056901}
  [\href{https://arxiv.org/abs/1805.05883}{{\ttfamily 1805.05883}}].

\bibitem{1998gaas.book.....B}
J.~{Binney} and M.~{Merrifield}, \emph{{Galactic Astronomy}}, {Princeton Series
  in Astrophysics}. {Princeton University Press}, 1998.

\bibitem{2008gady.book.....B}
J.~{Binney} and S.~{Tremaine}, \emph{{Galactic Dynamics: Second Edition}},
  {Princeton Series in Astrophysics}. {Princeton University Press}, 2008.

\bibitem{2014ApJ...795L...5S}
K.~{Spekkens}, N.~{Urbancic}, B.~S. {Mason}, B.~{Willman} and J.~E. {Aguirre},
  \emph{{The Dearth of Neutral Hydrogen in Galactic Dwarf Spheroidal
  Galaxies}}, \href{https://doi.org/10.1088/2041-8205/795/1/L5}{\emph{\apjl}
  {\bfseries 795} (2014) L5} [\href{https://arxiv.org/abs/1410.0028}{{\ttfamily
  1410.0028}}].

\bibitem{2004PhRvD..69l3501E}
N.~W. {Evans}, F.~{Ferrer} and S.~{Sarkar}, \emph{{A travel guide to the dark
  matter annihilation signal}},
  \href{https://doi.org/10.1103/PhysRevD.69.123501}{\emph{\prd} {\bfseries 69}
  (2004) 123501} [\href{https://arxiv.org/abs/astro-ph/0311145}{{\ttfamily
  astro-ph/0311145}}].

\bibitem{2016ApJ...832L...6W}
M.~{Winter}, G.~{Zaharijas}, K.~{Bechtol} and J.~{Vandenbroucke},
  \emph{{Estimating the GeV Emission of Millisecond Pulsars in Dwarf Spheroidal
  Galaxies}}, \href{https://doi.org/10.3847/2041-8205/832/1/L6}{\emph{\apjl}
  {\bfseries 832} (2016) L6}
  [\href{https://arxiv.org/abs/1607.06390}{{\ttfamily 1607.06390}}].

\bibitem{2016JCAP...02..039M}
{MAGIC Collaboration} et~al., \emph{{Limits to dark matter annihilation
  cross-section from a combined analysis of MAGIC and Fermi-LAT observations of
  dwarf satellite galaxies}},
  \href{https://doi.org/10.1088/1475-7516/2016/02/039}{\emph{\jcap} {\bfseries
  2016} (2016) 039} [\href{https://arxiv.org/abs/1601.06590}{{\ttfamily
  1601.06590}}].

\bibitem{2009ApJ...697.1071A}
{{\it Fermi}-LAT Collaboration}, \emph{{The Large Area Telescope on the Fermi
  Gamma-Ray Space Telescope Mission}},
  \href{https://doi.org/10.1088/0004-637X/697/2/1071}{\emph{\apj} {\bfseries
  697} (2009) 1071} [\href{https://arxiv.org/abs/0902.1089}{{\ttfamily
  0902.1089}}].

\bibitem{2012ApJS..203....4A}
{{\it Fermi}-LAT Collaboration}, \emph{{The {\it Fermi} Large Area Telescope on
  Orbit: Event Classification, Instrument Response Functions, and
  Calibration}}, \href{https://doi.org/10.1088/0067-0049/203/1/4}{\emph{\apjs}
  {\bfseries 203} (2012) 4} [\href{https://arxiv.org/abs/1206.1896}{{\ttfamily
  1206.1896}}].

\bibitem{2015PhRvL.115w1301A}
{{\it Fermi}-LAT Collaboration}, \emph{{Searching for Dark Matter Annihilation
  from Milky Way Dwarf Spheroidal Galaxies with Six Years of {\it Fermi} Large
  Area Telescope Data}},
  \href{https://doi.org/10.1103/PhysRevLett.115.231301}{\emph{\prl} {\bfseries
  115} (2015) 231301} [\href{https://arxiv.org/abs/1503.02641}{{\ttfamily
  1503.02641}}].

\bibitem{2013arXiv1303.3514A}
W.~{Atwood}, A.~{Albert}, L.~{Baldini}, M.~{Tinivella}, J.~{Bregeon},
  M.~{Pesce-Rollins} et~al., \emph{{Pass 8: Toward the Full Realization of the
  Fermi-LAT Scientific Potential}},
  \href{https://doi.org/10.48550/arXiv.1303.3514}{\emph{arXiv e-prints} (2013)
  arXiv:1303.3514} [\href{https://arxiv.org/abs/1303.3514}{{\ttfamily
  1303.3514}}].

\bibitem{2018arXiv181011394B}
P.~{Bruel}, T.~H. {Burnett}, S.~W. {Digel}, G.~{Johannesson}, N.~{Omodei} and
  M.~{Wood}, \emph{{Fermi-LAT improved Pass\raisebox{-0.5ex}\textasciitilde8
  event selection}},
  \href{https://doi.org/10.48550/arXiv.1810.11394}{\emph{arXiv e-prints} (2018)
  arXiv:1810.11394} [\href{https://arxiv.org/abs/1810.11394}{{\ttfamily
  1810.11394}}].

\bibitem{2015PhRvD..91l2002A}
{{\it Fermi}-LAT Collaboration}, \emph{{Updated search for spectral lines from
  Galactic dark matter interactions with pass 8 data from the $Fermi$ Large
  Area Telescope}},
  \href{https://doi.org/10.1103/PhysRevD.91.122002}{\emph{\prd} {\bfseries 91}
  (2015) 122002} [\href{https://arxiv.org/abs/1506.00013}{{\ttfamily
  1506.00013}}].

\bibitem{2020ApJS..247...33A}
{{\it Fermi}-LAT Collaboration}, \emph{{Fermi Large Area Telescope Fourth
  Source Catalog}},
  \href{https://doi.org/10.3847/1538-4365/ab6bcb}{\emph{\apjs} {\bfseries 247}
  (2020) 33} [\href{https://arxiv.org/abs/1902.10045}{{\ttfamily 1902.10045}}].

\bibitem{2017ApJ...834..110A}
{{\it Fermi}-LAT Collaboration} and {DES Collaboration}, \emph{{Searching for
  Dark Matter Annihilation in Recently Discovered Milky Way Satellites with
  Fermi-Lat}}, \href{https://doi.org/10.3847/1538-4357/834/2/110}{\emph{\apj}
  {\bfseries 834} (2017) 110}
  [\href{https://arxiv.org/abs/1611.03184}{{\ttfamily 1611.03184}}].

\bibitem{2021PhRvD.103l3005D}
M.~{Di Mauro} and M.~W. {Winkler}, \emph{{Multimessenger constraints on the
  dark matter interpretation of the \textit{Fermi}-LAT Galactic Center
  excess}}, \href{https://doi.org/10.1103/PhysRevD.103.123005}{\emph{\prd}
  {\bfseries 103} (2021) 123005}
  [\href{https://arxiv.org/abs/2101.11027}{{\ttfamily 2101.11027}}].

\bibitem{2020ApJ...905...76A}
{HAWC Collaboration}, \emph{{3HWC: The Third HAWC Catalog of Very-high-energy
  Gamma-Ray Sources}},
  \href{https://doi.org/10.3847/1538-4357/abc2d8}{\emph{\apj} {\bfseries 905}
  (2020) 76} [\href{https://arxiv.org/abs/2007.08582}{{\ttfamily 2007.08582}}].

\bibitem{2017ApJ...843...39A}
{HAWC Collaboration}, \emph{{Observation of the Crab Nebula with the HAWC
  Gamma-Ray Observatory}},
  \href{https://doi.org/10.3847/1538-4357/aa7555}{\emph{\apj} {\bfseries 843}
  (2017) 39} [\href{https://arxiv.org/abs/1701.01778}{{\ttfamily 1701.01778}}].

\bibitem{2015arXiv150708343V}
G.~{Vianello}, R.~J. {Lauer}, P.~{Younk}, L.~{Tibaldo}, J.~M. {Burgess},
  H.~{Ayala} et~al., \emph{{The Multi-Mission Maximum Likelihood framework
  (3ML)}}, \href{https://doi.org/10.48550/arXiv.1507.08343}{\emph{arXiv
  e-prints} (2015) arXiv:1507.08343}
  [\href{https://arxiv.org/abs/1507.08343}{{\ttfamily 1507.08343}}].

\bibitem{2018ApJ...853..154A}
{HAWC Collaboration}, \emph{{Dark Matter Limits from Dwarf Spheroidal Galaxies
  with the HAWC Gamma-Ray Observatory}},
  \href{https://doi.org/10.3847/1538-4357/aaa6d8}{\emph{\apj} {\bfseries 853}
  (2018) 154} [\href{https://arxiv.org/abs/1706.01277}{{\ttfamily
  1706.01277}}].

\bibitem{1994APh.....2..137F}
V.~P. {Fomin}, A.~A. {Stepanian}, R.~C. {Lamb}, D.~A. {Lewis}, M.~{Punch} and
  T.~C. {Weekes}, \emph{{New methods of atmospheric Cherenkov imaging for
  gamma-ray astronomy. I. The false source method}},
  \href{https://doi.org/10.1016/0927-6505(94)90036-1}{\emph{Astroparticle
  Physics} {\bfseries 2} (1994) 137}.

\bibitem{2006A&A...457..899A}
{H.E.S.S. Collaboration}, \emph{{Observations of the Crab nebula with HESS}},
  \href{https://doi.org/10.1051/0004-6361:20065351}{\emph{\aap} {\bfseries 457}
  (2006) 899} [\href{https://arxiv.org/abs/astro-ph/0607333}{{\ttfamily
  astro-ph/0607333}}].

\bibitem{2009APh....32..231D}
M.~{de Naurois} and L.~{Rolland}, \emph{{A high performance likelihood
  reconstruction of {\ensuremath{\gamma}}-rays for imaging atmospheric
  Cherenkov telescopes}},
  \href{https://doi.org/10.1016/j.astropartphys.2009.09.001}{\emph{Astroparticle
  Physics} {\bfseries 32} (2009) 231}
  [\href{https://arxiv.org/abs/0907.2610}{{\ttfamily 0907.2610}}].

\bibitem{2007A&A...466.1219B}
D.~{Berge}, S.~{Funk} and J.~{Hinton}, \emph{{Background modelling in
  very-high-energy {\ensuremath{\gamma}}-ray astronomy}},
  \href{https://doi.org/10.1051/0004-6361:20066674}{\emph{\aap} {\bfseries 466}
  (2007) 1219} [\href{https://arxiv.org/abs/astro-ph/0610959}{{\ttfamily
  astro-ph/0610959}}].

\bibitem{1983ApJ...272..317L}
T.-P. {Li} and Y.-Q. {Ma}, \emph{{Analysis methods for results in gamma-ray
  astronomy}}, \href{https://doi.org/10.1086/161295}{\emph{\apj} {\bfseries
  272} (1983) 317}.

\bibitem{2011APh....34..608H}
{H.E.S.S. Collaboration}, \emph{{H.E.S.S. constraints on dark matter
  annihilations towards the sculptor and carina dwarf galaxies}},
  \href{https://doi.org/10.1016/j.astropartphys.2010.12.006}{\emph{Astroparticle
  Physics} {\bfseries 34} (2011) 608}
  [\href{https://arxiv.org/abs/1012.5602}{{\ttfamily 1012.5602}}].

\bibitem{2014PhRvD..90k2012A}
{H.E.S.S. Collaboration}, \emph{{Search for dark matter annihilation signatures
  in H.E.S.S. observations of dwarf spheroidal galaxies}},
  \href{https://doi.org/10.1103/PhysRevD.90.112012}{\emph{\prd} {\bfseries 90}
  (2014) 112012} [\href{https://arxiv.org/abs/1410.2589}{{\ttfamily
  1410.2589}}].

\bibitem{2018JCAP...11..037A}
{H.E.S.S. Collaboration}, \emph{{Searches for gamma-ray lines and 'pure WIMP'
  spectra from Dark Matter annihilations in dwarf galaxies with H.E.S.S.}},
  \href{https://doi.org/10.1088/1475-7516/2018/11/037}{\emph{\jcap} {\bfseries
  2018} (2018) 037} [\href{https://arxiv.org/abs/1810.00995}{{\ttfamily
  1810.00995}}].

\bibitem{2014APh....56...26P}
R.~D. {Parsons} and J.~A. {Hinton}, \emph{{A Monte Carlo template based
  analysis for air-Cherenkov arrays}},
  \href{https://doi.org/10.1016/j.astropartphys.2014.03.002}{\emph{Astroparticle
  Physics} {\bfseries 56} (2014) 26}
  [\href{https://arxiv.org/abs/1403.2993}{{\ttfamily 1403.2993}}].

\bibitem{2016APh....72...76A}
{{MAGIC Collaboration}}, \emph{{The major upgrade of the MAGIC telescopes, Part
  II: A performance study using observations of the Crab Nebula}},
  \href{https://doi.org/10.1016/j.astropartphys.2015.02.005}{\emph{Astroparticle
  Physics} {\bfseries 72} (2016) 76}
  [\href{https://arxiv.org/abs/1409.5594}{{\ttfamily 1409.5594}}].

\bibitem{2014JCAP...02..008A}
{MAGIC Collaboration}, \emph{{Optimized dark matter searches in deep
  observations of Segue 1 with MAGIC}},
  \href{https://doi.org/10.1088/1475-7516/2014/02/008}{\emph{\jcap} {\bfseries
  2014} (2014) 008} [\href{https://arxiv.org/abs/1312.1535}{{\ttfamily
  1312.1535}}].

\bibitem{2018JCAP...03..009A}
{MAGIC Collaboration}, \emph{{Indirect dark matter searches in the dwarf
  satellite galaxy Ursa Major II with the MAGIC telescopes}},
  \href{https://doi.org/10.1088/1475-7516/2018/03/009}{\emph{\jcap} {\bfseries
  2018} (2018) 009} [\href{https://arxiv.org/abs/1712.03095}{{\ttfamily
  1712.03095}}].

\bibitem{2022PDU....3500912A}
{MAGIC Collaboration}, \emph{{Combined searches for dark matter in dwarf
  spheroidal galaxies observed with the MAGIC telescopes, including new data
  from Coma Berenices and Draco}},
  \href{https://doi.org/10.1016/j.dark.2021.100912}{\emph{Physics of the Dark
  Universe} {\bfseries 35} (2022) 100912}
  [\href{https://arxiv.org/abs/2111.15009}{{\ttfamily 2111.15009}}].

\bibitem{2016APh....72...61A}
{MAGIC Collaboration}, \emph{{The major upgrade of the MAGIC telescopes, Part
  I: The hardware improvements and the commissioning of the system}},
  \href{https://doi.org/10.1016/j.astropartphys.2015.04.004}{\emph{Astroparticle
  Physics} {\bfseries 72} (2016) 61}
  [\href{https://arxiv.org/abs/1409.6073}{{\ttfamily 1409.6073}}].

\bibitem{2013ICRC...33.2937Z}
R.~{Zanin}, E.~{Carmona}, J.~{Sitarek}, P.~{Colin}, K.~{Frantzen}, M.~{Gaug}
  et~al., \emph{{MARS, The MAGIC Analysis and Reconstruction Software}},  in
  \emph{International Cosmic Ray Conference}, vol.~33 of \emph{International
  Cosmic Ray Conference}, p.~2937, Jan., 2013.

\bibitem{2015ICRC...34..771P}
N.~{Park} and {VERITAS Collaboration}, \emph{{Performance of the VERITAS
  experiment}},  in \emph{34th International Cosmic Ray Conference (ICRC2015)},
  vol.~34 of \emph{International Cosmic Ray Conference}, p.~771, July, 2015,
  \href{https://doi.org/10.22323/1.236.0771}{DOI}
  [\href{https://arxiv.org/abs/1508.07070}{{\ttfamily 1508.07070}}].

\bibitem{2017PhRvD..95h2001A}
{{VERITAS Collaboration}}, \emph{{Dark matter constraints from a joint analysis
  of dwarf Spheroidal galaxy observations with VERITAS}},
  \href{https://doi.org/10.1103/PhysRevD.95.082001}{\emph{\prd} {\bfseries 95}
  (2017) 082001} [\href{https://arxiv.org/abs/1703.04937}{{\ttfamily
  1703.04937}}].

\bibitem{2013ICRC...33.2604Z}
B.~{Zitzer} and {VERITAS Collaboration}, \emph{{Dark Matter Annihilation Limits
  from Dwarf Galaxies using VERITAS}},  in \emph{International Cosmic Ray
  Conference}, vol.~33 of \emph{International Cosmic Ray Conference}, p.~2604,
  Jan., 2013, \href{https://arxiv.org/abs/1307.8367}{{\ttfamily 1307.8367}}.

\bibitem{2003A&A...410..389R}
G.~P. {Rowell}, \emph{{A new template background estimate for source searching
  in TeV gamma -ray astronomy}},
  \href{https://doi.org/10.1051/0004-6361:20031194}{\emph{\aap} {\bfseries 410}
  (2003) 389} [\href{https://arxiv.org/abs/astro-ph/0310025}{{\ttfamily
  astro-ph/0310025}}].

\bibitem{2008ICRC....3.1385C}
P.~{Cogan}, \emph{{VEGAS, the VERITAS Gamma-ray Analysis Suite}},  in
  \emph{International Cosmic Ray Conference}, vol.~3 of \emph{International
  Cosmic Ray Conference}, pp.~1385--1388, Jan., 2008,
  \href{https://doi.org/10.48550/arXiv.0709.4233}{DOI}
  [\href{https://arxiv.org/abs/0709.4233}{{\ttfamily 0709.4233}}].

\bibitem{2005PhR...405..279B}
G.~{Bertone}, D.~{Hooper} and J.~{Silk}, \emph{{Particle dark matter: evidence,
  candidates and constraints}},
  \href{https://doi.org/10.1016/j.physrep.2004.08.031}{\emph{\physrep}
  {\bfseries 405} (2005) 279}
  [\href{https://arxiv.org/abs/hep-ph/0404175}{{\ttfamily hep-ph/0404175}}].

\bibitem{2011JCAP...03..051C}
M.~{Cirelli}, G.~{Corcella}, A.~{Hektor}, G.~{H{\"u}tsi}, M.~{Kadastik},
  P.~{Panci} et~al., \emph{{PPPC 4 DM ID: a poor particle physicist cookbook
  for dark matter indirect detection}},
  \href{https://doi.org/10.1088/1475-7516/2011/03/051}{\emph{\jcap} {\bfseries
  2011} (2011) 051} [\href{https://arxiv.org/abs/1012.4515}{{\ttfamily
  1012.4515}}].

\bibitem{2015MNRAS.446.3002B}
V.~{Bonnivard}, C.~{Combet}, D.~{Maurin} and M.~G. {Walker}, \emph{{Spherical
  Jeans analysis for dark matter indirect detection in dwarf spheroidal
  galaxies - impact of physical parameters and triaxiality}},
  \href{https://doi.org/10.1093/mnras/stu2296}{\emph{\mnras} {\bfseries 446}
  (2015) 3002} [\href{https://arxiv.org/abs/1407.7822}{{\ttfamily 1407.7822}}].

\bibitem{2015ApJ...801...74G}
A.~{Geringer-Sameth}, S.~M. {Koushiappas} and M.~{Walker}, \emph{{Dwarf Galaxy
  Annihilation and Decay Emission Profiles for Dark Matter Experiments}},
  \href{https://doi.org/10.1088/0004-637X/801/2/74}{\emph{\apj} {\bfseries 801}
  (2015) 74} [\href{https://arxiv.org/abs/1408.0002}{{\ttfamily 1408.0002}}].

\bibitem{2015MNRAS.453..849B}
V.~{Bonnivard}, C.~{Combet}, M.~{Daniel}, S.~{Funk}, A.~{Geringer-Sameth},
  J.~A. {Hinton} et~al., \emph{{Dark matter annihilation and decay in dwarf
  spheroidal galaxies: the classical and ultrafaint dSphs}},
  \href{https://doi.org/10.1093/mnras/stv1601}{\emph{\mnras} {\bfseries 453}
  (2015) 849} [\href{https://arxiv.org/abs/1504.02048}{{\ttfamily
  1504.02048}}].

\bibitem{2020PhRvD.102b3029B}
K.~K. {Boddy}, J.~{Kumar}, A.~B. {Pace}, J.~{Runburg} and L.~E. {Strigari},
  \emph{{Effective J -factors for Milky Way dwarf spheroidal galaxies with
  velocity-dependent annihilation}},
  \href{https://doi.org/10.1103/PhysRevD.102.023029}{\emph{\prd} {\bfseries
  102} (2020) 023029} [\href{https://arxiv.org/abs/1909.13197}{{\ttfamily
  1909.13197}}].

\bibitem{2016MNRAS.462..223B}
V.~{Bonnivard}, D.~{Maurin} and M.~G. {Walker}, \emph{{Contamination of
  stellar-kinematic samples and uncertainty about dark matter annihilation
  profiles in ultrafaint dwarf galaxies: the example of Segue I}},
  \href{https://doi.org/10.1093/mnras/stw1691}{\emph{\mnras} {\bfseries 462}
  (2016) 223} [\href{https://arxiv.org/abs/1506.08209}{{\ttfamily
  1506.08209}}].

\bibitem{2020PhRvD.102f1302A}
S.~{Ando}, A.~{Geringer-Sameth}, N.~{Hiroshima}, S.~{Hoof}, R.~{Trotta} and
  M.~G. {Walker}, \emph{{Structure formation models weaken limits on WIMP dark
  matter from dwarf spheroidal galaxies}},
  \href{https://doi.org/10.1103/PhysRevD.102.061302}{\emph{\prd} {\bfseries
  102} (2020) 061302} [\href{https://arxiv.org/abs/2002.11956}{{\ttfamily
  2002.11956}}].

\bibitem{2024PhRvD.110f3034A}
{VERITAS Collaboration}, \emph{{Indirect search for dark matter with a combined
  analysis of dwarf spheroidal galaxies from VERITAS}},
  \href{https://doi.org/10.1103/PhysRevD.110.063034}{\emph{\prd} {\bfseries
  110} (2024) 063034} [\href{https://arxiv.org/abs/2407.16518}{{\ttfamily
  2407.16518}}].

\bibitem{2008APh....29...55A}
{H.E.S.S. Collaboration}, \emph{{Observations of the Sagittarius dwarf galaxy
  by the HESS experiment and search for a dark matter signal}},
  \href{https://doi.org/10.1016/j.astropartphys.2007.11.007}{\emph{Astroparticle
  Physics} {\bfseries 29} (2008) 55}
  [\href{https://arxiv.org/abs/0711.2369}{{\ttfamily 0711.2369}}].

\bibitem{2012ApJ...746...77V}
A.~{Viana}, M.~C. {Medina}, J.~{Pe{\~n}arrubia}, P.~{Brun}, J.~F.
  {Glicenstein}, K.~{Kosack} et~al., \emph{{Prospects for a Dark Matter
  Annihilation Signal toward the Sagittarius Dwarf Galaxy with Ground-based
  Cherenkov telescopes}},
  \href{https://doi.org/10.1088/0004-637X/746/1/77}{\emph{\apj} {\bfseries 746}
  (2012) 77} [\href{https://arxiv.org/abs/1103.2627}{{\ttfamily 1103.2627}}].

\bibitem{2020PhRvD.102f2001A}
{H.E.S.S. Collaboration}, \emph{{Search for dark matter signals towards a
  selection of recently detected DES dwarf galaxy satellites of the Milky Way
  with H.E.S.S.}},
  \href{https://doi.org/10.1103/PhysRevD.102.062001}{\emph{\prd} {\bfseries
  102} (2020) 062001} [\href{https://arxiv.org/abs/2008.00688}{{\ttfamily
  2008.00688}}].

\bibitem{2020PDU....2800529A}
{MAGIC Collaboration}, \emph{{A search for dark matter in Triangulum II with
  the MAGIC telescopes}},
  \href{https://doi.org/10.1016/j.dark.2020.100529}{\emph{Physics of the Dark
  Universe} {\bfseries 28} (2020) 100529}
  [\href{https://arxiv.org/abs/2003.05260}{{\ttfamily 2003.05260}}].

\bibitem{2015ApJ...814L...7K}
E.~N. {Kirby}, J.~G. {Cohen}, J.~D. {Simon} and P.~{Guhathakurta},
  \emph{{Triangulum II: Possibly a Very Dense Ultra-faint Dwarf Galaxy}},
  \href{https://doi.org/10.1088/2041-8205/814/1/L7}{\emph{\apjl} {\bfseries
  814} (2015) L7} [\href{https://arxiv.org/abs/1510.03856}{{\ttfamily
  1510.03856}}].

\bibitem{2017AJ....154..267C}
J.~L. {Carlin}, D.~J. {Sand}, R.~R. {Mu{\~n}oz}, K.~{Spekkens}, B.~{Willman},
  D.~{Crnojevi{\'c}} et~al., \emph{{Deep Subaru Hyper Suprime-Cam Observations
  of Milky Way Satellites Columba I and Triangulum II}},
  \href{https://doi.org/10.3847/1538-3881/aa94d0}{\emph{\aj} {\bfseries 154}
  (2017) 267} [\href{https://arxiv.org/abs/1710.06444}{{\ttfamily
  1710.06444}}].

\bibitem{2011EPJC...71.1554C}
G.~{Cowan}, K.~{Cranmer}, E.~{Gross} and O.~{Vitells}, \emph{{Asymptotic
  formulae for likelihood-based tests of new physics}},
  \href{https://doi.org/10.1140/epjc/s10052-011-1554-0}{\emph{European Physical
  Journal C} {\bfseries 71} (2011) 1554}
  [\href{https://arxiv.org/abs/1007.1727}{{\ttfamily 1007.1727}}].

\bibitem{rico_2022_7342721}
J.~Rico, C.~Nigro, D.~Kerszberg and T.~Miener, ``glike: numerical maximization
  of heterogeneous joint likelihood functions of a common free parameter plus
  nuisance parameters.''
  \href{https://doi.org/10.5281/zenodo.7342721}{10.5281/zenodo.7342721}, Nov.,
  2022.

\bibitem{tjark_miener_2021_4597500}
T.~Miener and D.~Nieto, ``Lklcom: Combining likelihoods from different
  experiments..''
  \href{https://doi.org/10.5281/zenodo.4597500}{10.5281/zenodo.4597500}, Mar.,
  2021.

\bibitem{2021arXiv211201818M}
T.~{Miener}, D.~{Kerszberg}, C.~{Nigro}, J.~{Rico} and D.~{Nieto},
  \emph{{Open-source Analysis Tools for Multi-instrument Dark Matter
  Searches}}, \href{https://doi.org/10.48550/arXiv.2112.01818}{\emph{arXiv
  e-prints} (2021) arXiv:2112.01818}
  [\href{https://arxiv.org/abs/2112.01818}{{\ttfamily 2112.01818}}].

\bibitem{2017ICRC...35..904Z}
B.~{Zitzer} and {VERITAS Collaboration}, \emph{{The VERITAS Dark Matter
  Program}},  in \emph{35th International Cosmic Ray Conference (ICRC2017)},
  vol.~301 of \emph{International Cosmic Ray Conference}, p.~904, July, 2017,
  \href{https://doi.org/10.22323/1.301.0904}{DOI}
  [\href{https://arxiv.org/abs/1708.07447}{{\ttfamily 1708.07447}}].

\bibitem{2020Galax...8...25R}
J.~{Rico}, \emph{{Gamma-Ray Dark Matter Searches in Milky Way
  Satellites{\textemdash}A Comparative Review of Data Analysis Methods and
  Current Results}},
  \href{https://doi.org/10.3390/galaxies8010025}{\emph{Galaxies} {\bfseries 8}
  (2020) 25} [\href{https://arxiv.org/abs/2003.13482}{{\ttfamily 2003.13482}}].

\bibitem{2024PhRvD.109f3024M}
A.~{McDaniel}, M.~{Ajello}, C.~M. {Karwin}, M.~{Di Mauro}, A.~{Drlica-Wagner}
  and M.~A. {S{\'a}nchez-Conde}, \emph{{Legacy analysis of dark matter
  annihilation from the Milky Way dwarf spheroidal galaxies with 14 years of
  Fermi -LAT data}},
  \href{https://doi.org/10.1103/PhysRevD.109.063024}{\emph{\prd} {\bfseries
  109} (2024) 063024} [\href{https://arxiv.org/abs/2311.04982}{{\ttfamily
  2311.04982}}].

\bibitem{2021JHEP...06..121B}
C.~W. {Bauer}, N.~L. {Rodd} and B.~R. {Webber}, \emph{{Dark matter spectra from
  the electroweak to the Planck scale}},
  \href{https://doi.org/10.1007/JHEP06(2021)121}{\emph{Journal of High Energy
  Physics} {\bfseries 2021} (2021) 121}
  [\href{https://arxiv.org/abs/2007.15001}{{\ttfamily 2007.15001}}].

\bibitem{2021ChPhC..45b5002A}
{LHAASO Collaboration}, \emph{{Observation of the Crab Nebula with LHAASO-KM2A
  - a performance study}},
  \href{https://doi.org/10.1088/1674-1137/abd01b}{\emph{Chinese Physics C}
  {\bfseries 45} (2021) 025002}
  [\href{https://arxiv.org/abs/2010.06205}{{\ttfamily 2010.06205}}].

\bibitem{2019scta.book.....C}
{Cherenkov Telescope Array Consortium}, \emph{{Science with the Cherenkov
  Telescope Array}}. World Scientific Publishing, 2019,
  \href{https://doi.org/10.1142/10986}{10.1142/10986},
  [\href{https://arxiv.org/abs/1709.07997}{{\ttfamily 1709.07997}}].

\bibitem{2019BAAS...51g.109H}
{SWGO Collaboration}, \emph{{The Southern Wide-Field Gamma-Ray Observatory
  (SWGO): A Next-Generation Ground-Based Survey Instrument}},  in
  \emph{Bulletin of the American Astronomical Society}, vol.~51, p.~109, Sept.,
  2019, \href{https://doi.org/10.48550/arXiv.1907.07737}{DOI}
  [\href{https://arxiv.org/abs/1907.07737}{{\ttfamily 1907.07737}}].

\bibitem{2017JInst..12P3012A}
{IceCube Collaboration}, \emph{{The IceCube Neutrino Observatory:
  instrumentation and online systems}},
  \href{https://doi.org/10.1088/1748-0221/12/03/P03012}{\emph{Journal of
  Instrumentation} {\bfseries 12} (2017) P03012}
  [\href{https://arxiv.org/abs/1612.05093}{{\ttfamily 1612.05093}}].

\bibitem{2011NIMPA.656...11A}
{ANTARES Collaboration}, \emph{{ANTARES: The first undersea neutrino
  telescope}}, \href{https://doi.org/10.1016/j.nima.2011.06.103}{\emph{Nuclear
  Instruments and Methods in Physics Research A} {\bfseries 656} (2011) 11}
  [\href{https://arxiv.org/abs/1104.1607}{{\ttfamily 1104.1607}}].

\bibitem{2016JPhG...43h4001A}
{KM3NeT Collaboration}, \emph{{Letter of intent for KM3NeT 2.0}},
  \href{https://doi.org/10.1088/0954-3899/43/8/084001}{\emph{Journal of Physics
  G Nuclear Physics} {\bfseries 43} (2016) 084001}
  [\href{https://arxiv.org/abs/1601.07459}{{\ttfamily 1601.07459}}].

\bibitem{1996MNRAS.278..488Z}
H.~{Zhao}, \emph{{Analytical models for galactic nuclei}},
  \href{https://doi.org/10.1093/mnras/278.2.488}{\emph{\mnras} {\bfseries 278}
  (1996) 488} [\href{https://arxiv.org/abs/astro-ph/9509122}{{\ttfamily
  astro-ph/9509122}}].

\bibitem{1911MNRAS..71..460P}
H.~C. {Plummer}, \emph{{On the problem of distribution in globular star
  clusters}}, \href{https://doi.org/10.1093/mnras/71.5.460}{\emph{\mnras}
  {\bfseries 71} (1911) 460}.

\bibitem{2014JCAP...02..023H}
D.~R. {Hunter}, \emph{{Derivation of the anisotropy profile, constraints on the
  local velocity dispersion, and implications for direct detection}},
  \href{https://doi.org/10.1088/1475-7516/2014/02/023}{\emph{\jcap} {\bfseries
  2014} (2014) 023} [\href{https://arxiv.org/abs/1311.0256}{{\ttfamily
  1311.0256}}].

\bibitem{2010MNRAS.405..340D}
B.~K. {Dhar} and L.~L.~R. {Williams}, \emph{{Surface mass density of the
  Einasto family of dark matter haloes: are they Sersic-like?}},
  \href{https://doi.org/10.1111/j.1365-2966.2010.16446.x}{\emph{\mnras}
  {\bfseries 405} (2010) 340}
  [\href{https://arxiv.org/abs/1112.3116}{{\ttfamily 1112.3116}}].

\bibitem{2007A&A...471..419B}
M.~{Baes} and E.~{van Hese}, \emph{{Dynamical models with a general anisotropy
  profile}}, \href{https://doi.org/10.1051/0004-6361:20077672}{\emph{\aap}
  {\bfseries 471} (2007) 419}
  [\href{https://arxiv.org/abs/0705.4109}{{\ttfamily 0705.4109}}].

\bibitem{2015MNRAS.448.2717W}
M.~G. {Walker}, E.~W. {Olszewski} and M.~{Mateo}, \emph{{Bayesian analysis of
  resolved stellar spectra: application to MMT/Hectochelle observations of the
  Draco dwarf spheroidal}},
  \href{https://doi.org/10.1093/mnras/stv099}{\emph{\mnras} {\bfseries 448}
  (2015) 2717} [\href{https://arxiv.org/abs/1503.02589}{{\ttfamily
  1503.02589}}].

\bibitem{2019MNRAS.482.3480P}
A.~B. {Pace} and L.~E. {Strigari}, \emph{{Scaling relations for dark matter
  annihilation and decay profiles in dwarf spheroidal galaxies}},
  \href{https://doi.org/10.1093/mnras/sty2839}{\emph{\mnras} {\bfseries 482}
  (2019) 3480} [\href{https://arxiv.org/abs/1802.06811}{{\ttfamily
  1802.06811}}].

\bibitem{2022PhRvD.106l3032D}
M.~{Di Mauro}, M.~{Stref} and F.~{Calore}, \emph{{Investigating the effect of
  Milky Way dwarf spheroidal galaxies extension on dark matter searches with
  Fermi-LAT data}},
  \href{https://doi.org/10.1103/PhysRevD.106.123032}{\emph{\prd} {\bfseries
  106} (2022) 123032} [\href{https://arxiv.org/abs/2212.06850}{{\ttfamily
  2212.06850}}].

\end{thebibliography}\endgroup

\newpage
\begin{appendix}

\section{dN/dE spectra\label{app:spectra}}

\begin{figure}[h]
\centering{
\includegraphics[width=1.0\textwidth]{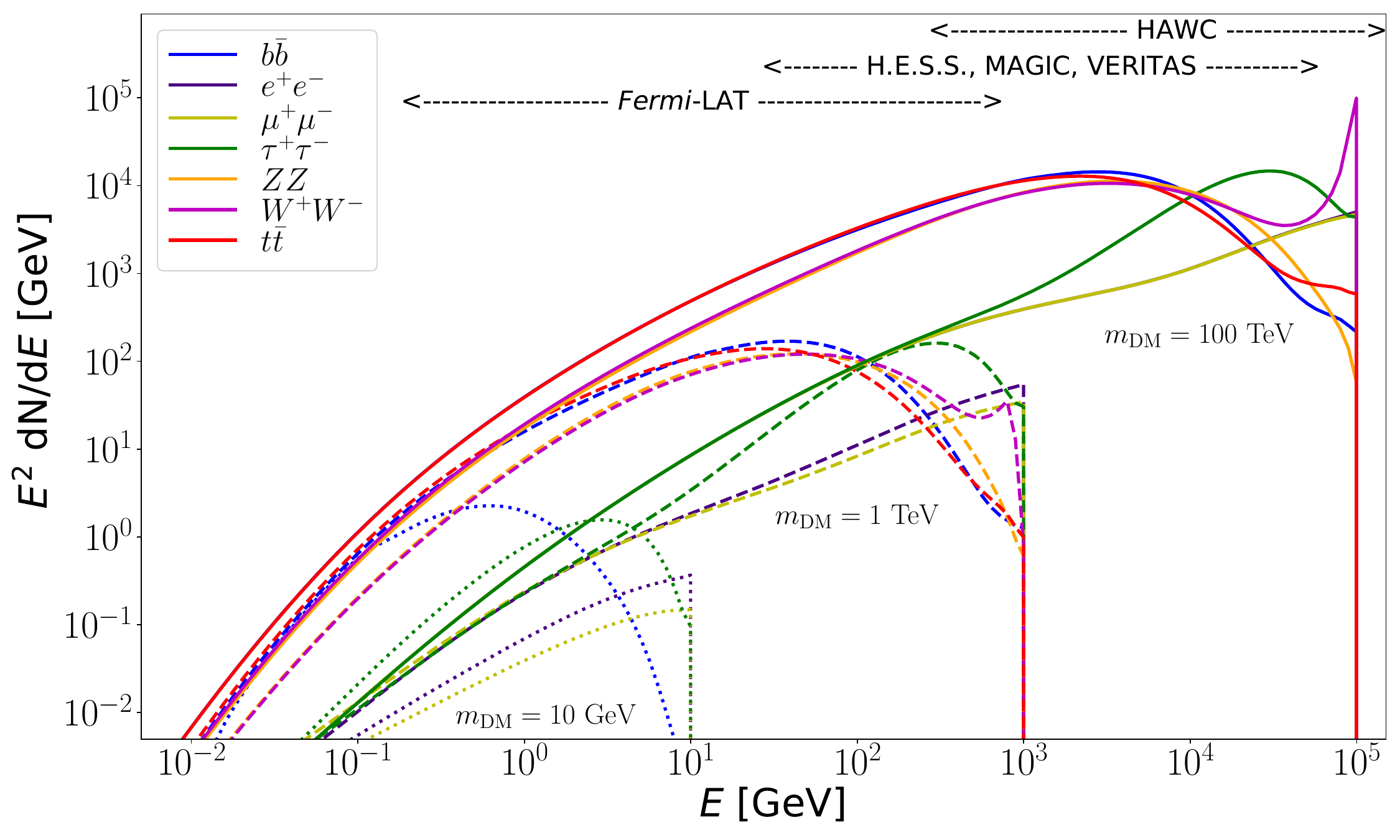}}
\caption{Differential photon yield per DM annihilation into SM pairs $ b\bar{b} $ (blue), $ e^{+}e^{-} $ (indigo), $ \mu^{+}\mu^{-} $ (light green), $ \tau^{+}\tau^{-} $ (dark green), $ ZZ $ (orange), $ W^{+}W^{-} $ (magenta), and $ t\bar{t} $ (red) for DM masses of 10 GeV (dotted), 1 TeV (dashed), and 100 TeV (solid). The approximate energy ranges of the gamma-ray instruments using in this work (see Section~\ref{sec:experiments}) are depicted at the top of the figure.}
\label{fig:spectra}
\end{figure}

We show in this appendix the differential photon yield per DM annihilation in Fig.~\ref{fig:spectra}, as computed by Cirelli et al.~\cite{2011JCAP...03..051C}. The computation of the photon yield includes electroweak corrections of the final state products. 

\section{\textit{J}-factor distributions\label{app:j-factors}}

We show in this appendix a comparison between the $J$-factors computed by Geringer-Sameth et al.~\cite{2015ApJ...801...74G} (the $\mathcal{GS}$ set) and the ones computed by Bonnivard et al.~\cite{2015MNRAS.446.3002B,2015MNRAS.453..849B} (the $\mathcal{B}$ set).

The $\mathcal{GS}$ $J$-factors are computed through a Jeans analysis of the kinematic stellar data of the selected dSphs, assuming a dynamic equilibrium and a spherical symmetry for the dSphs. In~\cite{2015ApJ...801...74G} the authors adopted the generalized DM density distribution, known as Zhao-Hernquist and introduced in~\cite{1996MNRAS.278..488Z}, which carries three additional index parameters to describe the inner and outer slopes, and the break of the density profile. Such a profile parametrization allows the reduction of the theoretical bias from the choice of a specific radial dependency on the kinematic data. In addition, a constant velocity anisotropy profile and a Plummer light profile~\cite{1911MNRAS..71..460P} for the stellar distribution were assumed. The velocity anisotropy profile depends on the radial and tangential velocity dispersions. However, its determination remains challenging since only the line-of-sight velocity dispersion can be derived from velocity measurements. Therefore, the parametrization of the anisotropy profile is obtained from simulated halos (see~\cite{2014JCAP...02..023H} for more details).

The $\mathcal{B}$ $J$-factors were computed through a Jeans analysis taking into account the systematic uncertainties induced by the DM profile parametrization, the radial velocity anisotropy profile, and the triaxiality of the halo of the dSphs. In~\cite{2015MNRAS.446.3002B,2015MNRAS.453..849B} the authors performed a study relying on different modelling choices and treatment of systematic uncertainties than $\mathcal{GS}$ for the determination of the $J$-factor. Conservative values of the $J$-factors were obtained using an Einasto DM density profile~\cite{2010MNRAS.405..340D}, a realistic anisotropy profile known as the Baes \& Van Hese profile~\cite{2007A&A...471..419B} which takes into account that the inner regions can be significantly non-isotropic, and a Zhao-Hernquist light profile~\cite{1996MNRAS.278..488Z} to fit the distribution of the luminous matter.

For both sets, $J$-factor values are provided for all dSphs as a function of the radius of the integration region~\cite{2015ApJ...801...74G,2015MNRAS.446.3002B,2015MNRAS.453..849B}. Table~\ref{tab:j-factor} shows the heliocentric distance and Galactic coordinates of the 20 dSphs, together with the two sets of $J$-factor values integrated up to the outermost observed star for $\mathcal{GS}$ and the tidal radius for $\mathcal{B}$. Both $J$-factor sets were derived through a Jeans analysis based on the same kinematic data, except for Draco, where the measurements of~\cite{2015MNRAS.448.2717W} have been adopted in the computation of the $\mathcal{B}$ value. The computations for producing the $\mathcal{GS}$ and $\mathcal{B}$ samples differ in the choice of the DM density, velocity anisotropy, and light profiles, for which the set $\mathcal{B}$ takes into account some sources of systematic uncertainties.

Figures~\ref{fig:comparison-j-factors-1} and~\ref{fig:comparison-j-factors-2} show the comparisons of the $J$-factor versus the angular radius for each of the 20 dSphs used in this study. The uncertainties provided by the authors are also indicated in the figures. As for Table~\ref{tab:j-factor}, for the $\mathcal{GS}$ set, the computation stops at the angular radius corresponding to the outermost observed star, while for the $\mathcal{B}$ set, the computation stops at the angular radius corresponding to the tidal radius. These $J$-factors are then translated into spatial templates for each experiment using the instrument point spread function. Note that the integral over $\Delta \Omega$ in Eq.~\ref{eq:signal-events} requires knowing the value of $\text{d}J/\text{d}\Omega$ for angular distance covering the whole DM halo, which is not the case in general for the $\mathcal{GS}$ sample as can be seen in Figs.~\ref{fig:comparison-j-factors-1} and~\ref{fig:comparison-j-factors-2}. This does not affect the computation for Cherenkov telescopes, since for all their observed dSph $\mathcal{GS}$ covers a range of angular distance above the size of the signal integration region. HAWC considers $\text{d}J/\text{d}\Omega=0$ for angular distances above the maximum ones provided by $\mathcal{GS}$, thus producing conservative upper limits on $\sv$. Finally, \textit{Fermi}-LAT has assumed the DM template to be point-like as for most of the dSphs the extension of the DM halos is smaller than the LAT PSF~\cite{2019MNRAS.482.3480P}. Ref.~\cite{2022PhRvD.106l3032D} investigated the effect of including the spatial extension of the DM template in the analysis, which could weaken the \textit{Fermi}-LAT upper limits by at most a factor $1.5-1.8$ depending on the annihilation channel.

\begin{figure}[p]
\centering{
\vspace{-2cm}
\includegraphics[scale=0.32]{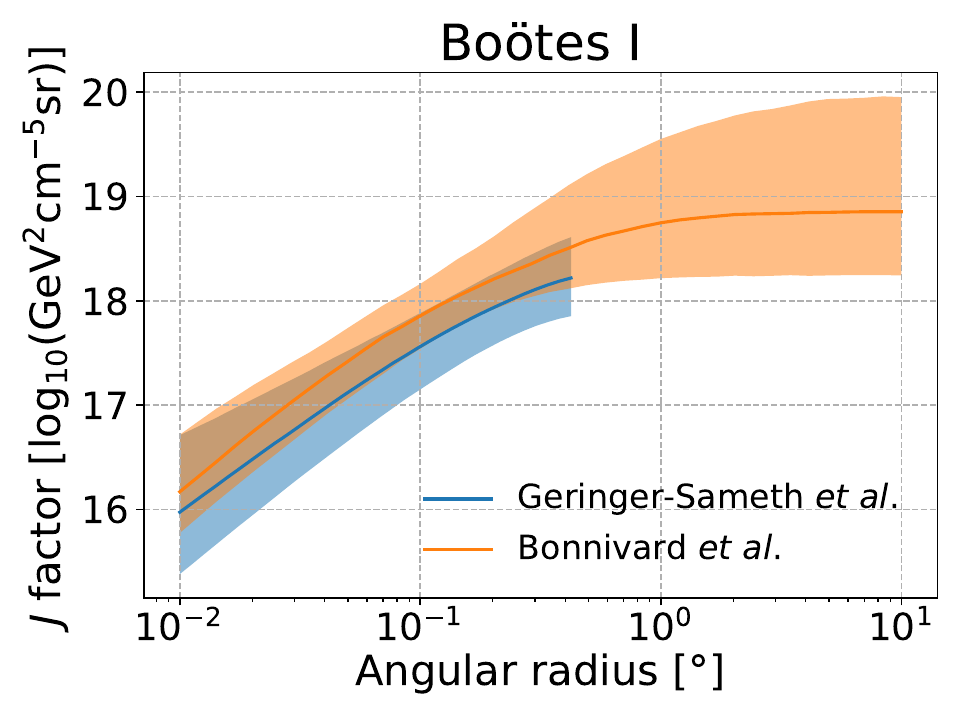}
\includegraphics[scale=0.32]{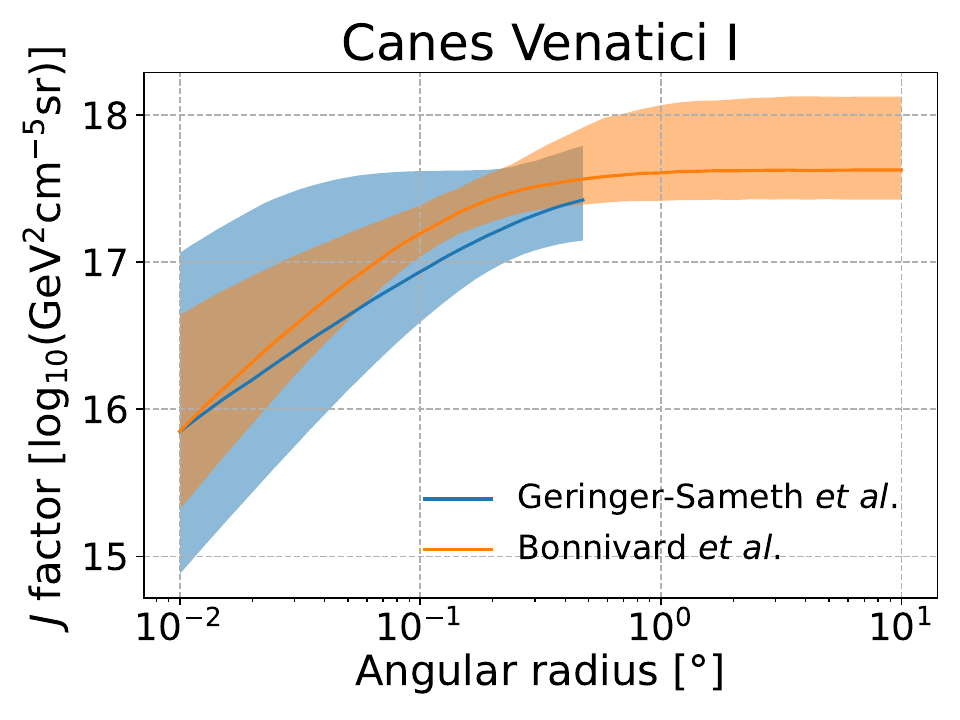}
\includegraphics[scale=0.32]{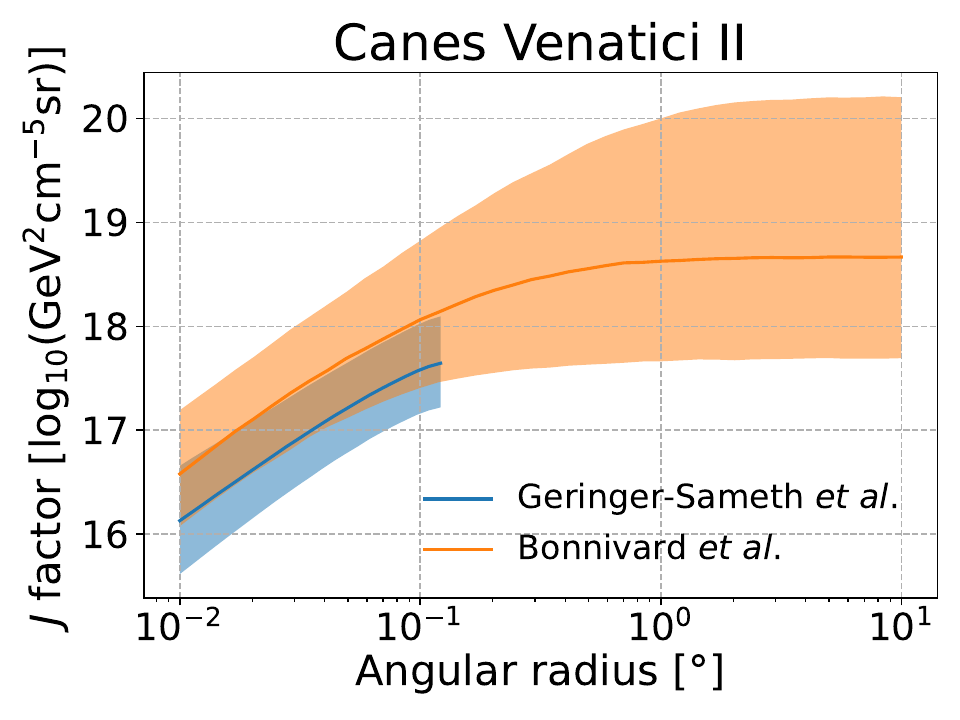}
\includegraphics[scale=0.32]{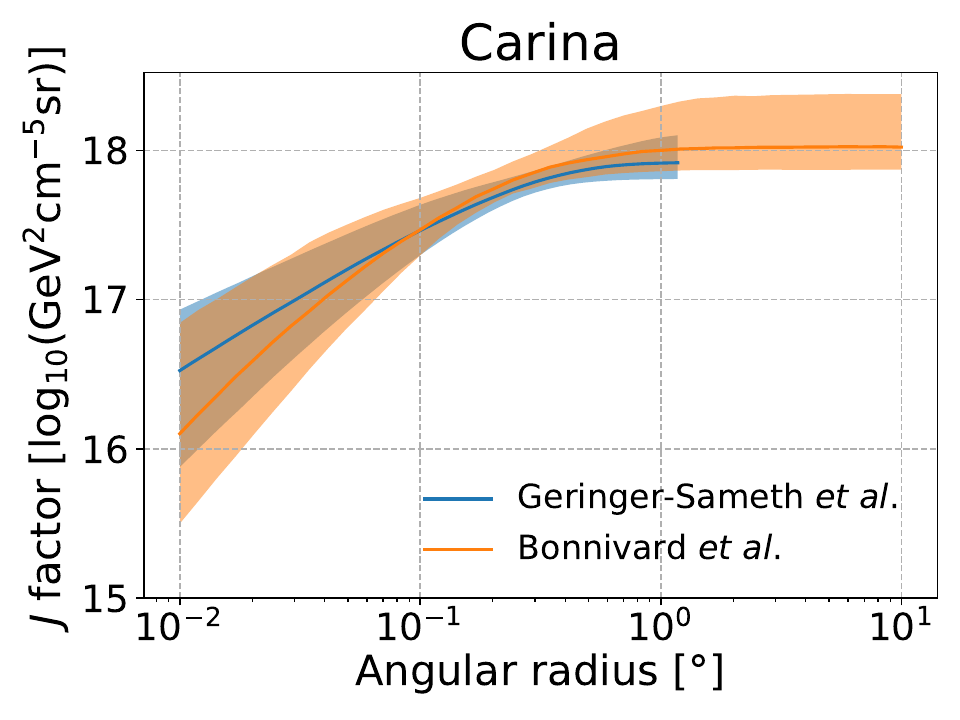}
\includegraphics[scale=0.32]{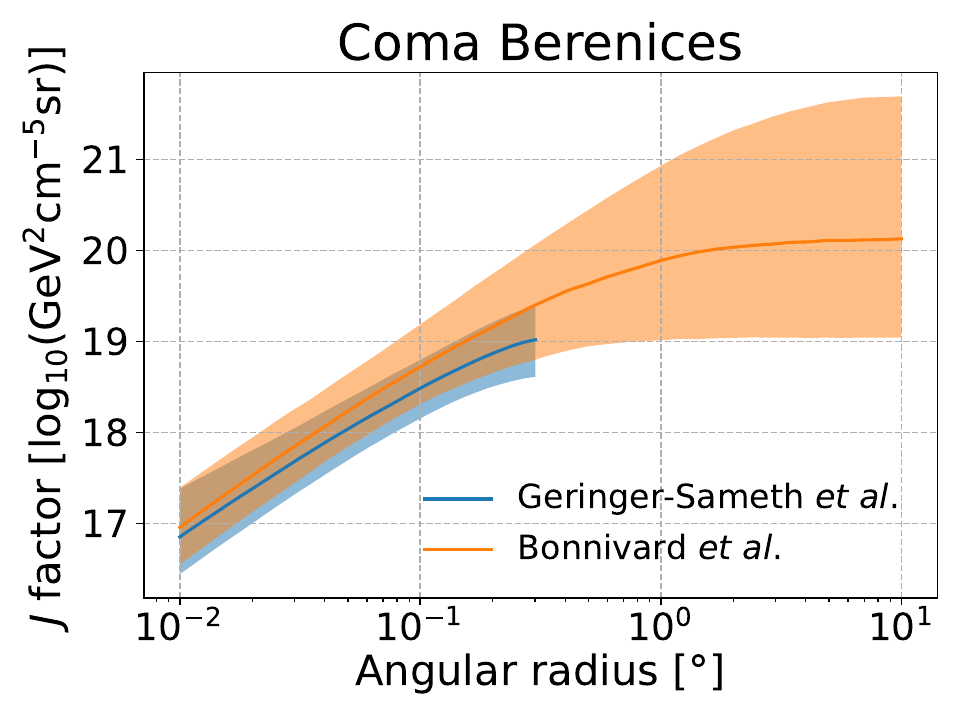}
\includegraphics[scale=0.32]{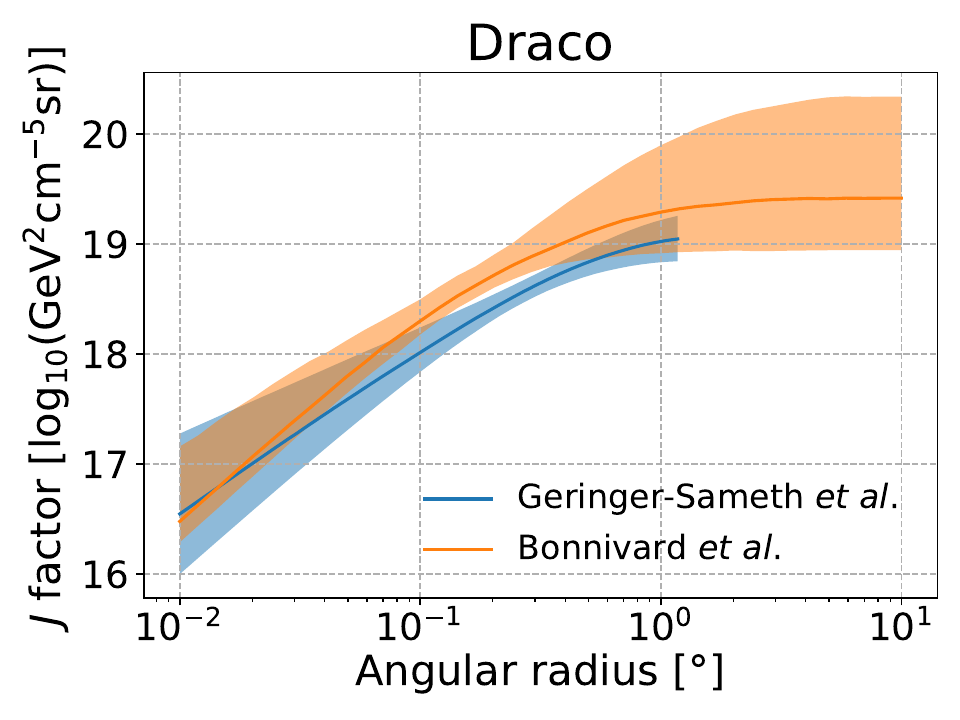}
\includegraphics[scale=0.32]{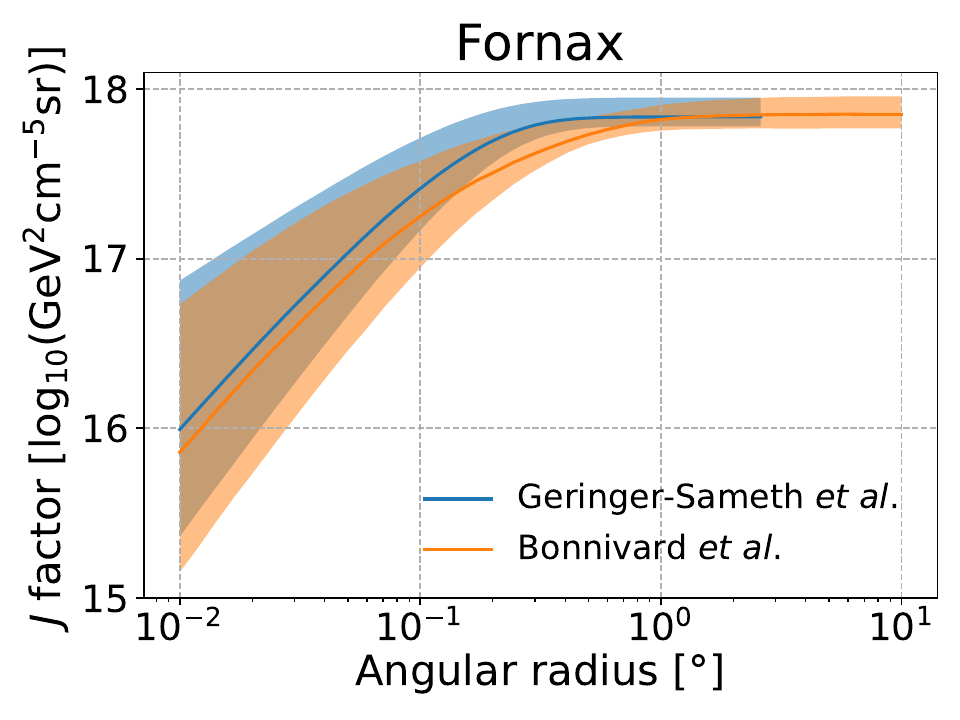}
\includegraphics[scale=0.32]{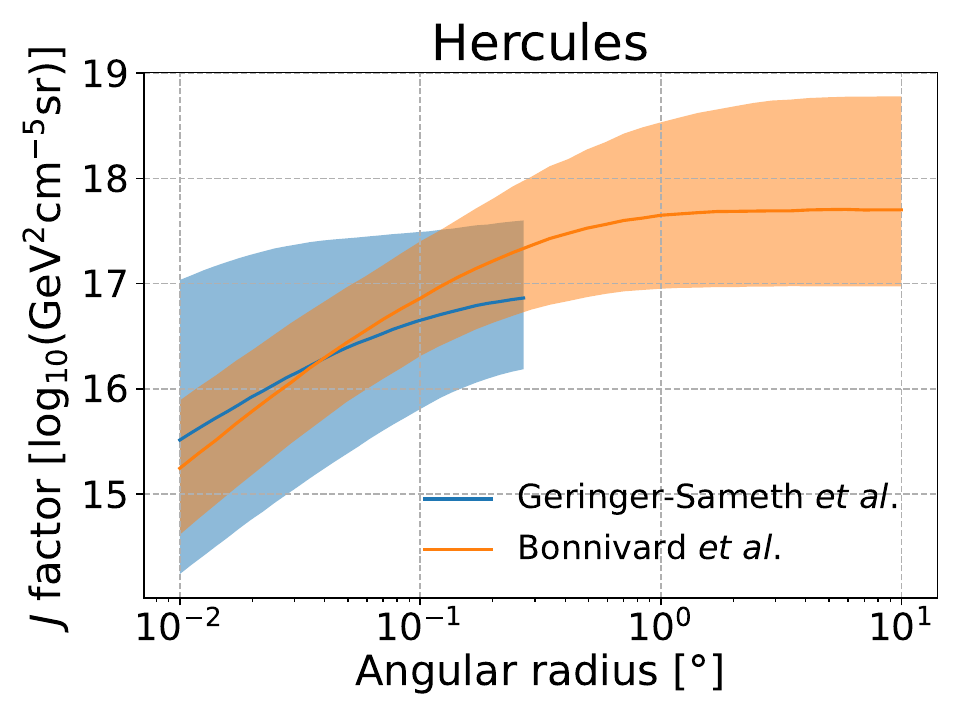}
\includegraphics[scale=0.32]{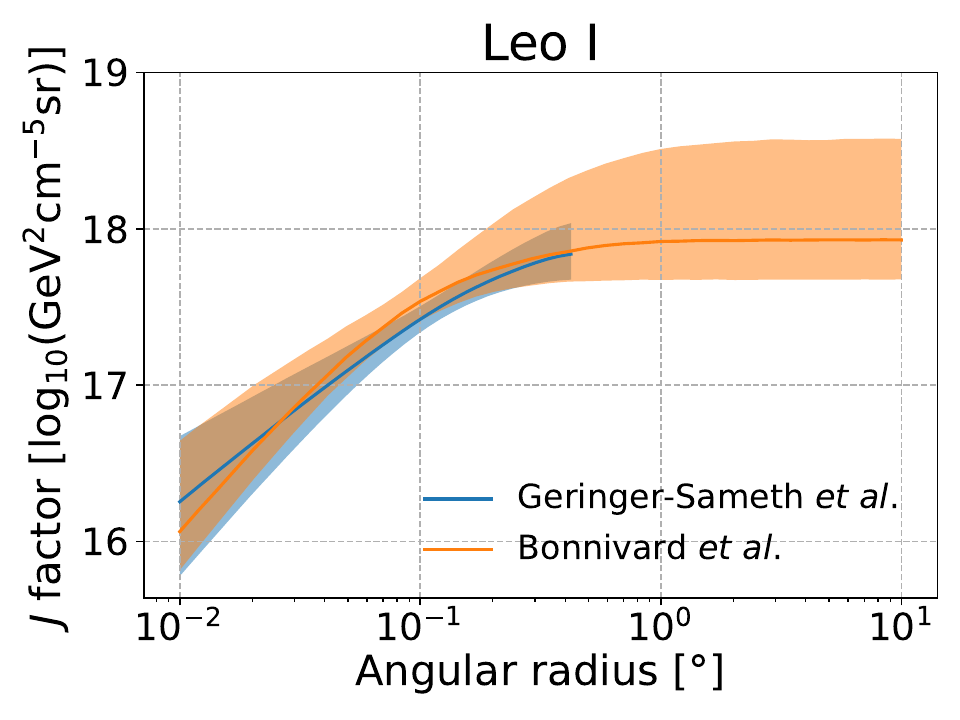}
\includegraphics[scale=0.32]{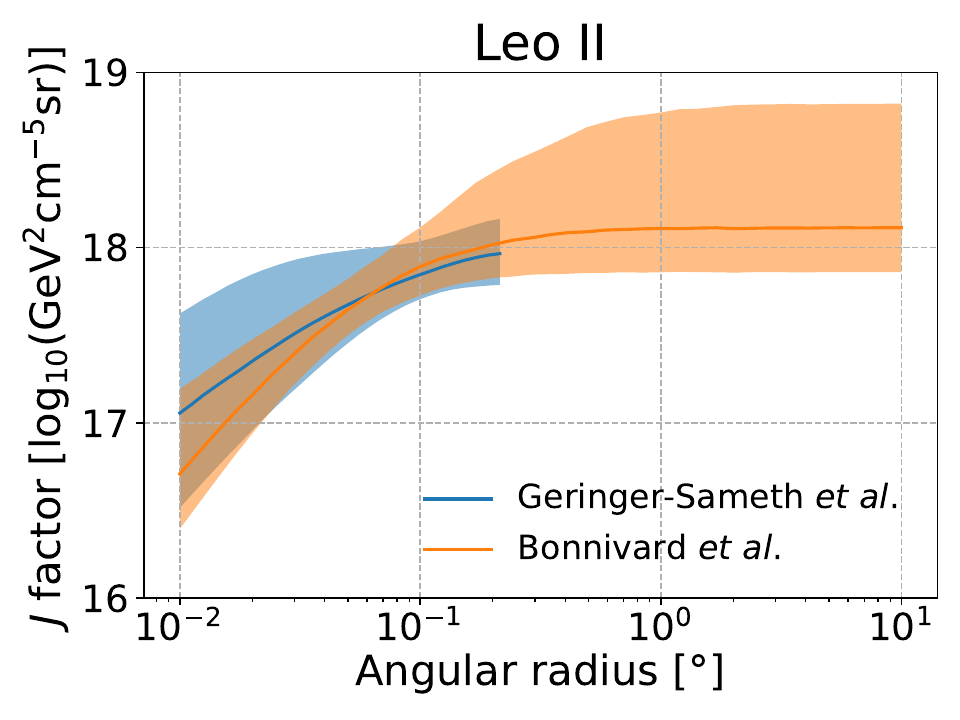}
}
\caption{Comparisons between the $J$-factors used for this study versus the angular radius. Computed $J$-factors from~\cite{2015ApJ...801...74G} ($\mathcal{GS}$ set in Table~\ref{tab:j-factor}) are extended up to the outermost visible star and are in blue. The $J$-factors computed from~\cite{2015MNRAS.446.3002B,2015MNRAS.453..849B} ($\mathcal{B}$ set in Table~\ref{tab:j-factor}) are extended up to the tidal radius of the dwarfs and are in orange. The solid lines represent the most probable value of the $J$-factors while the shaded regions correspond to the 1$\sigma$ standard deviation.}
\label{fig:comparison-j-factors-1}
\end{figure}

\begin{figure}[p]
\centering{
\includegraphics[scale=0.32]{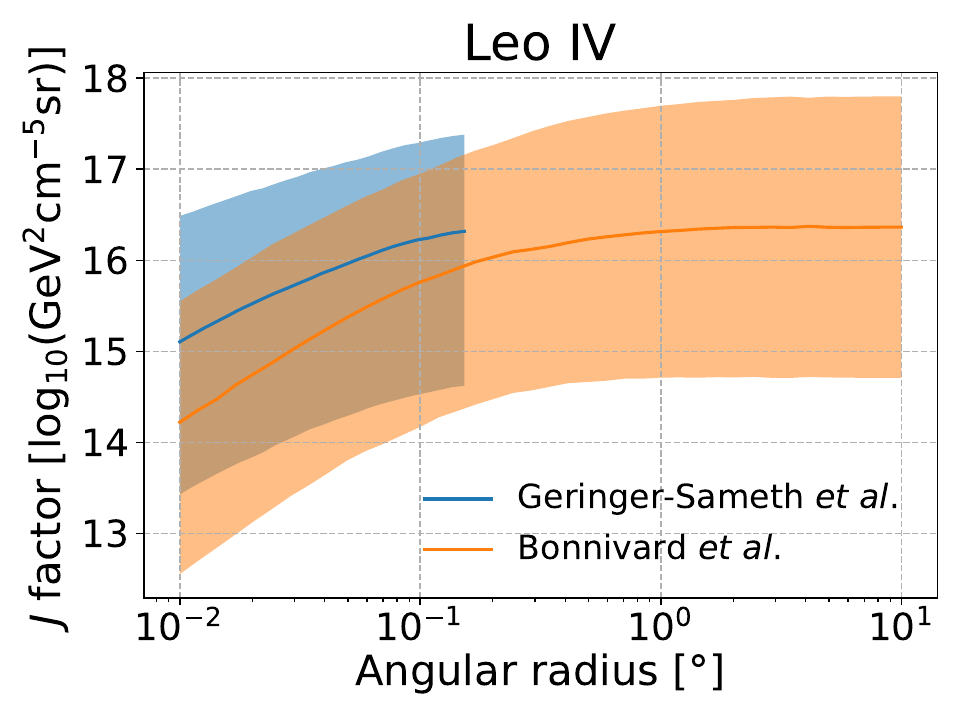}
\includegraphics[scale=0.32]{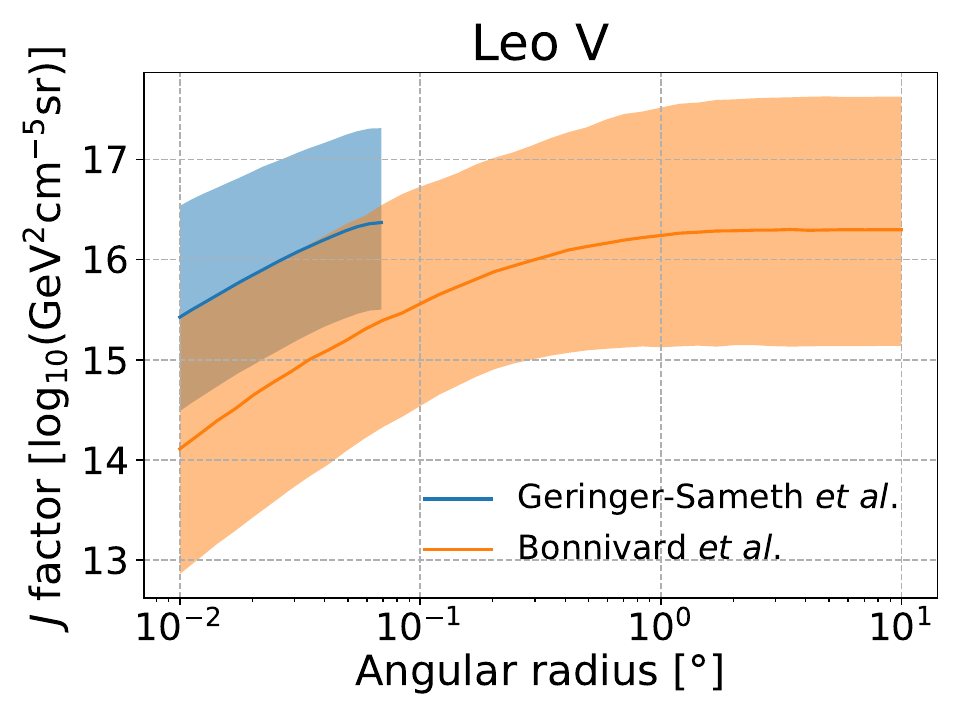}
\includegraphics[scale=0.32]{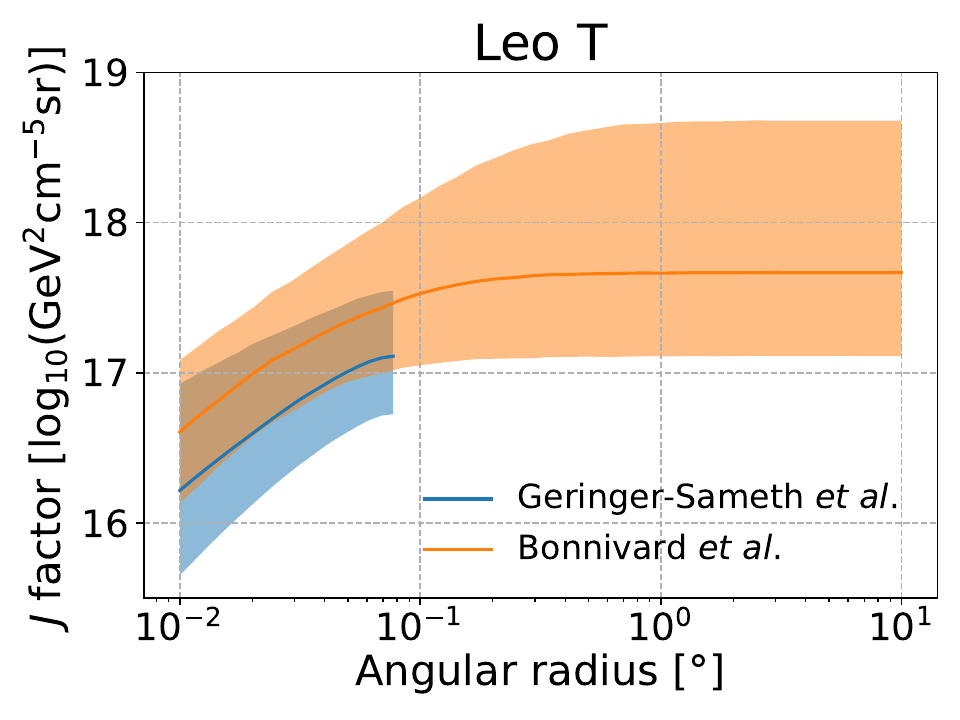}
\includegraphics[scale=0.32]{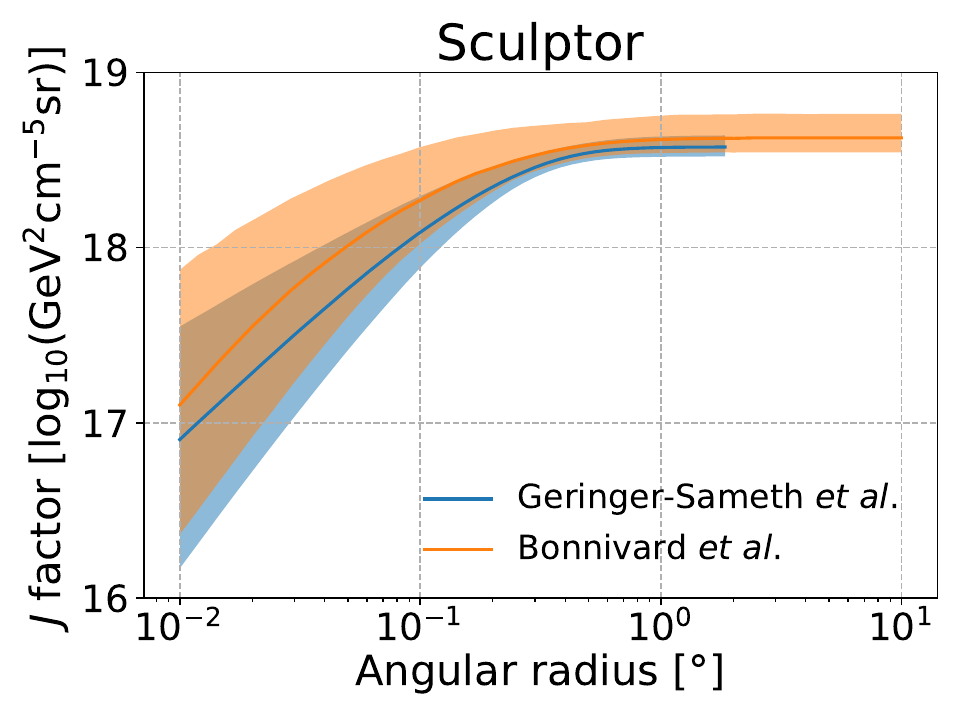}
\includegraphics[scale=0.32]{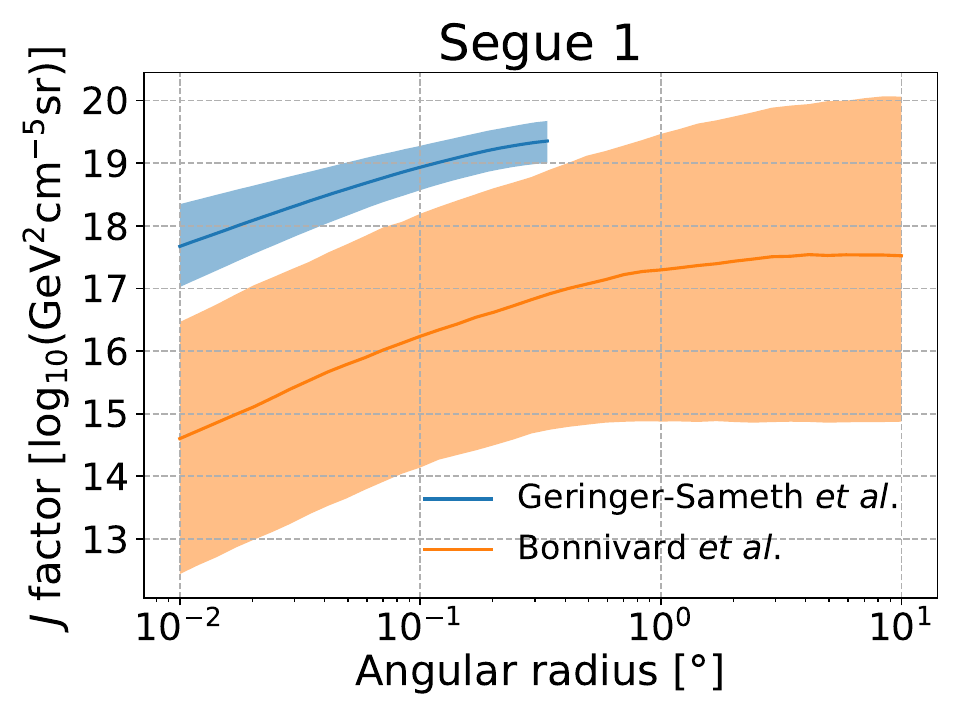}
\includegraphics[scale=0.32]{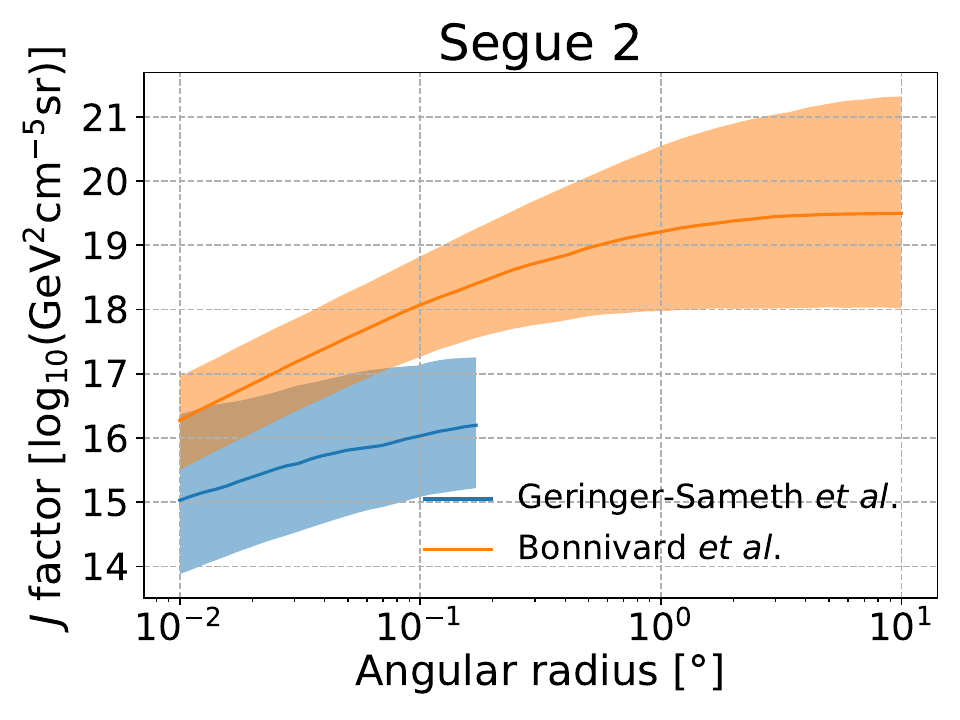}
\includegraphics[scale=0.32]{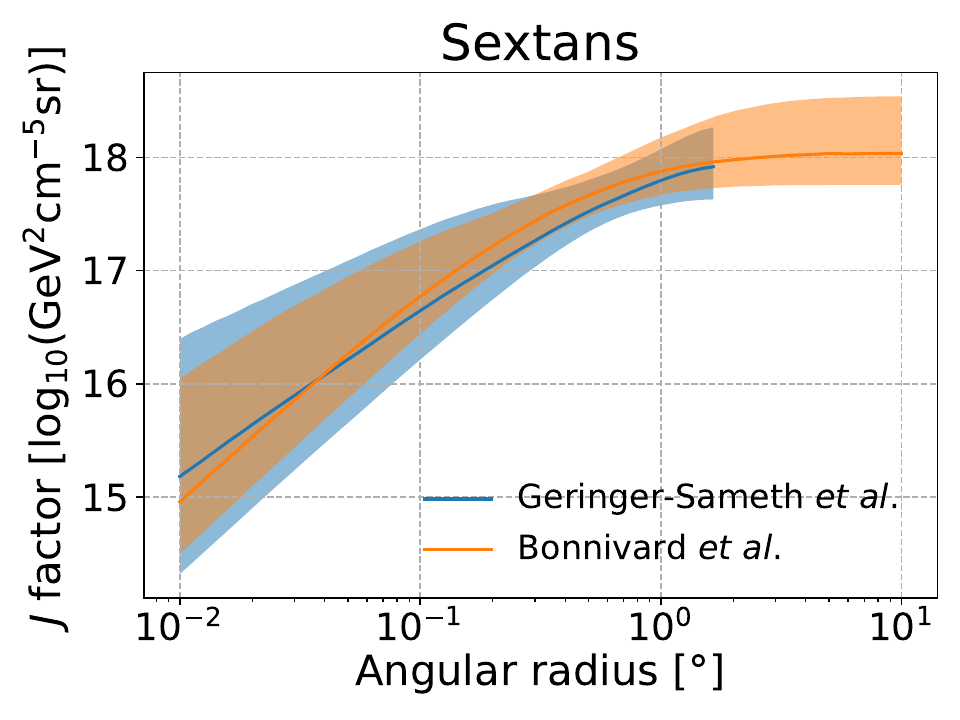}
\includegraphics[scale=0.32]{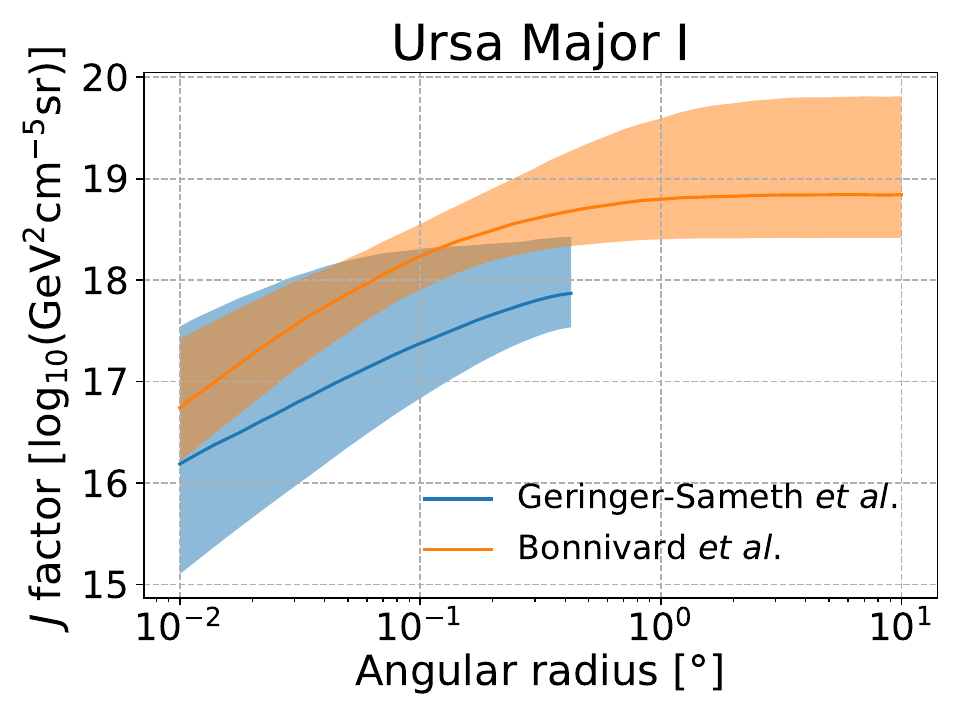}
\includegraphics[scale=0.32]{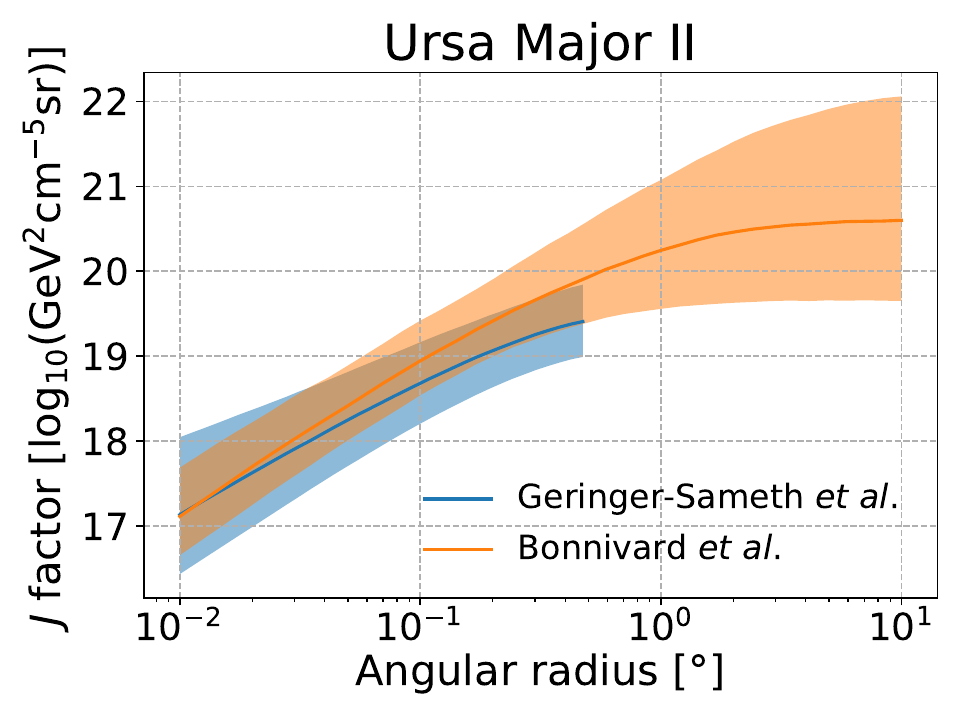}
\includegraphics[scale=0.32]{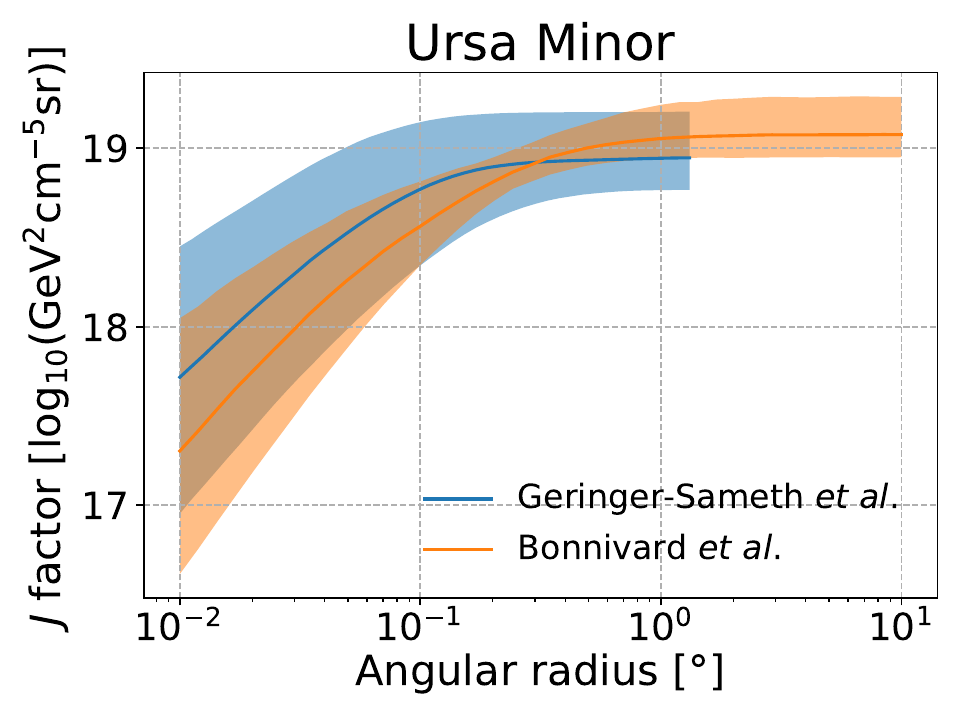}
}
\caption{Continuation of Fig.~\ref{fig:comparison-j-factors-1}.}
\label{fig:comparison-j-factors-2}
\end{figure}

\end{appendix}

\end{document}